%% file: dynamics.tex
\documentstyle[prl,aps,multicol,epsfig,pstricks,amsmath,amssymb]{revtex}
\tighten
\begin{document}
\newcommand{\onecolm}{
  \end{multicols}
  \vspace{-3.5ex}
  \noindent\rule{0.5\textwidth}{0.1ex}\rule{0.1ex}{2ex}\hfill
}
\newcommand{\twocolm}{
  \hfill\raisebox{-1.9ex}{\rule{0.1ex}{2ex}}\rule{0.5\textwidth}{0.1ex}
  \vspace{-4ex}
  \begin{multicols}{2}
}

\def\RR{{\rm
         \vrule width.04em height1.58ex depth-.0ex
         \kern-.04em R}}
\def\i{{\rm i}\,}
\def\trace{{\rm tr}\;}
\def\up{\uparrow}
\def\down{\downarrow}
\newcommand{\bigfrac}[2]{\mbox {${\displaystyle \frac{ #1 }{ #2 }}$}}
\newcommand{\Matrix}[2]{\left( \begin{array}{#1} #2 \end{array}
  \right)}

\newcommand{\bra}[1]{\left\langle #1 \right |}
\newcommand{\ket}[1]{\left | #1 \right\rangle}
\newcommand{\braket}[2]{\left\langle #1 | #2 \right\rangle}
\newcommand{\expect}[1]{\left\langle #1 \right\rangle}
\newcommand{\beq}{\begin{equation}}
\newcommand{\beqa}{\begin{eqnarray}}
\newcommand{\eeq}{\end{equation}}
\newcommand{\eeqa}{\end{eqnarray}}
\newcommand{\nbeqa}{\begin{eqnarray*}}
\newcommand{\neeqa}{\end{eqnarray*}}

\def\I {{\rm 1} \hspace{-1.1mm} {\rm I} \hspace{0.5mm}}

\title{Dynamics of Entanglement in One-Dimensional Spin Systems}
\author{Luigi Amico, Andreas Osterloh}
\address{MATISS-INFM $\&$ Dipartimento di Metodologie Fisiche e
    Chimiche (DMFCI), viale A. Doria 6, 95125 Catania, ITALY}
\author{Francesco Plastina, Rosario Fazio}
\address{NEST-INFM $\&$ Scuola Normale Superiore, I-56127 Pisa, ITALY}
\author{G. Massimo Palma}
\address{NEST-INFM $\&$ Dipartimento di Tecnologie dell'Informazione, Universit\`{a} degli studi di Milano
via Bramante 65, I-26013 Crema(CR), ITALY}

\maketitle

\begin{abstract}
We study the dynamics of quantum correlations in a class of
exactly solvable  Ising-type models. We analyze in particular the
time evolution of initial Bell states created in a fully polarized
background and on the ground state. We find that the pairwise entanglement
propagates with a velocity proportional to the reduced interaction 
for all the four Bell states.
Singlet-like states are favored during the propagation, in the sense that
triplet-like states change their character during the propagation
under certain circumstances.
Characteristic for the anisotropic models is the instantaneous
creation of pairwise entanglement from a fully polarized state;  
furthermore, the propagation of pairwise entanglement is suppressed 
in favor of a creation of different types of entanglement. 
The ``entanglement wave'' evolving from a Bell state on the ground state
turns out to be very localized in space-time.
Further support to a recently formulated conjecture on entanglement 
sharing is given.

\end{abstract}

\pacs{PACS numbers: }

\begin{multicols}{2}

\input{intro}

\input{themodels}

\input{concurrence}

\input{results}

\input{conclusions}

\acknowledgments
The authors would like to thank G. Falci and J. Siewert
for helpful discussions. This work was supported by the EU (IST-SQUBIT), 
RTN2-2001-00440, HPRN-CT-2002-00144.

\input{appendix2}

\begin{multicols}{2}

\end{multicols}

\end{document}

%% file: intro.tex
\section{Introduction}
A quantum mechanical system possesses additional correlations that do 
not have a classical counterpart. This phenomenon, called 
entanglement~\cite{Bell87}, is probably one of the most
astonishing features of quantum mechanics. Understanding the 
nature of these non-local correlations has been a central issue in the 
discussion of the foundation of quantum mechanics. More recently,
with the burst of interest in quantum information processing, entanglement 
has been identified as an ingredient for the speed-up in quantum 
computation and quantum communication protocols~\cite{Nielsen00} as 
compared with their classical counterparts. Therefore it is of crucial 
importance to be able to generate, manipulate and detect entangled states
and experimental efforts in this direction have been put forward.
Notable results have been obtained with photons~\cite{Zeilinger}, 
cavity QED systems~\cite{Haroche}, ion traps~\cite{Wineland}, 
and coupled quantum dots~\cite{Bayer}.
Also encouraged by the advances in the
field of nanoscience, there has been already a number of proposals to 
detect signatures of entanglement. 
Systems under current study are multiterminal mesoscopic 
devices~\cite{Loss} and Josephson junctions~\cite{Plastina}.
\\
In many-body systems correlated states naturally appear. 
In fact a research activity investigating 
entanglement in condensed matter systems has been emerging very recently.
Many aspects still need further investigation though. From a conceptual point 
of view, it is not always obvious for example, how 
to distill out the quantum part of the correlations since interaction between 
the subsystems is always on. 
Nevertheless many-body systems might become useful for the development of 
new computation schemes and/or communication protocols. 
As an example we mention the recent proposal by Bose~\cite{Bose01}
to use the spin dynamics in Heisenberg rings to transfer 
quantum states. It is conceiveable that along the same lines other 
quantum information tasks can be implemented as well.

An important motivation for us to study the interconnection between 
condensed matter and quantum information is to investigate whether it 
is possible to  better characterize condensed matter states by 
looking at the entanglement properties of their wavefunction.
To realize such a program, tools of quantum information theory can be 
applied~\cite{Preskill00}. Already a  number of interesting results in this 
direction have been obtained for systems of interacting spins.    
In the Heisenberg chain the maximization of the entanglement at 
zero temperature 
is related to the energy minimization \cite{Oconnors01} while at finite 
temperature violation of Bell's inequality can be directly related to the 
properties of the internal energy~\cite{Wang02}. It is known that 
Werner states~\cite{Werner89} can be generated in a one dimensional 
XY model~\cite{Wang01} and that temperature and/or magnetic 
field can increase the entanglement of the systems as shown for the 
Ising and Heisenberg model in Refs.~\cite{Arnesen01,Gunlycke01}.
The peculiar aspects of non-local correlations become particularly 
evident when many bodies behave collectively, a prominent example being
a system near a quantum phase transition~\cite{Sachdev00}. 
It was found that near a quantum phase transition entanglement 
can be classified in the framework of 
scaling theory~\cite{Osterloh02,Osborne02,Vidal02,Korepin03}, but
also profound differences between non-local quantum and 
classical correlations have been highlighted.
In a very recent paper the problem of decoherence in a 
near-critical one-dimensional system  was addressed~\cite{Khveshchenko03}.
\\
The analysis of the entanglement was not confined to spin systems. 
Attention has also been devoted to the BCS model~\cite{Zanardi02,Delgado02}, 
quantum Hall~\cite{Zeng02} and Boson systems~\cite{Hines02}.

The link between statistical mechanics and quantum information theory 
has been carried ahead in refining established methods 
for examining many body interacting systems. 
An example is the entanglement preserving Density Matrix Renormalization 
Group introduced in Ref.~\onlinecite{Osborne01}.
In this paper we are interested in the 
dynamics of entanglement in one-dimensional spin systems.
In condensed matter physics it is customary to analyze the  
behaviour of the system by creating an excitation 
in a given point and studying its time evolution. 
We follow this idea, creating a localized entangled state (typically a Bell 
state) and following the evolution of entanglement generated by 
the dynamics of the spin chain. 
There are a number of questions that can be addressed in this way.
In particular we would like to see if there is a well defined velocity 
of propagation of entanglement, how it is related to the  
propagation of the elementary excitations 
of the spin system and what are the time scales for 
damping the entanglement created at the initial time. 

The questions addressed in this paper may be relevant for the study of 
quantum limits to dynamical evolution~\cite{Giovannetti02}. 
Entanglement, indeed, allows to achieve the quantum mechanically 
accessible speed maximum
in the evolution of a composite system towards orthogonal states.
Finally, an important aspect we address in our analysis is
to discriminate pairwise from different types of entanglement.
This problem was quantified by Coffman, Kundu, and Wootters (CKW)
recently in terms of a conjecture~\cite{Coffman00}
on a measure for residual entanglement that connects the pairwise 
entanglement between a certain
spin and all others with the von Neumann entropy of this spin.
 
The paper is organized as follows: In the next 
section we will introduce the model Hamiltonian and describe briefly 
the techniques used to obtain the results (spectrum and 
correlation functions) discussed in the subsequent sections.
In section III we discuss the measures of one- and two-site entanglement 
of formation and recall the CKW conjecture on the residual entanglement. 
The results for the isotropic and anisotropic models are presented separately 
in section IV. We will consider different initial states and 
different parameter ranges of the chosen model. 
The final section is devoted to the conclusions. 
Some details of the calculation are presented in the appendices.

%% file: themodels.tex
\section{The model}
The system under consideration is a spin-1/2 ferromagnetic chain with
an exchange coupling $\lambda$ in a transverse
magnetic field of strength $h$.
The Hamiltonian is $H=h H_s$ with the dimensionless Hamilton operator
$H_s$ being
\begin{equation}
H_s=-\lambda \sum_{i=1}^N (1+\gamma)S^x_i S^x_{i+1}+
(1-\gamma)S^y_i S^y_{i+1} - \sum_{i=1}^N S^z_i
\label{model}
\end{equation}
where $S^a$ are the spin-$1/2$ matrices ($a=x,y,z$) and $N$ is
the number of sites. We assume periodic boundary conditions.
The anisotropy parameter $\gamma$ connects the quantum Ising model
for $\gamma =1$ with the isotropic XY model for $\gamma = 0$.
In the interval  $0<\gamma\le 1$ the model belongs to the Ising
universality class and for $ N =\infty$ it undergoes a quantum phase
transition at the critical value $\lambda_c=1$. The order parameter is
the magnetization in $x$-direction, $\langle
S^x\rangle $, which is different from zero for $\lambda >1$ and
vanishes at and below the transition.
On the contrary the magnetization along the $z$-direction,
$\langle S^z\rangle $, is different from zero for any value of $\lambda$.

This class of models can be diagonalized by means of the
Jordan-Wigner transformation~\cite{Lieb61,Pfeuty70,Mccoy70}  that maps spins
to one dimensional spinless fermions with creation and annihilation operators
$c^\dagger_l$ and $c^{}_l$. It is convenient to use the operators
$A_l\doteq c_l^\dagger + c^{}_l$, $B_l\doteq c_l^\dagger - c^{}_l$,
which fulfill the anticommutation rules
\begin{eqnarray}
\{A_l, A_m\}&=&-\{B_l, B_m\}=2 \delta_{lm} \; ,\nonumber \\
\{A_l, B_m\}&=&0 \; .
\label{A-B-algebra}
\end{eqnarray}
In terms of these operators the Jordan-Wigner transformation reads
\begin{eqnarray}
S_l^x&=&\frac{1}{2}A_l \prod_{s=1}^{l-1} A_s B_s \nonumber \\
S_l^y&=&-\frac{i}{2}B_l \prod_{s=1}^{l-1} A_s B_s \nonumber \\
S_l^z&=&-\frac{1}{2} A_l B_l \;.
\label{jordanwigner}
\end{eqnarray}
The Hamiltonian defined in Eq.(\ref{model}) is bilinear in the
fermionic degrees of freedom and therefore can be diagonalized
by means of the transformation
\begin{equation}
\eta_k=\frac{1}{\sqrt{N}}\sum_l e^{ikl}\left (\alpha_k c^{}_l+i \beta_k c_l^\dagger \right )
\label{eta}
\end{equation}
with coefficients
\begin{eqnarray}
\alpha_k &=& \frac{\Lambda_k-(1+\lambda \cos k)}{\sqrt{2 [\Lambda_k^2
-(1+\lambda \cos k) \Lambda_k ]}} \nonumber \\
\beta_k &=& \frac{\gamma \lambda \sin k}{\sqrt{2 [\Lambda_k^2
-(1+\lambda \cos k) \Lambda_k ]}} \; .
\end{eqnarray}
The Hamiltonian thereafter assumes the form
\begin{equation}
H=\sum_k \Lambda_k \eta_k^\dagger \eta_k
- \bigfrac{1}{2}\sum_k \Lambda_k
\end{equation}
and the associated energy spectrum is
$$\
\Lambda_k= \sqrt{ \left (1+\lambda \cos k\right )^2 +
\lambda^2 \gamma^2 \sin^2 k}\; .
$$
The time evolution of the original spin operators
(or equivalently of the spinless fermions
introduced in Eq.(\ref{jordanwigner})) are the key to
determine the time evolution of the entanglement
and can be found by means of the inverse of Eq.(\ref{eta})
$$
c^{}_j(t)= \sum_l [\tilde{a}_{l-j}(t) c^{}_l - \tilde{b}_{l-j}(t)c^\dagger_l ]
$$
where the new coefficients are
\begin{eqnarray}
\tilde{a}_{x}(t) &=& \frac{1}{\sqrt{N}}\sum_{k} \cos{k x}
    \left ( e^{i \Lambda_k t}-2i \beta^2_k \sin \Lambda_k t\right ) \\
\tilde{b}_{x}(t) &=& \frac{2\i}{\sqrt{N}}\sum_{k} \sin{k x}\,
    \alpha_k \beta_k \sin \Lambda_k t \;\; .
\end{eqnarray}

In the limit $\gamma =0$ the previous expressions
simplify considerably. In this case the magnetization, i.e. the
$z$-component of the total spin $S^z=\sum_j S^z_j$, is a conserved
quantity. In terms of fermions this corresponds to the
conservation of the total number of particles, $N=\sum_j
n_j=\sum_j c^\dagger_j c^{}_j$. For $\gamma\longrightarrow 0$ and
$|\lambda|\leq 1$ we find that $\alpha_k\longrightarrow 0$ and
$\beta_k\longrightarrow {\rm sign}\,k$. The energy spectrum is
$\Lambda_k=\left|1+\lambda \cos k\right|$ and the eigenstates are
plane waves (the Hamiltonian corresponds to a tight binding model)
\beq \label{c(t):gamma0} c^{}_j(t) =
\frac{1}{\sqrt{N}}\sum_{k}\sum_{l} \cos{k(l-j)} e^{-i \Lambda_k t}
c^{}_l \eeq
\begin{equation}
\eta^\dagger_k=\frac{1}{\sqrt{N}}\sum_l e^{-ikl} c^{}_l\; .
\label{eta:gamma0}
\end{equation}

\subsection{Correlation functions}

As specified within the next section, one- and two site- entanglement
measures are obtained from the (one- and two-body) reduced density matrix
whose entries can be related to various spin correlation functions
\begin{eqnarray}
M^{\alpha}_l(t) &=&\langle \psi | S_l^\alpha(t) |\psi  \rangle \\
g^{\alpha\beta}_{lm}(t) &=&\langle \psi | S_l^\alpha(t) S_m^\beta(t)| \psi  \rangle \; .
\label{correlators}
\end{eqnarray}
These can be recast in the form of
Pfaffians~\cite{Mccoy70,Caianello52}.
Correlators defined in Eq.(\ref{correlators}) have been calculated
for this class of models in the case of thermal
equilibrium~\cite{Lieb61,Pfeuty70,Mccoy70,footnote1}. In this
case, the expression for the correlators reduces to the
calculation of Toeplitz determinants (i.e. determinants, whose
entries depend only on the difference of its row and column
number). In the present work we are interested
in calculating averages and correlation functions in the chain as
a function of time for a given initial state $| \psi  \rangle$ at
time $t_0=0$, which is not an eigenstate of the Hamiltonian of
Eq.(\ref{model}). Therefore, the Pfaffians~\cite{footnote2} do not
reduce to Toeplitz determinants. Since the Hamiltonian is
time-independent all the observables depend on the difference
$t-t_0=t$. Details of the calculations are reported in Ref.~\cite{AOXY}.

In order to grasp the properties associated to the dynamics of
entanglement, we will always imagine
that an initial configuration is prepared in which the
entanglement is concentrated between two sites in the chain. The
simplest situation to imagine is that two sites are in a Bell
state and the rest of the chain is factorized with all spins in
the state $|\downarrow \rangle $ (or $|\uparrow \rangle $) 
\begin{eqnarray}
{ \phantom{c} \atop | \Psi_{i,j}^\varphi \rangle \equiv}&& {\hspace{4mm} i \atop
\displaystyle \frac{1}{\sqrt{2}}  \left (| \downarrow, \dots
\downarrow \uparrow \downarrow \dots \downarrow
\rangle_N\right.} { \phantom{c} \atop + e^{i\varphi}} {\hspace{-1mm} j \atop \left .
| \downarrow, \dots \downarrow\uparrow \downarrow
\dots \downarrow \rangle_N\right )}  \nonumber \\
&& = \frac{1}{\sqrt{2}} \left (c_i^\dagger + e^{i\varphi} c_j^\dagger  \right ) 
\ket{\Downarrow}
\label{wavefunction}
\end{eqnarray}
where the vacuum state is defined as $\ket{\Downarrow}=\ket{\down\dots\down}$.
We notice again that the translational invariance is
broken since the result for the wavefunction (\ref{wavefunction})
in (\ref{correlators}) depends on the positions $i$ and $j$.
The correlators $\bra{\Psi_{i,j}^\varphi} S_l^\alpha S_m^\beta
	\ket{\Psi_{i,j}^\varphi}$
can be expressed as a {\it sum of}\/ Pfaffians~\cite{AOXY}

In the case $\varphi=0,\; \pi$ 
($\ket{\Psi_{i,j}^\pi}=:\ket{-}$ and $\ket{\Psi_{i,j}^0}=:\ket{+}$)
it turns out that 
\begin{equation}
\langle \pm | S_l^\alpha S_m^\beta| \pm  \rangle=\sum_{s=0}^{2R-1}
{\rm pf} {\cal P}_{2R}^{s}\, .
\label{corr}
\end{equation}
with the pfaffian  ${\rm pf}{\cal P}_{2R}^{s}$ defined as in
appendix~\ref{pfaffians},
Eq.(\ref{pfaffian}), where the entries of the $s-th$ column 
are replaced by zeros and the entries of the $s-th$ row are
to be replaced by 
$$
\bra{0}A_l B_m\ket{0} \longrightarrow \expect{A_l B_m}^{(\pm)}
$$
with
$$
\expect{A_l B_m}^{(\pm)}\doteq \bra{\pm} A_l \ket{0} \bra{0} B_m\ket{\pm} -
    \bra{\pm} B_m \ket{0} \bra{0} A_l\ket{\pm}\;.
$$
Differently from the case where the average is performed over an
equilibrium state, the correlators $\expect{A_l A_m}$, $
\expect{B_l B_m}$ do not vanish. The relevant correlation
functions to be computed are then (see~\cite{AOXY} for details)
\end{multicols}
\begin{eqnarray}
\expect{A_l(t) B_{m}(t)}^{(\pm)}= & -&\left [ V(i-m) \pm V(j-m)\right]
    \left [V(i-l) \pm V(j-l)\right]  \nonumber \\
    & + & \left [U^{o}(i-l)\right.+ \left. U^{e}(i-l)\pm
    \left (U^{o}(j-l)+U^{e}(j-l) \right )\right ] \times \nonumber \\
    &&
    \left [
    U^{o}(i-m)-U^{e}(i-m)\pm \left (U^{o}(j-m)-U^{e}(j-m) \right ) \right ] \\
    \nonumber \\
\expect{A_l(t) A_{m}(t)}^{(\pm)} &=& i \left \{ \left [ V(i-m) \pm V(j-m)\right]  \right .
    \left [ U^{o}(i-l)+U^{e}(i-l)\pm \left (U^{o}(j-l)+U^{e}(j-l) \right ) \right ]
    \nonumber \\
    &&\left . - \left [V(i-l) \pm V(j-l)\right] \left [
    U^{o}(i-m)+U^{e}(i-m)\pm \left (U^{o}(j-m)+U^{e}(j-m) \right ) \right ] \right \}
    \\ \nonumber \\
\expect{B_l(t) B_{l+R}(t)}^{(\pm)} &=& i \left \{ \left [ V(i-m) \pm V(j-m)\right] \right.
    \left [ U^{o}(i-l)-U^{e}(i-l)\pm \left (U^{o}(j-l)-U^{e}(j-l) \right ) \right ]
    \nonumber \\
    &&\left . - \left [V(i-l) \pm V(j-l)\right] \left [
    U^{o}(i-m)-U^{e}(i-m)\pm \left (U^{o}(j-m)-U^{e}(j-m) \right ) \right ] \right \} \; ,
\end{eqnarray}
\twocolm where \beqa U^{o}(r)&:=&\frac{\lambda
\gamma}{\pi}\int_0^\pi \d k
    \sin k \frac{\sin (\Lambda_k t)}{\Lambda_k} \sin k r\\
U^{e}(r)&:=&\frac{1}{\pi}\int_0^\pi \d k
    (1+\lambda \cos k) \frac{\sin (\Lambda_k t)}{\Lambda_k} \cos k r\\
V(r)&:=&\frac{1}{\pi}\int_0^\pi \d k \cos (\Lambda_k t) \cos k r
\eeqa
These  two-point functions together with Eq.(\ref{corr}) allow
to evaluate all the spin-correlation functions necessary for the
calculation of the two-body reduced density matrix.
Inaccessible by the above technique is the correlation 
$\bra{\pm} S_l^{x}\ket{\pm}$. 
However, if $\bra{\pm} S_l^{x,y}(t=0) \ket{\pm} =0$, then it will remain
zero during the subsequent evolution. This correlation is zero if
the parity symmetry of the Hamiltonian is not broken by the initial
and final state.
We will consider exclusively this case in the present work.

In the case $\gamma=0$ the particle number conservation leads to a
considerable simplification since the vacuum is an eigenstate. For the
XY-model we consider with some detail the case of the propagation of a singlet.
In this case, because of the conservation of the total magnetization, only
states with one reversed spin will enter the dynamics.
In terms of corresponding
fermions only one particle states are involved in the dynamics.
Correlation functions
can be evaluated directly without resorting to the techniques described above.

%% file: concurrence.tex
\section{Measures of entanglement}

On a qualitative basis entanglement is well understood. Both
for distinguishable particles (e.g. spins on a lattice) and
identical particles (e.g. free fermions/bosons) it is known that if
the wavefunction can be written as a (tensor-) product state and
as the (anti-)~symmetrization of such product states, respectively, then
the corresponding degrees of freedom are not entangled~\cite{Ghirardi02}.
In order to analyze entanglement properties of spin chains, we need to characterize
entanglement at a quantitative level and therefore we have to use some
measure for the entanglement stored in the chain.
In general this is a formidable task and it is not yet known how to
completely accomplish it in a many-body system. As it was done in Refs.
\cite{Arnesen01,Gunlycke01,Osterloh02,Osborne02} we will consider here
the entanglement of formation as a measure of entanglement.
We confine our attention
to the analysis of the entanglement of a given subsystem with the rest of the
chain and to the entanglement between two arbitrary sites in the chain
(for simplicity we will refer to such quantities as the
amount entanglement). In the former case, accessible
entanglement measures have been found when the total system is
in a pure state. In the latter, i.e. for two-qubit entanglement,
a measure is known also for the total system being in a mixed state.

\subsection{Entanglement of a subsystem with its complement}

In order to identify whether a pure state of the total system $S$
is a product state of a part $A$ and the rest of the system $B=S\setminus A$
it is sufficient to trace out $A$ or $B$; only in case of a product state,
the rank of the outcoming reduced density matrix will be one.
If the rank is larger than one, then we know that $A$ and $B$
are entangled.~\cite{Ghirardi02}
A measure for the entanglement between these two subsystems is
$
S[\rho_A]:=-{\rm tr} \rho_A \log_{{\rm dim} A} \rho_A
$,
which is the von Neumann entropy of the subsystem A.
${\rm dim} A$ is the dimension of the Hilbert space corresponding to $A$
and bounds the quantity to the interval $[0,1]$.
We analyze the case when the subsystem $A$ is one site and the
corresponding entanglement measure quantifies the entanglement
of one site with the rest of the chain\cite{note:A-entropy}
$$
S[\rho^{(1)}]:=- {\rm tr} \rho^{(1)} \log_2 \rho^{(1)}
$$
where $\rho_A=\rho^{(1)}_j$ is the one-site reduced density matrix. In
this case, if there is no broken symmetry, this parameter is
related to the average magnetization along the $z$ component
\beq\label{rho1} 
\rho^{(1)}_j = \Matrix{cc}{
\bigfrac{1}{2}+\expect{S^z_j}&0\\
0&\bigfrac{1}{2}-\expect{S^z_j}} 
\eeq 
The one-site density matrix
contains only one real unitarily invariant parameter: its
eigenvalue $p\leq 1/2$. 
One can then choose other measures of the entanglement which 
are related to the von Neumann entropy.
A relevant example is the (one-)tangle~\cite{Coffman00}, whose
importance stems from the fact that it was proved to be an
additive measure of entanglement, including the two- and 
three-tangle for pure states of up to three qubits and there is some
evidence for additivity also for larger ensembles. We will
therefore in particular study the tangle in this work and will find further
evidence for this surmised additivity formulated in
Ref.~\cite{Coffman00}. The one-tangle is given by
$$\tau^{(1)}[\rho^{(1)}]:=4 {\rm det} \rho^{(1)}
$$
and is connected to the von Neumann entropy through the relation
$$
S[\rho^{(1)}]=h\left(\bigfrac{1}{2}\left(1+\sqrt{1-\tau_1[\rho_1]}\right) \right).
$$
where
$h(x)=: -x \log_2 x - (1-x) \log_2 (1-x)$~\cite{FOOTNOTETANGLE}.
Quantified by the tangle, the entanglement of the site $j$ with the
rest of the chain for the anisotropic transverse XY models is given by
\beq
\tau^{(1)}[\rho^{(1)}_j]=4\det \rho^{(j)}_1 = \bigfrac{1}{4}-\expect{S^z_j}^2\; .
\eeq
For what follows we omit the index of the one-tangle and write
$\tau:=\tau^{(1)}$. 

It must be stressed that the entanglement measures discussed so far
are applicable only if the total system is found in a pure state;
being in a mixed state one had to consider instead
the minimum average of the above quantities evaluated on all possible
decompositions of the density matrix of the total system.
Without this minimization procedure an upper bound for the entanglement
is obtained, since the entanglement is a convex function in the
space of density matrices~\cite{Wootters98,Hill97}.

\subsection{Entanglement between two sites - Concurrence}

All the information needed to analyze bipartite entanglement is
contained in the two-qubit reduced density matrix $\rho^{(2)}$,
obtained from the wave-function of the state after all the spins
except those at positions $i$ and $j$ have been traced out. The
resulting $\rho^{(2)}$ represents a mixed state of a bipartite system
for which a good deal of work has been devoted to quantify its
entanglement \cite{2-site_entanglement}. As a measure for
arbitrary mixed states of two qubits, we use the
concurrence~\cite{Wootters98} $C[\rho^{(2)}]$
\beqa\label{def:concurrence}
C[\rho^{(2)}] &:=& \max\{0,2\lambda_{max} - {\rm tr} \sqrt{R}\},\\
R &:=& \rho^{(2)} \sigma_y\otimes\sigma_y \rho^{(2)\,*}
\sigma_y\otimes\sigma_y \label{def:R} \eeqa where $\lambda_{max}$
is the largest eigenvalue of the matrix $\sqrt{R}$, which is the
product of $\rho^{(2)}$ with its time-reversed; $\sigma_y$ is a
Pauli matrix. For pure two-qubit states it is $\tau^{(1)}\equiv
C^2=:\tau^{(2)}$. The great advantage of the concurrence is that it is
directly defined in terms of the density matrix $\rho^{(2)}$
without any minimization procedure; therefore it can be calculated
from correlation functions of the corresponding Hamiltonian.

We derive the structure of the two-qubit reduced density matrix
$\rho^{(2)}$ for the model Hamiltonian of Eq.(\ref{model}).
Certain symmetries of the Hamiltonian do restrict the
structure of the reduced density matrix as long as they are not broken.

As discussed more in detail in Appendix~\ref{app:reducedrho}, the
concurrence can be expressed in terms of spin correlation functions
(see Eqs.(\ref{app:C-of-corrs}))
\onecolm
\beqa\label{C-of-corrs}
C_{lm}=2\max\left\{
\phantom{\sqrt{(\bigfrac{1}{4}-g^{zz}_{lm})^2-
\bigfrac{1}{4}(M^z_l+M^z_m)^2}} \hspace*{-4.7cm} \right.
0&,&\sqrt{(g^{xx}_{lm}-g_{lm}^{yy})^2
    +(g_{lm}^{xy}+g_{lm}^{yx})^2}-
    \sqrt{(\bigfrac{1}{4}-g_{lm}^{zz})^2-\bigfrac{1}{4}(M^z_l-M^z_m)^2}
\nonumber \\
&,&\left.   \sqrt{(g_{lm}^{xx}+g_{lm}^{yy})^2
    +(g_{lm}^{xy}-g_{lm}^{yx})^2}-
    \sqrt{(\bigfrac{1}{4}+g_{lm}^{zz})^2-\bigfrac{1}{4}(M^z_l+M^z_m)^2}\right\}\;.
\eeqa
\twocolm
In the isotropic case ($\gamma =0$) the conservation of
magnetization, or equivalently of the total number of particles,
the structure of the $2$-site reduced density matrix simplifies 
considerably (provided that the initial state belongs to a
definite spin sector of the overall Hilbert space). In this case
the concurrence is given by
\begin{equation}
C^{iso}=\max\left\{0, 4 |g_{lm}^{xx}|-
    \sqrt{(\bigfrac{1+4 g_{lm}^{zz}}{2})^2-1(M^z_l+M^z_m)^2}\right\}\;.
\label{concurrence:non-equilibrium:gamma0}
\end{equation}

As we will study the time evolution of a singlet
(one of the Bell states) on top of the vacuum, the structure of $\rho^{(2)}$
further simplifies because one has to deal with one-particle states only.
This leads to a concurrence
\beqa\label{concurrence:non-equilibrium:gamma0-1P}
C^{iso}_{singlet}=4|g_{lm}^{xx}|\; .
\eeqa

\subsection{The conjecture by Coffman, Kundu, and Wootters}

One- and two- site entanglement do not furnish a complete characterization
of the entanglement present in spin chains. In this direction it is
interesting to study a conjecture put forward by Coffman, Kundu,
and Wootters that states that the sum of the two-tangles
(the square of the concurrences) be at most equal to the one-tangle.
\beq
\sum_{j\neq n} C^2_{n,j} \leq 4 \det \rho^{(1)}_n = \tau^{(1)}_n\; .
\eeq
A non-zero difference in the above inequality has been
interpreted as ``residual tangle'', i.e. entanglement not
stored in two-qubit entanglement.
This inequality was proved in Ref.~\cite{Coffman00}
for a three qubit system, giving rise to the definition
of the three-tangle as a measure of three-qubit entanglement.

%% file: results.tex
\section{Results}

\subsection{$\gamma=0$: The isotropic model}
\label{section:gammazero}
\input{gammazero}

\subsection{$\gamma \ne 0$: Singlet onto the vacuum}
\label{section:generic}
\input{gammavacuum}

\subsection{$\gamma \ne 0$: Singlet onto the ground state}
\label{section:gammaground}
\input{gammaground}

%% file: gammazero.tex
In this section, we describe the dynamics of entanglement for
$\gamma=0$. The Hamiltonian of Eq.(\ref{model}) is then reduced to
the $XY$ model. We will discuss the case in which the initial
condition is given by one of the maximally entangled Bell states
as defined in Eq.(\ref{wavefunction}), where the
description of the entanglement dynamics is amenable to a quite
simple analytical solution. Already for this case 
we will get many results that will emerge also  in the
analysis of the models for generic $\gamma$. As already mentioned,
the feature that distinguishes this model from the general case is
that the $z$-component of the total spin, $S^z$, is conserved.
Consequently the Jordan-Wigner transformed fermionic Hamiltonian
becomes a tight binding model for each sector with fixed $S^z$.
The dynamics cannot generate entanglement from the vacuum state
$\ket{\Downarrow} = \ket{\downarrow}^{\otimes N}$ (as it turns out to occur
for $\gamma \ne 0$).

We first consider the case of a chain initially prepared in a
maximally entangled singlet-like state $\ket{\Psi_{i,j}^{\varphi}}$ on sites 
$i$ and $j$ as defined in Eq. (\ref{wavefunction}). 
By using the
evolution of the fermion operators obtained in
Eq.(\ref{c(t):gamma0}), the state vector at later times is found
to be
\begin{equation}
\ket{\Psi_{i,j}^\varphi} =\sum_l w_l(t) \, c_l^{\dagger} \, \ket{\Downarrow}
\label{statounap}
\end{equation}
with
\begin{equation}
w_l(t) = \frac{1}{\sqrt{2N}} \sum_{k} \left [e^{\frac{2 \pi
i k }{N}(i-l)} + e^{i \varphi} e^{\frac{2 \pi i k }{N}(j-l)}
\right ] e^{i \Lambda _k t} \;\; .
\end{equation}
In the thermodynamic limit, $N\rightarrow \infty$, the coefficients become
\begin{equation}
w_l(t) = \frac{1}{\sqrt{2}} \left \{
J_{i-l}(\lambda t) + e^{i \varphi} \,
 (-i)^{(j-i)} J_{j-l}(\lambda t) \right \} \; ,
\end{equation}
where $J_n(x)$ is the Bessel function of order $n$
(we omitted an irrelevant global time dependent phase).

\subsubsection{Concurrence}
For the initial singlet-like state of Eq.(\ref{statounap}), it can
be shown that the concurrence between two selected sites $n$ and
$m$ is given by
\begin{equation}
C_{n,m}(t) = 2 \Bigl | w_n(t) w_m^*(t) \Bigr | \; .
\label{concnm}
\end{equation}
This expression is plotted in Figs.
\ref{priconc}--\ref{quaconc} for some particular cases.

The time evolution dictated by the Hamiltonian, amounts to a simultaneous
spin flip between sites $l$ and $l\pm1$, i.e.: a propagation of the
single flipped spin in either directions.
Expressed in terms of the fermion operators the Hamiltonian 
becomes a tight-binding model.
In particular, this implies that each one of the initially
entangled sites (say $i$) tends to become entangled with the
nearest neighbors of the other ($j \pm 1$). The
time scale is set up by the interaction strength, so that the information
exchange or ``entanglement propagation" over a distance of $d$
lattice spacings, approximately takes a time $t \sim d/\lambda$.
The external local field $h$ does not enter
anyway, due to the fact that all of the components of the state
(\ref{statounap}) are in the same spin sector of the Hilbert
space.

The concurrence defined in  Eq.(\ref{concnm}) depends on four
sites: apart from  $n$ and $m$, whose entanglement we want to
measure, also the initially entangled ones, $i$ and $j$, enter the
general expression for $C_{n,m}$. We first look at the decay of the
concurrence shared by the initially maximally entangled
sites themselves depending on their distance $x$
(i.e. $j=i+x$). The result is seen in Fig. \ref{priconc}.
\begin{equation}
C_{i,i+x} = \Bigl | J_0^2 + 2 i^x J_0
J_x \cos \varphi + (-1)^x J_x^2 \Bigr |\; ,
\end{equation}
where the Bessel functions $J$ have to be evaluated at $\lambda t$.
After a first recurrence of the entanglement at the time $t \sim
x/\lambda$ (due to the swap between the two sites), the
concurrence decays at longer times as $t^{-1}$, modulated by the
subsequent revivals of the entanglement at times which are integer
multiples of the first revival time, and by oscillations of period
$\lambda ^{-1}$.

\end{multicols}

\begin{minipage}[h]{\linewidth}
\begin{center}
\begin{minipage}[h]{.47\linewidth}
\begin{figure}\centering
\includegraphics[width=\linewidth]{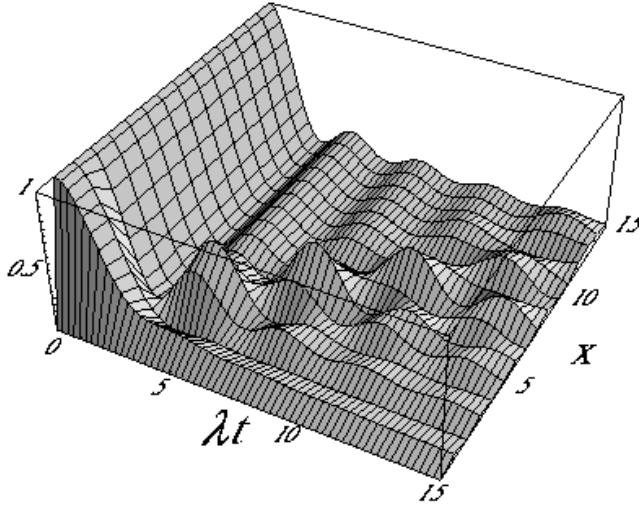}
\caption{Concurrence $C_{i,i+x}$ as a function of time and of the
distance between the initially entangled sites. The initial state
is a singlet ($\varphi=\pi$ in the initial condition) involving
the very same sites $i$ and $i+x$.} \label{priconc}
\end{figure}
\end{minipage}
\hspace{4mm}
\begin{minipage}[h]{.47\linewidth}
\begin{figure}
\includegraphics[width=.9\linewidth]{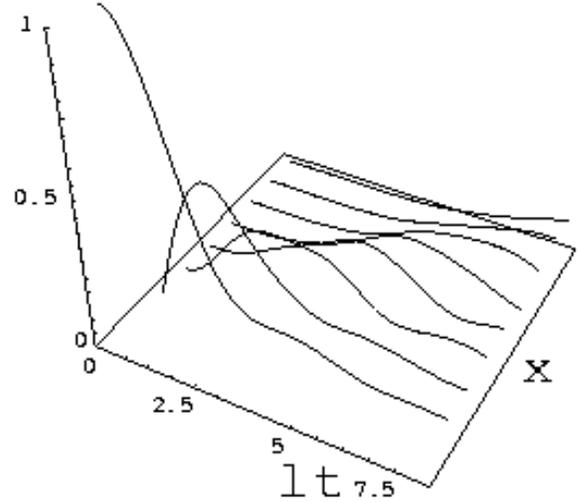}
\caption{Concurrence between sites at a distance $x$ (namely,
$n=i$, and $n=i+x$ in Eq. \ref{concnm}), for the case of an
initial $0$-triplet state ($\varphi=0$) shared by two nearest
neighbor sites, $i=0, j=1$. The various plots correspond to $x=1,
\ldots, 8$.} \label{seconc}
\end{figure}
\end{minipage}
\\
\begin{minipage}[h]{.47\linewidth}
\begin{figure}
\vspace*{2mm}
\includegraphics[width=.87\linewidth]{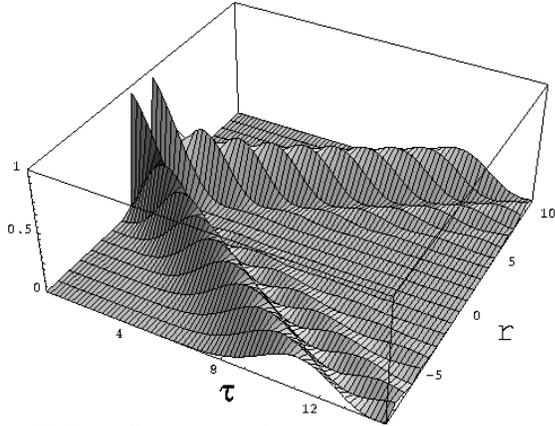}
\caption{Concurrence between sites $n=-x, m=x$, symmetrically
displaced from their initial position $i=-1$ and $j=1$
($\varphi=\pi$).} \label{treconc}
\end{figure}
\end{minipage}
\hspace{5mm}
\begin{minipage}[h]{.47\linewidth}
\begin{figure}
\vspace*{9mm}
\includegraphics[width=\linewidth]{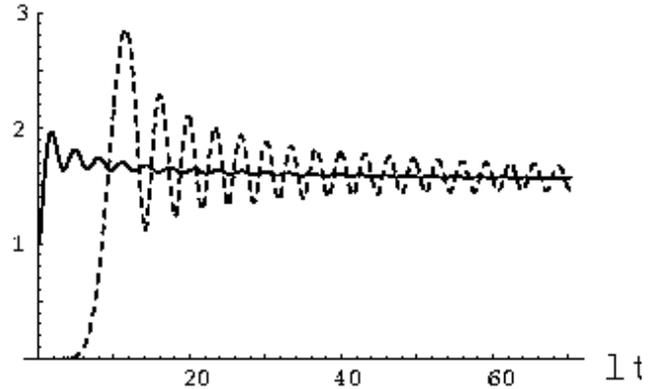}
\caption{Summed concurrences for an initially entangled site
($n=0$, full line) and an initially unentangled site ($n=10$,
dashed line). The initial state is a singlet ($\varphi=\pi$)
created on sites $i=0$ and $j=1$.} \label{quaconc}
\end{figure}
\end{minipage}
\end{center}
\end{minipage}

\twocolm

We now concentrate on the distribution and propagation of the entanglement.
Supposing that the initially selected sites are nearest neighbors,
$j=i+1$, we look at the concurrence between one of these two
(namely, $i$) and another site of the chain at distance $x$ (i.e.
$n=i$ and $m=i+x$ in Eq.(\ref{concnm}) ). The result is shown in
Fig.\ref{seconc}, where one can see the propagation of the maximum
with velocity $\lambda$.
As already pointed out, the dynamics of the model consists in
flipping the spin of neighbor sites, leading to a propagation
of the initially flipped spin along the chain both in the same
and in opposite directions.
This comprises a propagation of the concurrence as confirmed by
Fig.\ref{treconc}, where the concurrence is shown between two
sites symmetrically displaced with respect to the initial
excitation (namely, $i=-1$ and $j=1$ as initially entangled pair
of sites, and $n=x$, $m=-x$ in Eq. (\ref{concnm})). 
This can be regarded as the propagation of an EPR-like pair;
also in this case the speed of propagation is $\lambda $.
In Fig.\ref{quaconc}, the total concurrence
$$
C_{tot,n} = \sum_m C_{n,m}
$$
of the single site $n$ is shown, representing the total amount of
pairwise entanglement involving the selected site. We found that
the same stationary value (around 1.6) is reached for all values
of $n$, independently of the initial conditions. This demonstrates
that the initial state becomes homogeneously spread at long times
and so does the concurrence.

\subsubsection{Entropy}
For a deeper investigation of the entanglement propagation
observed for the concurrence $C_{n,m}$, we evaluate the entropy
$S^{(2)}_{n,m}$ of the reduced density matrix $\rho^{(2)}_{n,m}$.
Since the chain is in  a pure state, this entropy  gives a measure
of the entanglement of the two sites with the rest of the chain.
In particular, if $S^{(2)}_{n,m}=0$, then the state of the two
sites is pure and no entanglement exists with the rest of the
chain, while  $S^{(2)}_{n,m}\ne 0$ means that $\rho^{(2)}_{n,m}$
is a mixed state, and consequently the pair $(n,m)$ is
entangled with the rest of the system.

From the two-site reduced density matrix for this case, Eq.
(\ref{app:1rho1}) in the Appendix, we deduce 
\beq\label{S2} 
S^{(2)}_{n,m}(t) = -
(1-p) \log_2 (1-p) - p \log_2 p \; ,
\eeq 
with  $p = |w_n(t)|^2 + |w_m(t)|^2 $. $S^{(2)}_{-x,x+1}$ is shown in Fig.
\ref{entro} unveiling the same qualitative structure found for the
concurrence. In particular, the propagation velocity is the same.
Initially, this entropy is zero and for $x>0$ it remains zero
until the ``entanglement wave" has arrived. This is because a pure
state $\ket{\downarrow}_n\ket{\downarrow}_m$ is found before.
Then the pair of spins becomes
entangled not only between themselves, but also with the rest of
the system. This is understood in the sense of entanglement
sharing~\cite{Coffman00} from the fact that also the nearest
neighbors $x$ and $x+1$ become entangled at that time. After that,
the two sites are left partially entangled with the rest of the
chain.
\onecolm

\begin{minipage}[h]{\linewidth}
\begin{figure}\centering
\includegraphics[width=.9\linewidth]{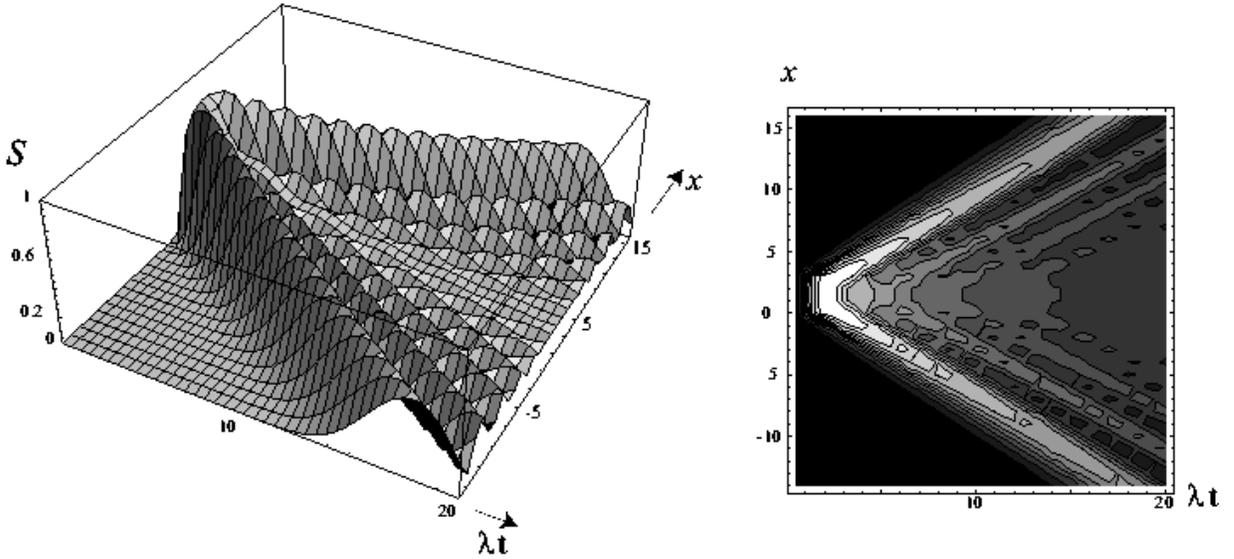}
\caption{Entropy $S^{(2)}_{(-x,x+1)}(\tau)$ for pairs of sites
symmetrically displaced with respect to the initial singlet
position at $(i,j)= (0,1)$. This is for $\varphi = \pi$.}
\label{entro}
\end{figure}

\begin{figure}\centering
\includegraphics[width=.9\linewidth]{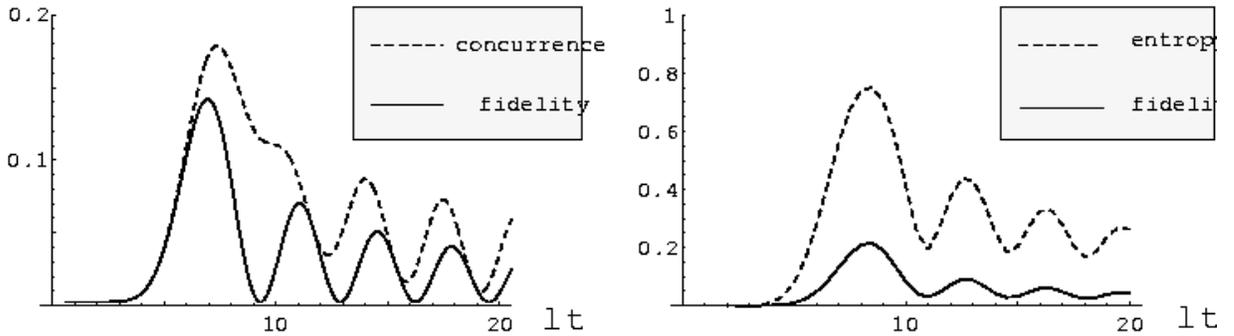}
\caption{Comparison between the temporal behaviors of the fidelity
and the concurrence (left) or entropy (right), for the cases
$(a,b)=(-1,1)$, $(n,m)=(-6,6)$ for $\varphi=\pi/2$(left) and
$(a,b)=(0,1)$, $(n,m)=(-6,7)$ for $\varphi=\pi$ (right).}
\label{confrofidelity}
\end{figure}
\end{minipage}

\begin{multicols}{2}
The above discussion can be extended to the von Neumann entropy for
an $N$-site subsystem at positions $\vec{j}:=(j_1,\ \dots\; ,\ j_N)$:
since the reduced density matrix has still rank $2$ we still have
$S^{(N)}_{\vec{j}}=- p \log_2{p} - (1-p) \log_2{1-p}$ but with
$p=\sum_{i=1}^N |w_{j_i}|^2$.
This gives access to the time dependence of the entropy also 
for a block of spins in the system. Such a quantity was studied
at equilibrium in Ref.~\cite{Vidal02}.

\subsubsection{One-site entanglement and the CKW conjecture}
\begin{figure}
\includegraphics[width=.9\linewidth]{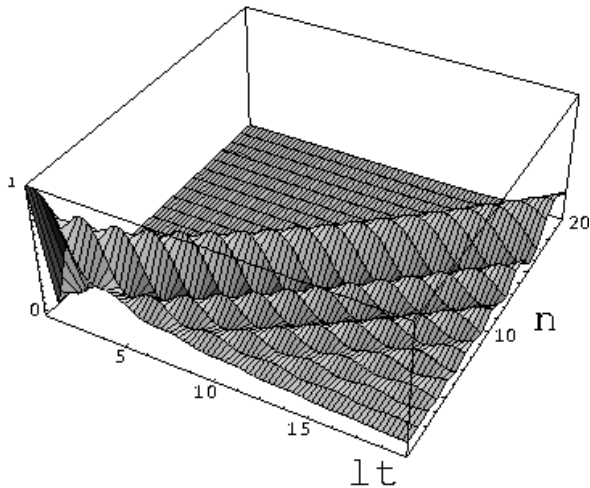}
\caption{Entanglement of the site $n$ with the rest of the chain
quantified by the entropy of the one-site reduced density matrix.
The initial state was a singlet at $(a,b)=(0,1)$.} \label{entrosing}
\end{figure}

The same structure as for the two-site entropy
is also found in the single site entropy,
$S^{(1)}_n(t)$, quantifying the amount of entanglement of one site with
all the other spins (see Fig. \ref{entrosing}).
Also this can be understood from the CKW conjecture~\cite{Coffman00}
as explained in what follows.

For the isotropic model it is possible to analytically
check this conjecture. All needed quantities are known explicitly
and we obtain
$$
4 \det \rho^{(1)}_j=4 |w_j|^2 (1-|w_j|^2)
$$
and
$$
\sum_{n\neq j} C^2_{j,n}=4\sum_{n\neq j} |w_j w_n^*|^2
=  4 |w_j|^2 (1-|w_j|^2)\; .
$$
Thus, the two quantities coincide, meaning that the entanglement
present in the system is restricted to the class of pairwise
entanglement, and no higher order entanglement is created. Since,
as mentioned in the previous section, the one-tangle is a
monotonic function of the von Neumann entropy, the information
gained from studying the entropies of the system is already
(implicitly) contained in the concurrence.
This has already been observed for these Werner-type states in
Ref.~\cite{Coffman00}.

\subsubsection{Fidelity}
In order to gain information on what type of Bell state propagates
and on eventual state mutation, we study how similar is the mixed
state $\rho^{(2)}_{n,m}$ to the initial 
$\ket{\Psi^{\varphi}}$~\cite{note:noindex}. 
This similarity is quantified by the fidelity
\begin{equation}
F_{n,m}(t) = \mbox{Tr} \left \{\rho^{(2)}_{n,m}(t) \,
\ket{\Psi^{\varphi}}\bra{\Psi^{\varphi}} \right \}\; ,
\end{equation}
which is found to be
\begin{equation}
F_{n,m}(t) = \frac{1}{2} \left | w_n(t) + e^{-i
\varphi} w_m(t) \right |^2 \; .
\end{equation}
We compare this quantity to both concurrence and entropy in Fig.
\ref{confrofidelity}, where it is shown that the fidelity displays
in-phase oscillation with respect to both the entanglement
measures. Thus, when the entanglement wave arrives,
$\rho^{(2)}_{n,m}$ becomes more similar to the initially prepared
state. We can interpret this by saying that the state itself is
propagating along the chain, although this propagation is far from
a perfect transmission, due to the entanglement sharing with many
sites at a time.

\subsection{Other maximally entangled states}
Apart from the states $\ket{\Psi^{\varphi}}$ of Eq.
(\ref{wavefunction}), we also analyzed the propagation of entanglement
starting from other maximally entangled states, of the form
\begin{eqnarray}
\ket{\Phi_{i,j}^{\varphi}} &=&\frac{1}{\sqrt{2}} \left (
\ket{\downarrow_i, \downarrow_j} + e^{i \varphi} \ket{\uparrow_i,
\uparrow_j} \right ) \otimes \ket{\downarrow}^{\otimes (N-2)} \nonumber\\
&=&
\frac{1}{\sqrt{2}} ( \I + e^{i \varphi} c_i^{\dagger}
c_j^{\dagger} ) \, \ket{\Downarrow} \; . \label{statiPhi}
\end{eqnarray}
These are not single-particle states and, furthermore, since they
are superpositions of two components pertaining to different 
spin sectors of the global Hilbert space, one cannot take full advantage
of the conservation of the magnetization. As a result, 
the two-site reduced density matrix for this case is of 
the form of Eq.~(\ref{app:rho:non-equilibrium}),
and the concurrence is given by $C=\max\{0,C^{(1)},
C^{(2)}\}$ (see Eq.~(\ref{app:concurrence:non-equilibrium}));
a concurrence of the form $C^{(2)}$ indicates that
$\ket{\Psi}$-like correlation arises, while for $C^{(1)} > C^{(2)}$
the correlations between the two selected sites is more
$\ket{\Phi}$-like.

As we will show, the entanglement propagates with velocity $\lambda$
along the chain also for the initial condition (\ref{statiPhi}). 
Under certain conditions, however, a clear difference arises 
with respect to the situation analyzed in the previous section; 
namely, when the two initially entangled sites are separated by an 
odd number of spins.
It turns out that then, the propagating 
quantum correlations change their character after a while, 
and from that point on a singlet-like concurrence continues 
propagating even if the initial state was not a singlet. 
To demonstrate this, we study the time evolution of an initial 
Bell state $\ket{\Phi^\varphi_{i,j}}$ and the fidelity of the
Bell states $\ket{\Phi^{\varphi'}_{m,n}}$ and 
$\ket{\Psi^{\varphi'}_{m,n}}$ in the actual two-site states. 

The time evolution of the various coefficients involved in the
evaluation of $C$ can be obtained using once again Eq.
(\ref{c(t):gamma0}). In the thermodynamic limit ($N\rightarrow
\infty$) and by looking at the sites $n,m$ (with $m>n$, for
definiteness), we find
\begin{eqnarray}
a &=& \frac{1}{2} \left [ J_{n-i} J_{m-j} -
J_{n-j} J_{m-i} \right ]^2 \; , \\
b &=& 1 + a - \frac{1}{2} \left [J_{m-i}^2 + J_{n-i}^2 + J_{m-j}^2
+ J_{n-j}^2 \right ] \; ,
\eeqa
\beqa
|c| &=&\frac{1}{2} \Bigl | J_{n-i} J_{m-j} - J_{m-i} J_{n-j} \Bigr
| \; \\
x&=& \frac{1}{2} \left [ J_{n-i}^2 + J_{n-j}^2 \right ] - a \; ,\\
y&=& \frac{1}{2} \left [ J_{m-i}^2 + J_{m-j}^2 \right ] - a \; ,
\end{eqnarray}
\onecolm
\begin{equation}
|z| =  \Bigl | \frac{1}{2} \Bigl (J_{m-i} J_{n-i} + J_{m-j}
J_{n-j} \Bigr ) -
\sum_{r=n+1}^{m-1} \Bigl ( J_{n-i}J_{r-j} - J_{n-j}J_{r-i} \Bigr
)\Bigl ( J_{m-i}J_{r-j} - J_{m-j}J_{r-i} \Bigr ) \Bigr | \;,
\end{equation}

\begin{minipage}[h]{\linewidth}
\begin{figure}\centering
\includegraphics[width=.9\linewidth]{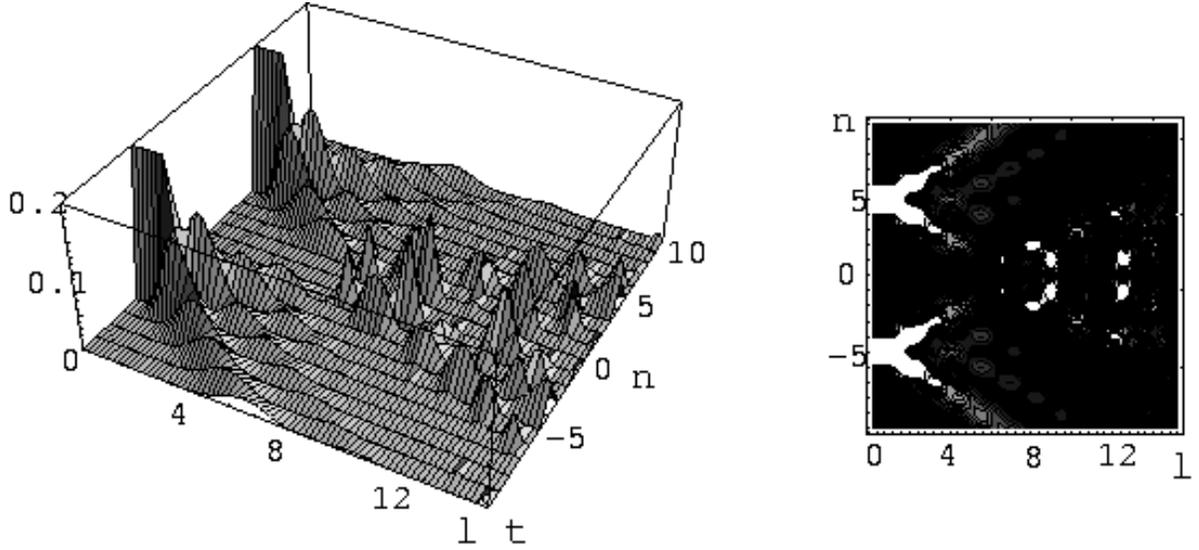}
\caption{Time evolution of the concurrence between sites $-n$ and
$n$ for the initial state $\ket{\Phi^\varphi_{-5,5}}$. The plot
is cut at $0.2$ in order to make the revival after the crossing
visible.} \label{concPhi}
\end{figure}

\begin{figure}\centering
\includegraphics[width=.9\linewidth]{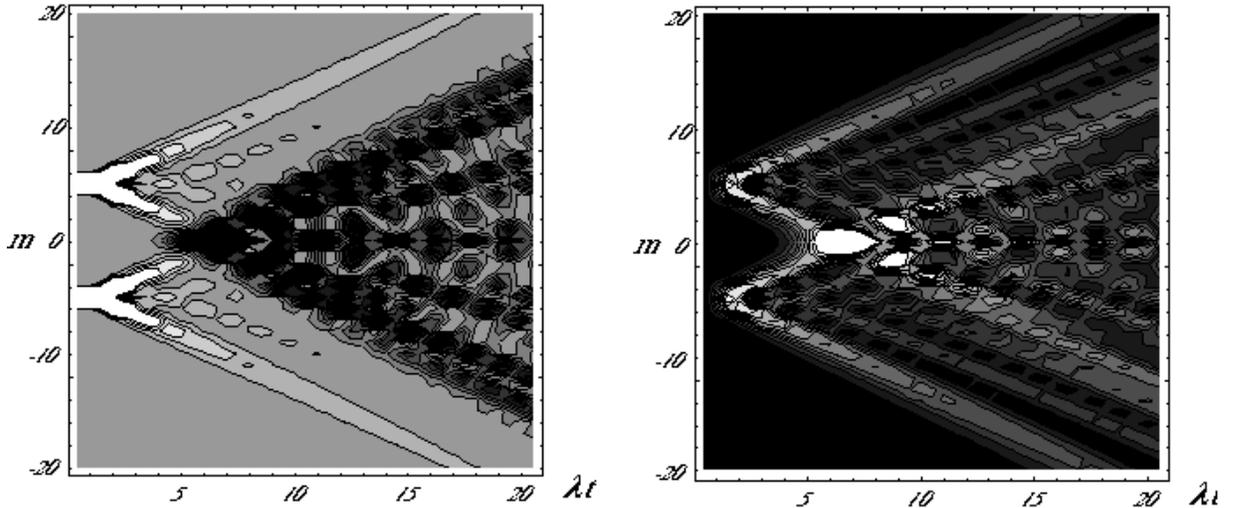}
\caption{Degree of similarity (white:high, black:low)
between the density matrix
$\rho^{(2)}_{-n,n}(t)$ and the states $\ket{\Phi^{\varphi_{opt}}}$ (left)
and $\ket{\Psi^{\varphi_{opt}}}$ (right)
(quantified by the corresponding fidelities).} \label{fidePhi}
\end{figure}
\end{minipage}

\begin{multicols}{2}
where all the Bessel functions are evaluated at $\lambda t$. We
note that any dependence on the phase $\varphi$ of the initial
superposition has disappeared (only the phase of $c$
depends on $\varphi$).

The concurrence is then readily obtained from these expressions.
An example is shown in Fig. (\ref{concPhi}), where $C_{-n,n}$ is
displayed for an initial state $\ket{\Phi^\varphi_{-5,5}}$. 
One can see that the concurrence between the original
Bell state positions suddenly decays
while the maximum propagates along the chain. At the
intersection between the two peaks coming from the initial
Bell state, however, a revival is present. From this time on,
the entanglement spreads out from this middle point.
A detailed analysis reveals that at the crossing the
concurrence switches from $C^{(1)}$ to $C^{(2)}$. 
This indicates that quantum correlations of type $\ket{\Phi}$ 
do not propagate much along the chain before changing to
singlet-like correlations (of type $\ket{\Psi}$) in this case.
In order to make visible this switching from the initial 
triplet to a singlet-like state, we analyzed the degree of resemblance 
of the two-site density matrix with both the states $\ket{\Phi^{\varphi'}}$ 
and $\ket{\Psi^{\varphi'}}$. These two fidelities are given by
\onecolm
\begin{eqnarray}
\mbox{Tr} \left \{
\rho^{(2)}_{n,m}(t) \, \ket{\Phi^{\varphi'}} \bra{\Phi^{\varphi'}}
\right \}
 &=& \frac{a+b}{2} + |c| \cos \left [ (\varphi' - \varphi) +
\frac{\pi}{2} (m+n-i-j) \right ] \\
 \mbox{Tr} \left \{
\rho^{(2)}_{n,m}(t) \, \ket{\Psi^{\varphi'}} \bra{\Psi^{\varphi'}}
\right \} 
 &=& \frac{x+y}{2} + |z| \cos \left [ \varphi' +
\frac{\pi}{2} (n-m) \right ]
\end{eqnarray}
\twocolm
It can be seen from these relations that there exist ``optimal''
phases that maximize the two degrees of resemblance, namely
$$
\varphi_{opt}^{\Phi} = \varphi + \frac{\pi}{2} (a+b-m-n) \; , \;
\varphi_{opt}^{\Psi} = \frac{\pi}{2} (m-n) \; , 
$$ 
respectively,
which depend only on the positions along the chain ($m$ and $n$)
and on the initially excited sites ($i$ and $j$) but do not depend
on time.
In Fig. (\ref{fidePhi}), we show the fidelity of
$\ket{\Phi^{\varphi_{opt}}}$ (left plot) and 
$\ket{\Psi^{\varphi_{opt}}}$ (right plot)
in $\rho^{(2)}_{-m,m}(t)$.
We used the same initial conditions as in the evaluation of 
the concurrence in Fig.~(\ref{concPhi}). 
Taking into account that the degree of similarity
with a state $\ket{\Phi^\varphi}$ is $0.5$ before the arrival of the
entanglement wave (due to its $\ket{\downarrow, \downarrow}$
component), it is seen from these plots
that after the crossing point the state of the pair of sites
becomes $\ket{\Psi}$-like.

As stated above, this phenomenon of a $\ket{\Phi}$-like state turning
into a $\ket{\Psi}$-like state during the propagation only occurs in
certain cases, and in particular every time two ``entanglement
waves" coming from the two sides of the chain intersect on a given
spin (the site $0$ in the example above). At the crossing, the
amplitude for two parallel spins does not survive and the outgoing
states of this scattering event only contain antiparallel spins.
On the contrary, when the initially entangled spins are separated 
by an even number of sites, the crossing involves two sites and the
character of the state is preserved.

%% file: gammavacuum.tex
We now move to consider the case of generic $\gamma$ for an 
infinitely long chain. 
In contrast to the previous section, where the
isotropic model (i.e. $\gamma=0$) was examined,
here the complexity of the computation grows with the distance $d$ of
the sites because the dimension of the Pfaffian expression
for the correlation functions is $2d\times 2d$.
Therefore we focus our investigation on $d\in\{1,2,3\}$.
For the critical Ising model (i.e. $\gamma=\lambda=1$)
it turns out to be sufficient considering
$d=1$, since the concurrence for the larger distances vanishes.
 
The initial state (at time $t=0$) is the singlet 
created at the sites with number $1$ and $2$ onto the vacuum 
$\ket{\Downarrow}$:
$\ket{\Psi_{1,2}^\pi}=1/\sqrt{2} 
(\ket{\uparrow\downarrow}-\ket{\downarrow\uparrow})
\hat{=}1/\sqrt{2} (c^\dagger_1 - c^\dagger_2)\ket{\Downarrow}$.
We start from small nonzero $\gamma$, then going to medium anisotropy,
$\gamma=0.5$, toward the transverse Ising model, $\gamma=1$.
The initial state, as well as the Hamiltonian, is invariant under
a reflection by a mirror placed between the sites $1$ and $2$ up
to a global prefactor (which does not affect the concurrence).
Because of this mirror symmetry, we only show the propagation
in direction of increasing site number. The propagation to
the other side is the mirror image as in the previous figures.

\subsubsection{Concurrence}

Figure \ref{C1-Vac} shows the nearest neighbor concurrence
for $\gamma=0.1,\ 0.5,\ 1.0$ (from the top to the bottom)
and $\lambda=0.5,\ 1.0$ (left and right, respectively). 
A rough estimate of the propagation velocity can be taken from 
the contour lines in the plot.
For not too large $\gamma$ we find concurrence 
propagation roughly with velocity $\lambda$
as for the isotropic model. For increasing $\gamma$ 
(see the right column -- $\lambda=1$ -- of Fig.~\ref{C1-Vac}) 
it slightly increases. The concurrence on the
original singlet position decays quickly and oscillations are 
increasingly damped.

We notice (Fig.\ref{C1-Vac}) an instantaneous signal which, 
sufficiently far away from the initial singlet position is 
spatially uniform. 
This phenomenon reflects the creation of entanglement from the vacuum,
which is characteristic for the anisotropic XY models. 
It is originated from the double spin
flip operators $\gamma\lambda/2\, \sum_i s_i^+ s_{i+1}^+$ 
(and its Hermitean conjugate), which are absent in the isotropic model.
Estimating the effect of this operator to second  order in  $\lambda t$ 
we argue that it transforms the vacuum into
an entangled state corresponding to the bell states 
$\ket{\up\up}\pm\ket{\down\down}$; the nearest neighbor concurrence,
originated from these terms only, turns  out to be 
\begin{equation}
C_{k,k+1}=\gamma\lambda t-\gamma^2\lambda^2t^2/2 \;\; .
\label{C-PERT}
\end{equation}
That is, a linear increase with slope $\lambda\gamma$
and maximum value $C^{max}_{k,k+1}=0.5$ at $t_{max}=1/\lambda\gamma$.
\onecolm
\begin{minipage}[h]{\linewidth}
\begin{figure}\centering
\includegraphics[width=.29\linewidth,angle=-90]{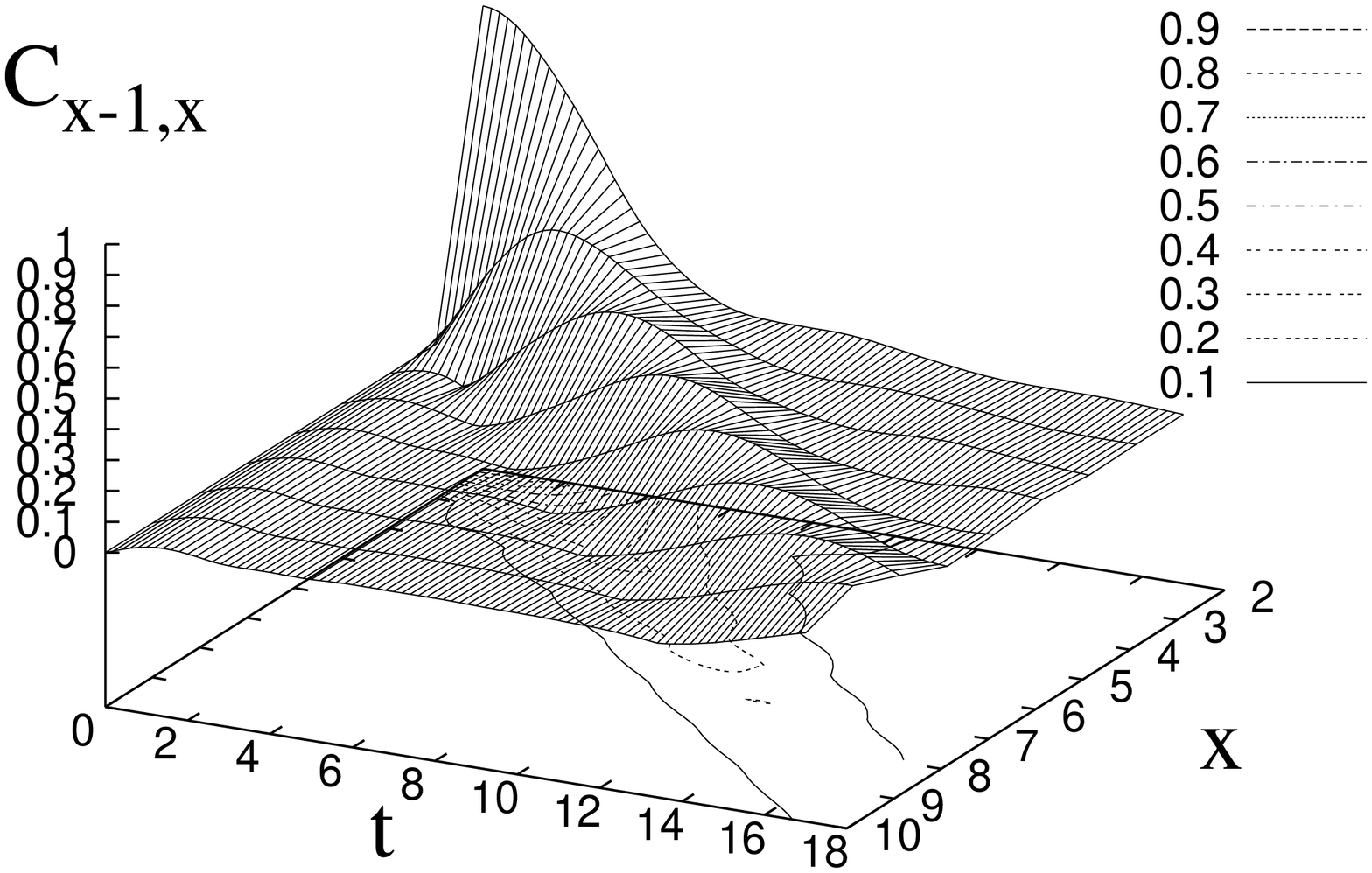}
\includegraphics[width=.29\linewidth,angle=-90]{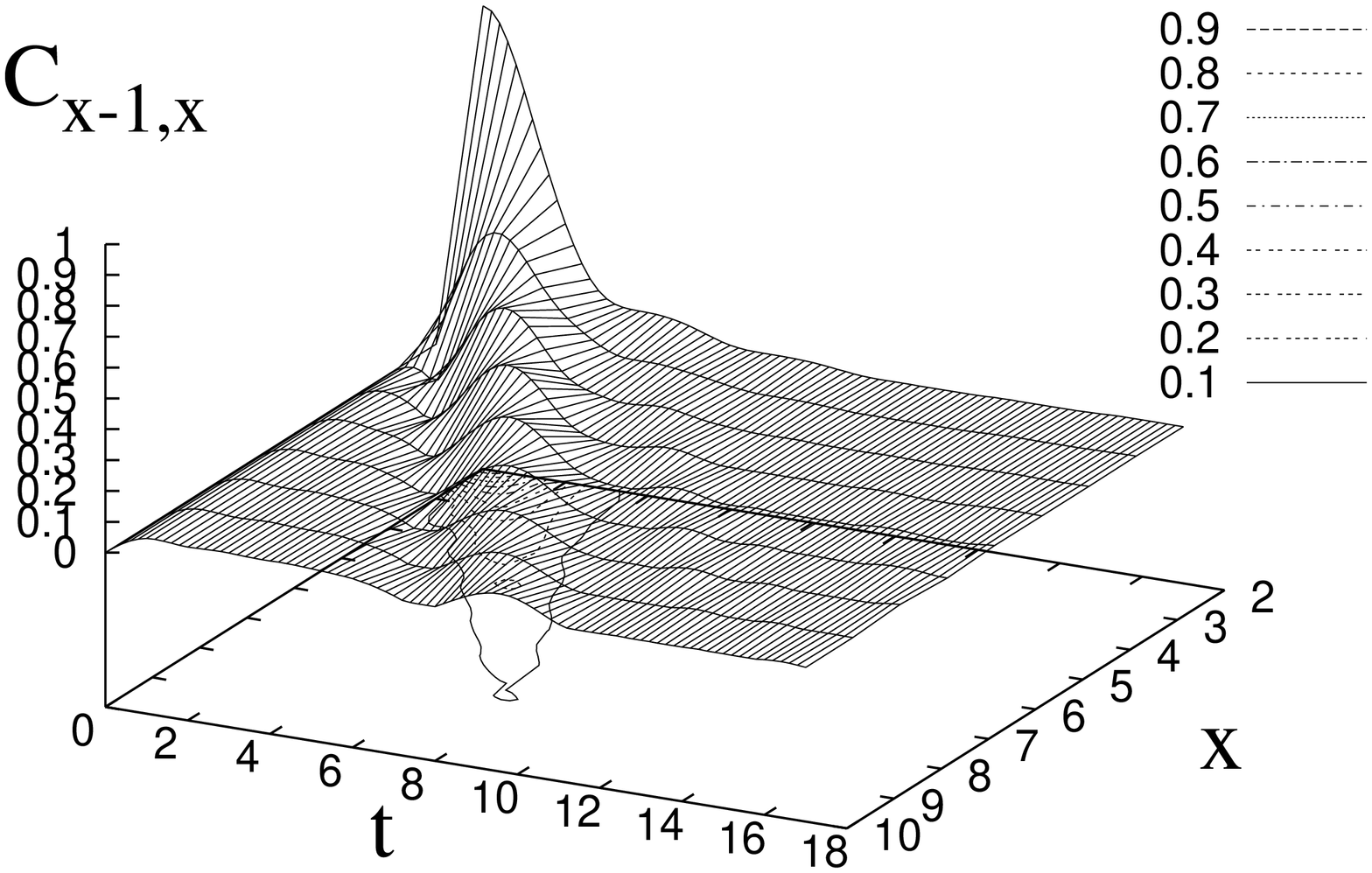}
\includegraphics[width=.29\linewidth,angle=-90]{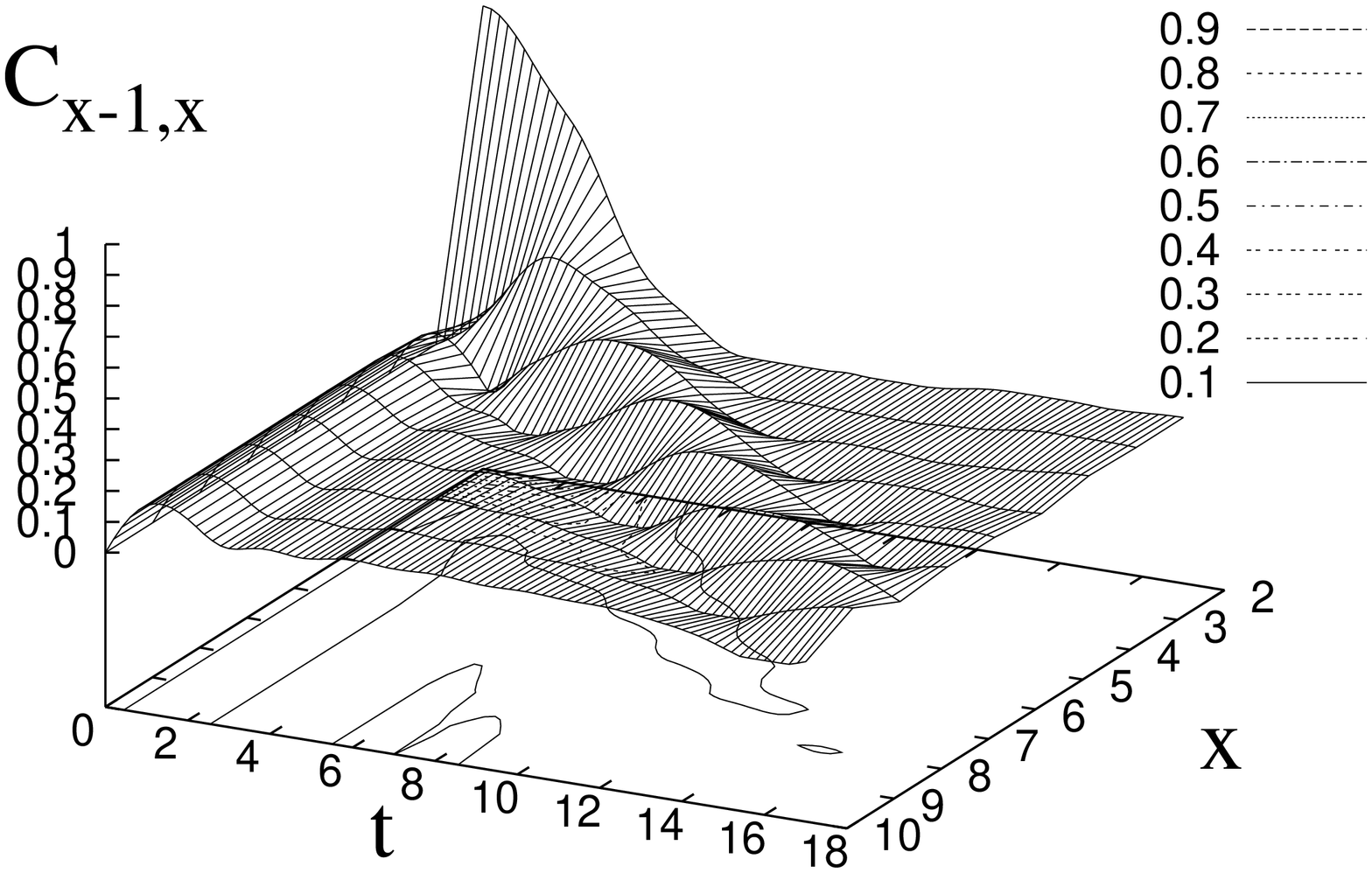}
\includegraphics[width=.29\linewidth,angle=-90]{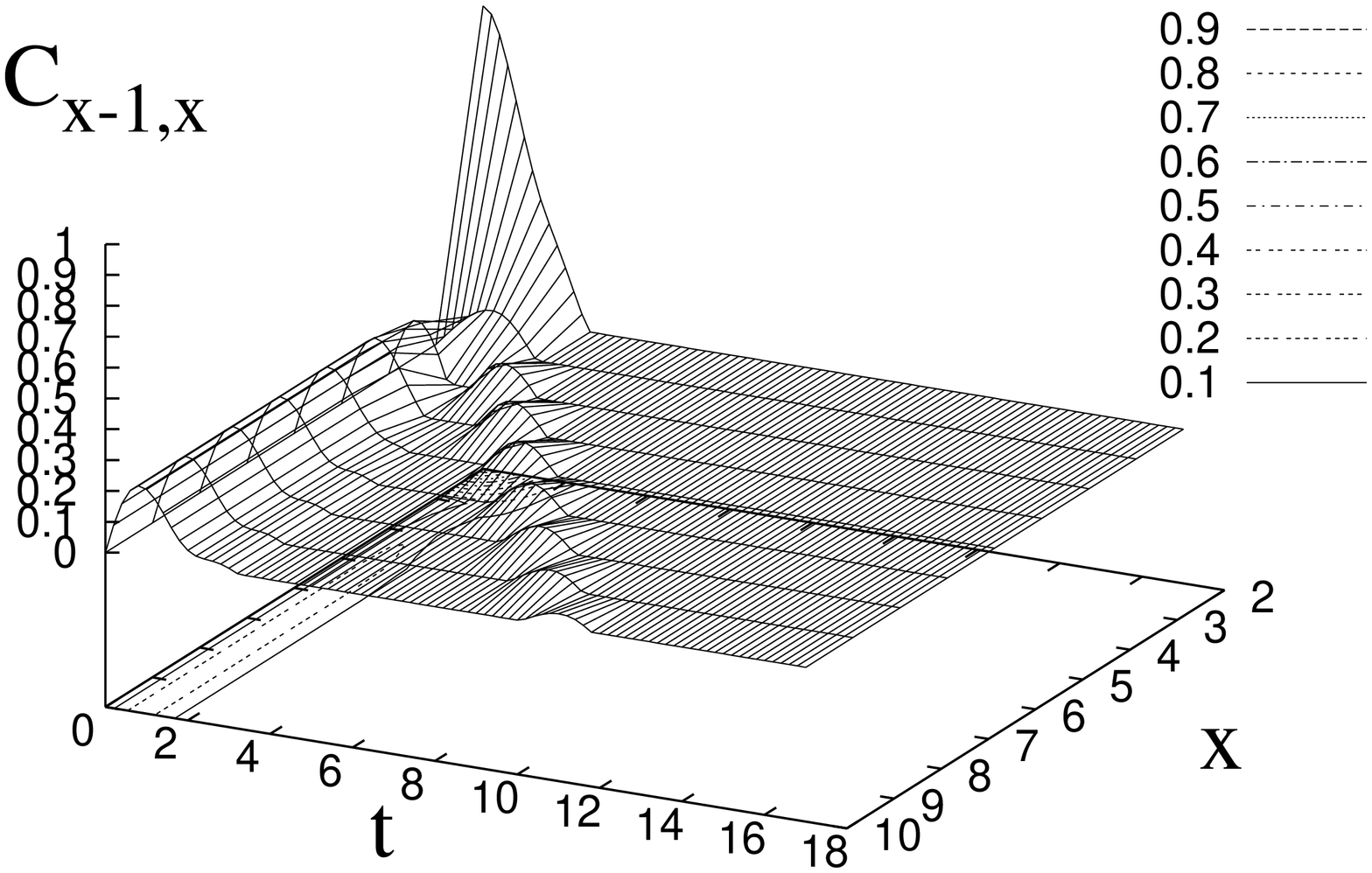}
\includegraphics[width=.29\linewidth,angle=-90]{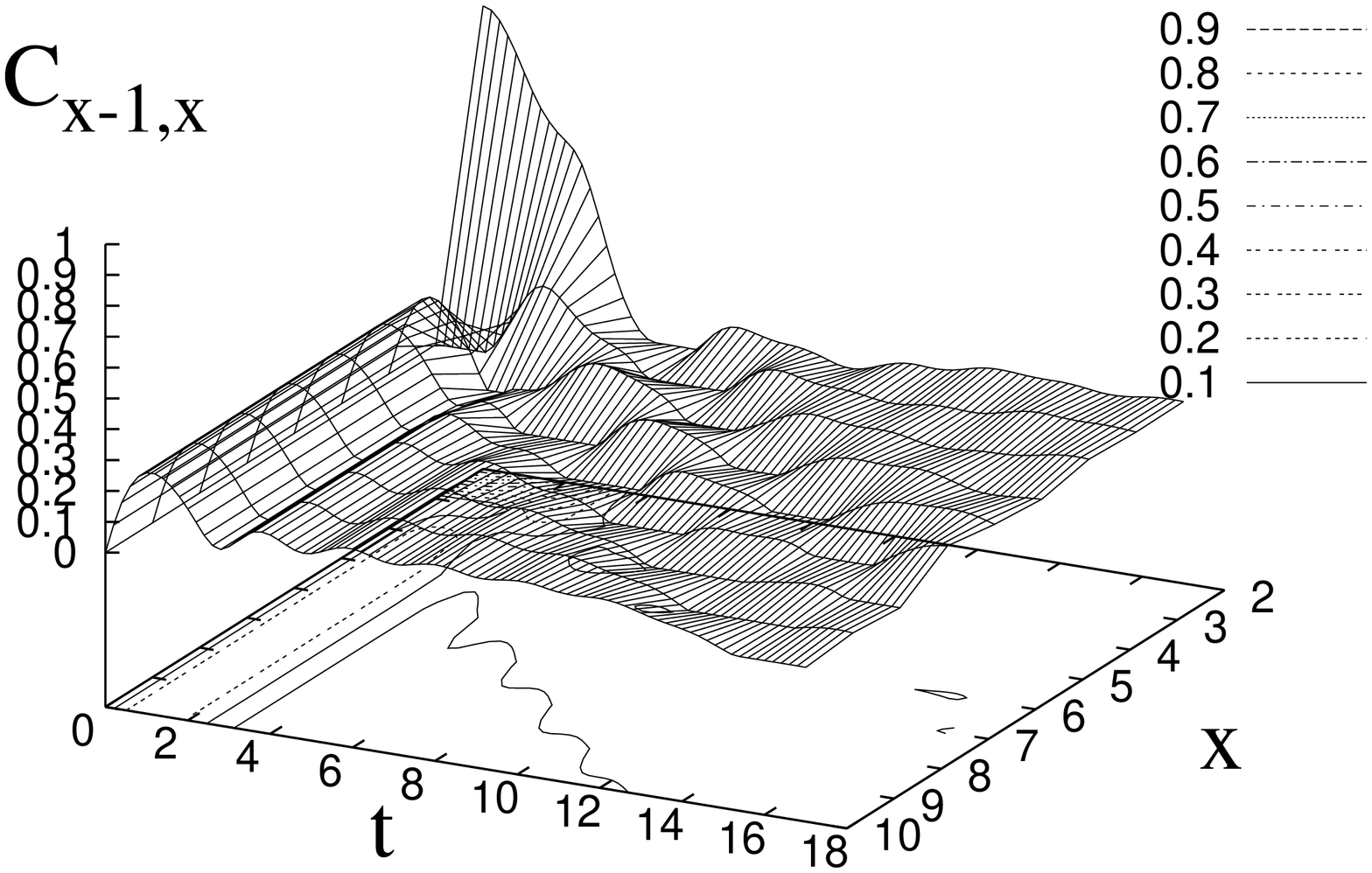}
\includegraphics[width=.29\linewidth,angle=-90]{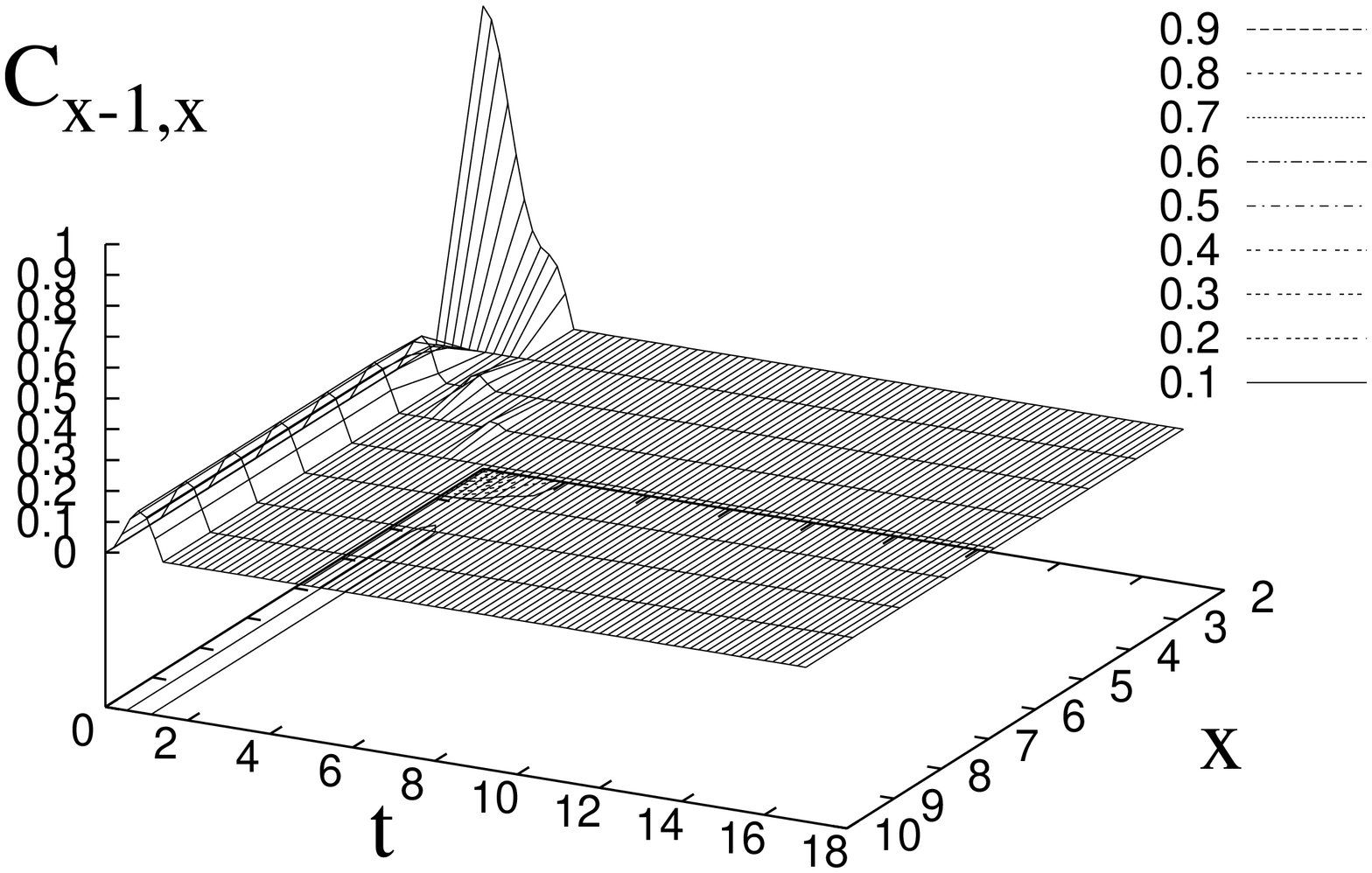}
\caption{
{\em Top panel - }
The nearest neighbor concurrence for the anisotropy $\gamma=0.1$
and two different values of $\lambda$ ($\lambda=0.5$ left,$\lambda=1$ right). 
As for the isotropic model, also here the singlet gets propagated 
roughly with velocity $v=\lambda$.
There is a new feature, though, which is the small concurrence wall 
created from the vacuum.
At the critical coupling $\lambda=1$ the propagation
is damped slightly stronger, but there are no drastic changes.
{\em Middle panel -} The nearest neighbor concurrence for $\lambda=0.5$ 
and medium anisotropy $\gamma=0.5$; that is where the 
spin-wave approximation breaks down.
Also here the singlet gets propagated roughly with velocity 
$\lambda$ in proper units, but it is stronger damped than for $\gamma=0.1$.
The creation from the vacuum is instead sharply enhanced and survives
at least until it reaches the propagating pulse.
At the critical coupling, the nearest neighbor
concurrence from the initial singlet dies out immediately and so does
the vacuum creation. Only few bumps are indicating a glimpse of 
a propagation.}
\label{C1-Vac}
\end{figure}
\end{minipage}
\begin{multicols}{2}
Whereas the initial slope of concurrence creation agrees 
with the exact calculation, the maximum  differs 
significantly even for $\gamma=1$, indicating that the kinetic term
dispersively suppresses the concurrence evolving from the vacuum.
In fact, the pure vacuum signal dies out very quickly. 
Towards the critical coupling and the Ising model, 
the damping of the concurrence propagation gets stronger. 
The vacuum signal survives much longer for
medium $\lambda$ and $\gamma\longrightarrow 1$ such 
that for $\gamma=0.5$ and $\gamma=1$ it interferes with the 
propagating singlet. 
Although the damping of the propagation becomes 
stronger at the critical coupling,
nevertheless it is of pure dynamic origin 
and not related to the quantum phase transition since,
for generic $\gamma$, the energy of the vacuum is higher 
than the ground state energy (where the 
critical behaviour is encoded in). Consistently, the damping turned out
to be independent of the size of the chain.\cite{Sachdev00,remark:QPT}

For $\gamma=0.1$ the most notable effect is the propagation of
the singlet as for the isotropic model. 
The concurrence for this $\gamma$ is shown in the mid pannel of 
Fig. \ref{C1-Vac}:
We still see a clear propagation of the concurrence, which is only
slightly stronger damped than for the isotropic model, but the creation
from the vacuum is much enhanced and survives longer.
A ``shoulder'' appears in the singlet peak of
$C_{i,i+1}$, i.e. on the original singlet position.
A better understanding of the evolution of the initial entangled state 
can be reached by evaluating the overlap of the reduced density matrix 
with the four Bell states. This is shown in  Fig. \ref{fidel:g0-5l0-5})
where the four plots refer to the fidelity to the corresponding Bell 
states indicated above.
We see that the singlet-fidelity
(shown in the upper leftmost picture in Fig. \ref{fidel:g0-5l0-5})
does not have a shoulder, meaning that this feature is an effect associated 
to a formation of a triplet-type Bell state.
In the plot of the fidelity for $\gamma=\lambda=0.5$ 
(figure \ref{fidel:g0-5l0-5}) we
see that the main contribution to the propagation is coming
from the singlet and the triplet 
$1/\sqrt{2}(\ket{\up\down}+\ket{\down\up})$,
which we call the $0$-triplet.
The initially predominant signal is the singlet;
it seems though, that the singlet signal decays quicker than
the $0$-triplet. Hence could it be that after some time
the singlet switched to the $0$-triplet. Furthermore, we extract from
the rightmost pictures of figure \ref{fidel:g0-5l0-5}, showing the fidelities 
of the $\pm$-triplets $1/\sqrt{2}(\ket{\up\up}\pm\ket{\down\down}$,
that the entanglement created from the vacuum is of triplet type
as estimated above~(\ref{C-PERT}).
\onecolm
\begin{minipage}[h]{\linewidth}
\begin{figure}\centering
\includegraphics[width=.3\linewidth,angle=-90]{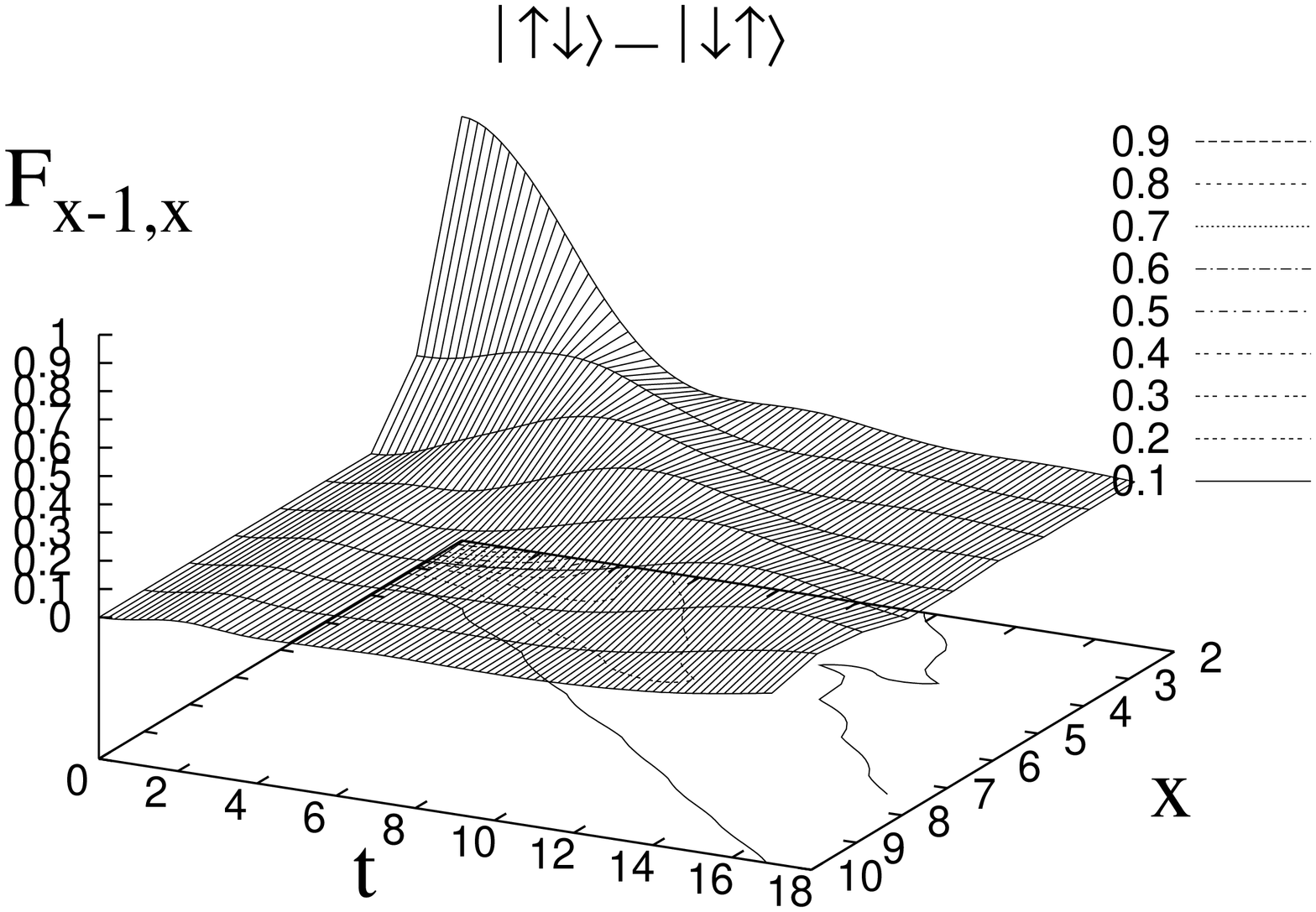}
\includegraphics[width=.3\linewidth,angle=-90]{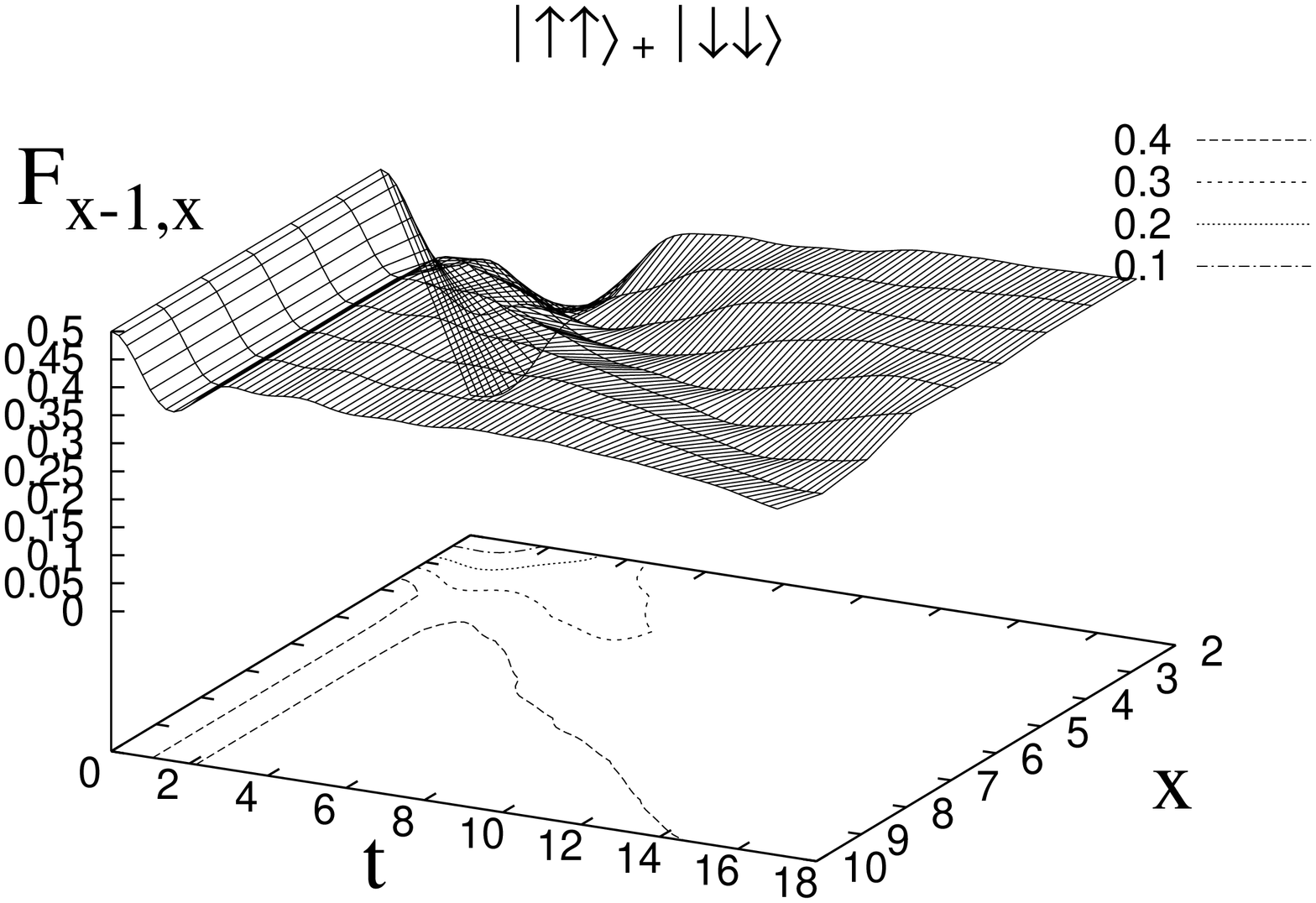}
\includegraphics[width=.3\linewidth,angle=-90]{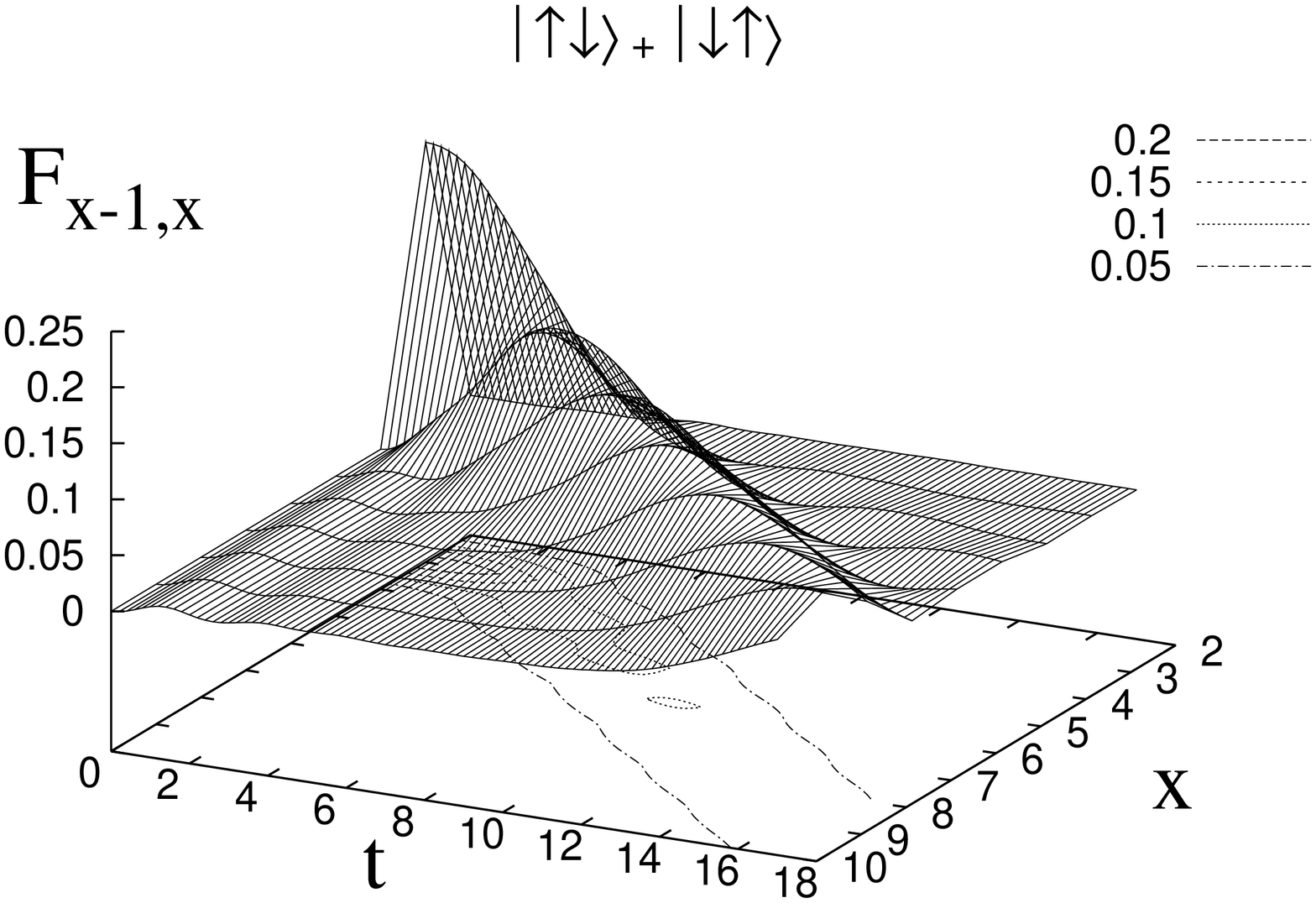}
\includegraphics[width=.3\linewidth,angle=-90]{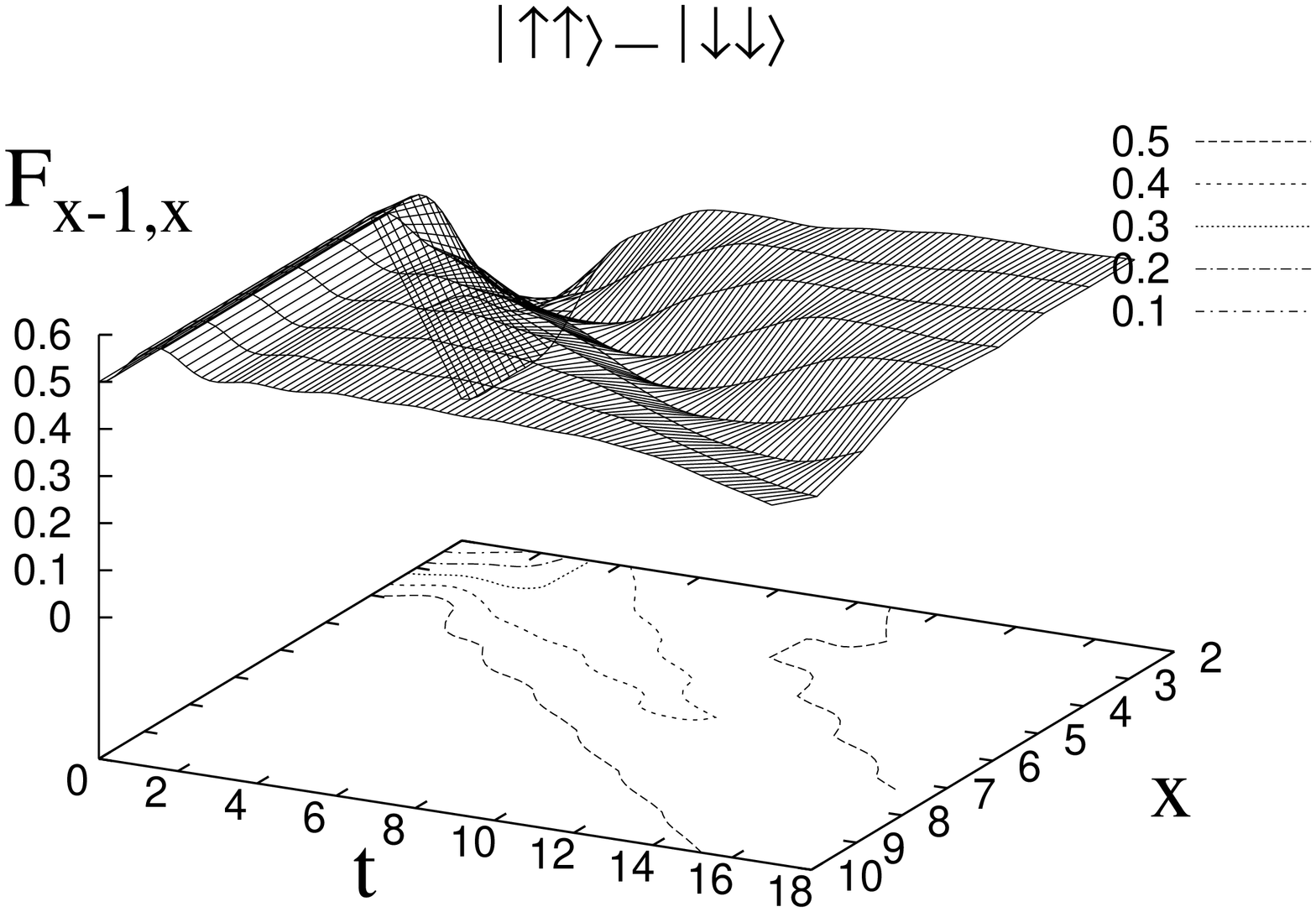}
\caption{The fidelity, here the portion of the  four Bell states,
i.e a singlet and three triplets, in the reduced density matrix is drawn for 
$\gamma=\lambda=0.5$. Since an equal mixture of two Bell states
is a disentangled state, the pictures show that
the predominant Bell state in the propagation is indeed the 
singlet (upper leftmost picture),
even though the triplet with zero $S^z$ component (lower leftmost picture)
decays slower and eventually could become dominent after sufficient time.
It can also be seen that the vacuum creation is dominated by the 
Bell state antisymmetric under $\sigma^x$ spin flip, i.e. 
$1/\sqrt{2}(\ket{\up\up}-\ket{\down\down}$ (lower rightmost picture).
In fact it grows on cost of its symmetric ``brother''.}
\label{fidel:g0-5l0-5}
\end{figure}
\end{minipage}

\begin{minipage}[h]{\linewidth}
\begin{figure}\centering
\includegraphics[width=.3\linewidth,angle=-90]{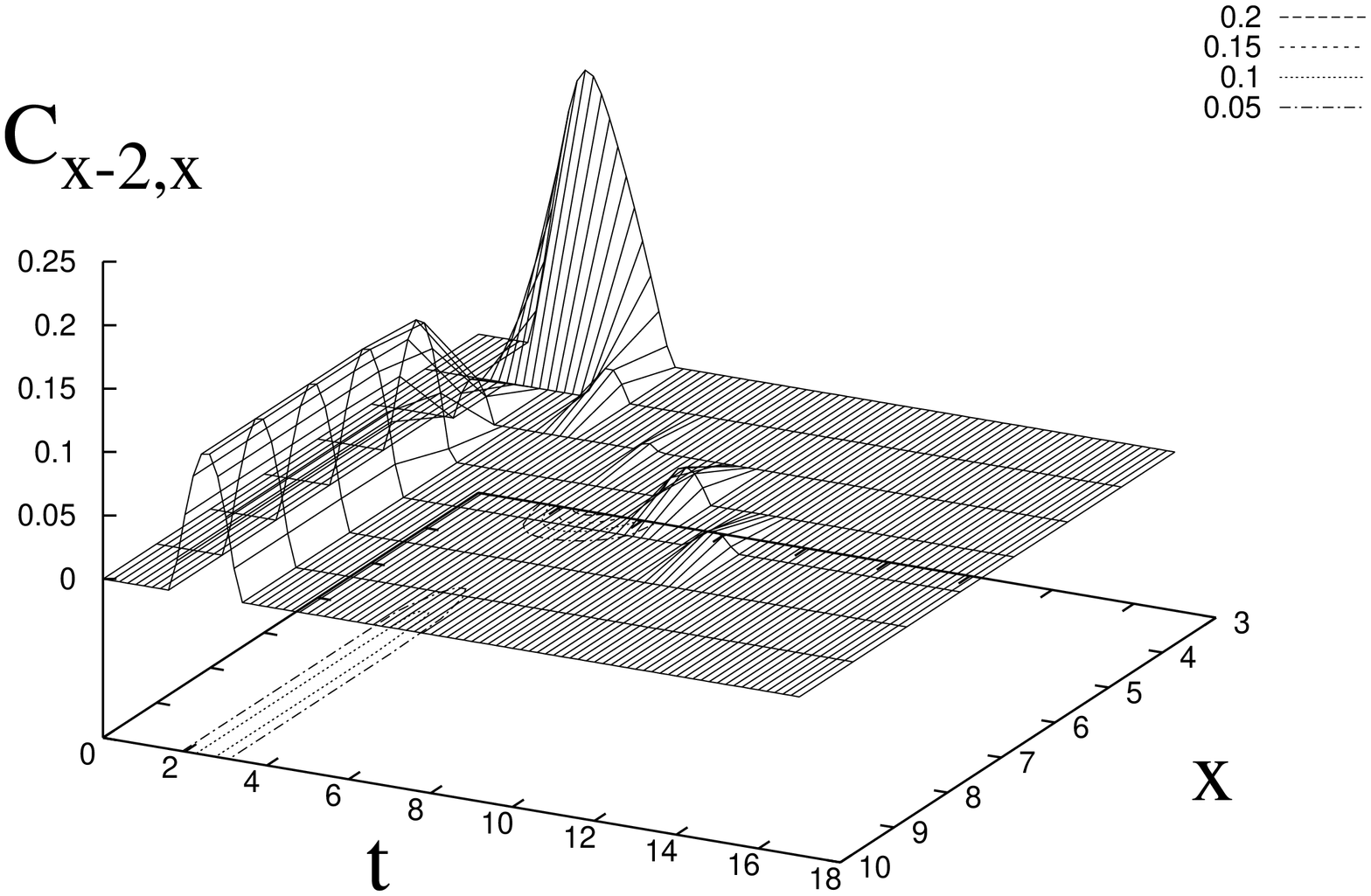}
\includegraphics[width=.3\linewidth,angle=-90]{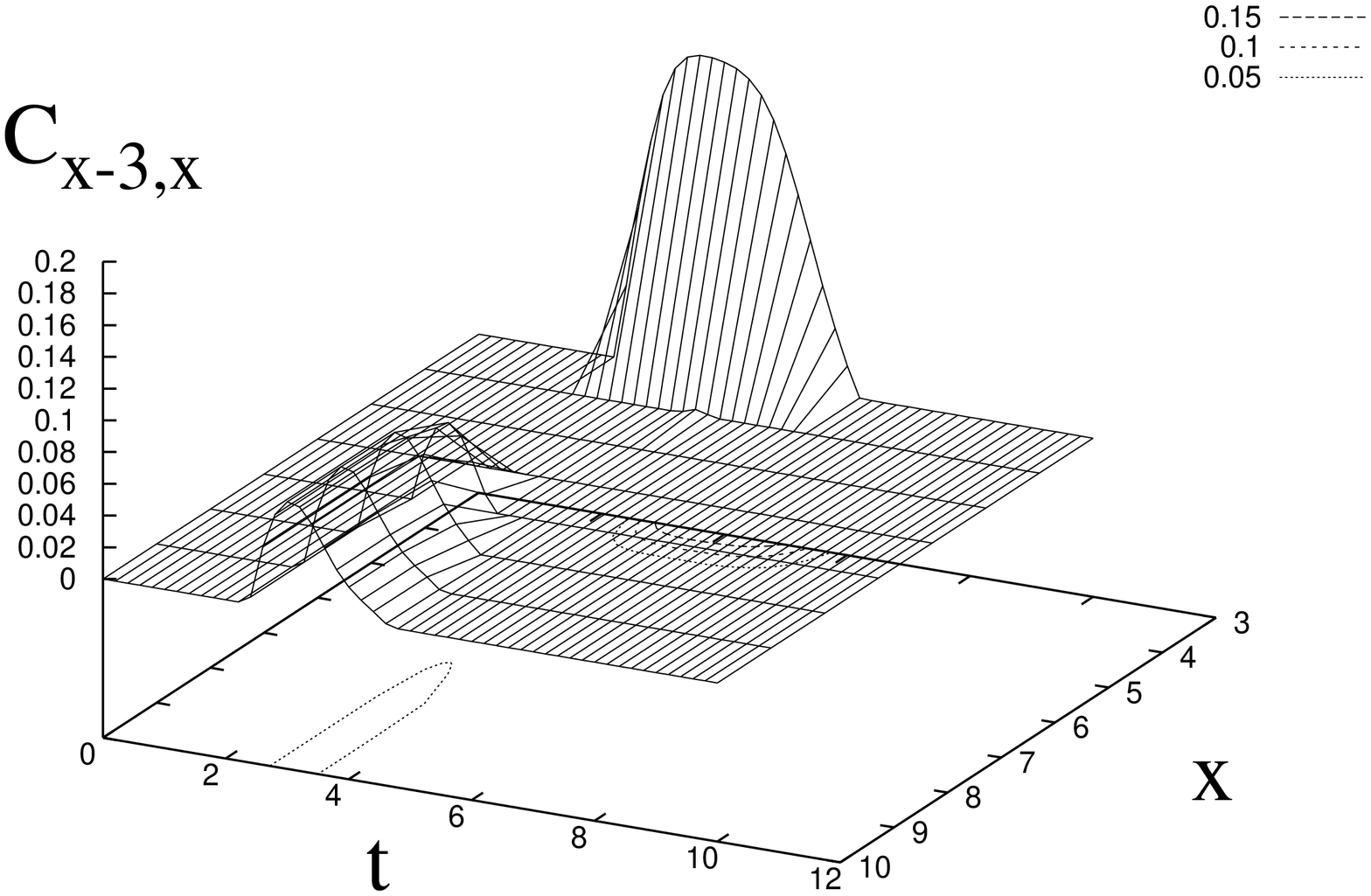}
\caption{
{\em Left -} The next nearest neighbor concurrence $C_2$ for the Ising model
far from the critical coupling. At the critical coupling it is zero.
{\em Right -} The next next nearest neighbor concurrence $C_3$ for the Ising model
far from the critical coupling. At the critical coupling, also $C_3$ vanishes.
At $\lambda=0.5$ we see a considerably large signal, which is due to
an EPR-type propagation of a ``split'' singlet in the sense that
the singlet somewhat splits up into two which separately propagate
in opposite direction. Much as the two EPR photons originated from the 
relaxation of an $s$ state of an atom at rest. 
The first EPR-signal is that one obseved in this figure
and corresponds to the concurrence between site $0$ and $3$, 
which are located one site to the left and right, respectively, of the 
original singlet sitting on sites $1$ and $2$. This concurrence propagation
was also observed for the isotropic $XY$ model and we expect that this
signal should continue to larger distances, though stronger suppressed
for generic $\gamma$. It is absent for the Ising model at the critical coupling.
}
\label{C2andC3}
\end{figure}
\end{minipage}

\begin{minipage}[h]{\linewidth}
\begin{figure}\centering
\includegraphics[width=.30\linewidth,angle=-90]{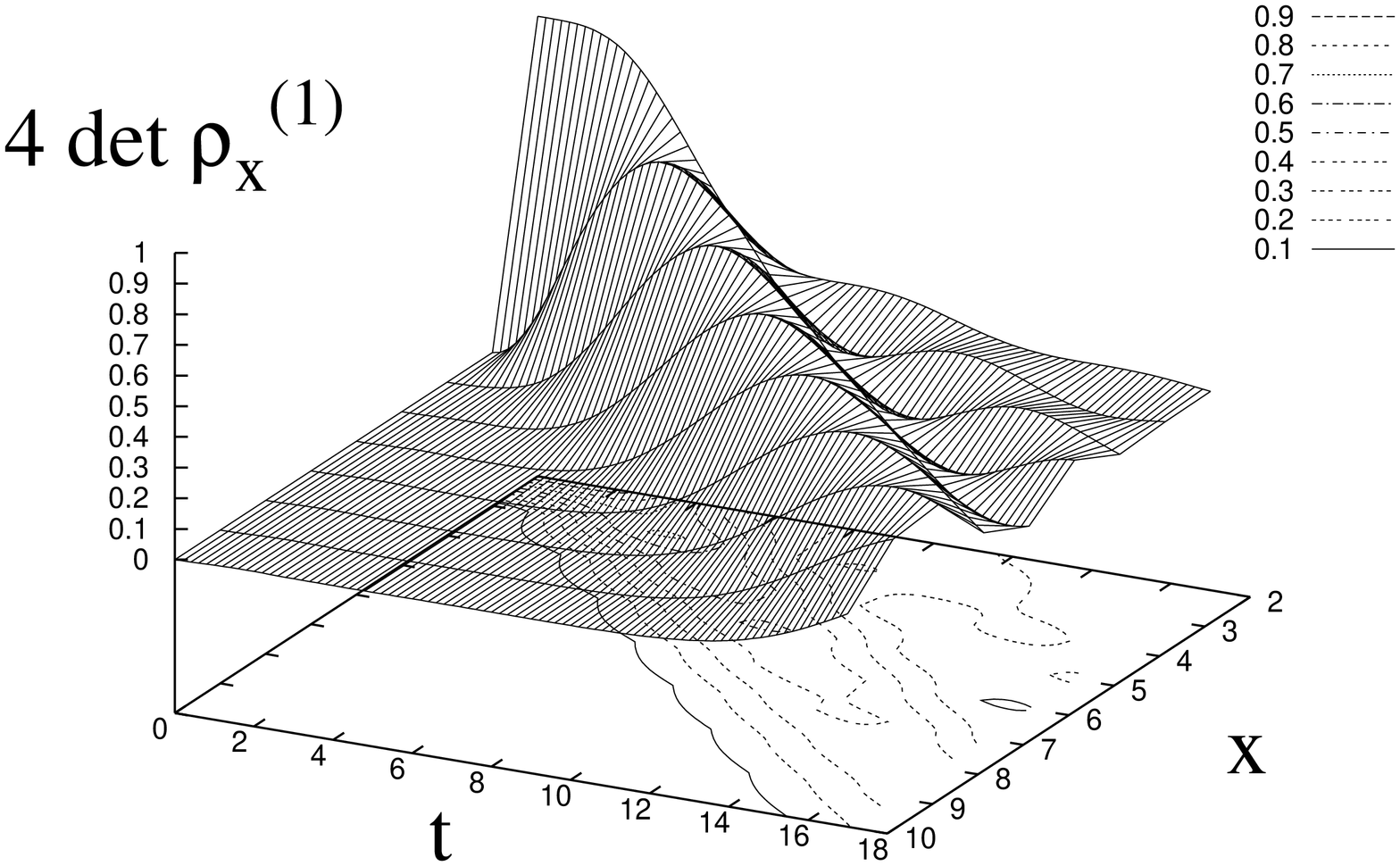}
\includegraphics[width=.30\linewidth,angle=-90]{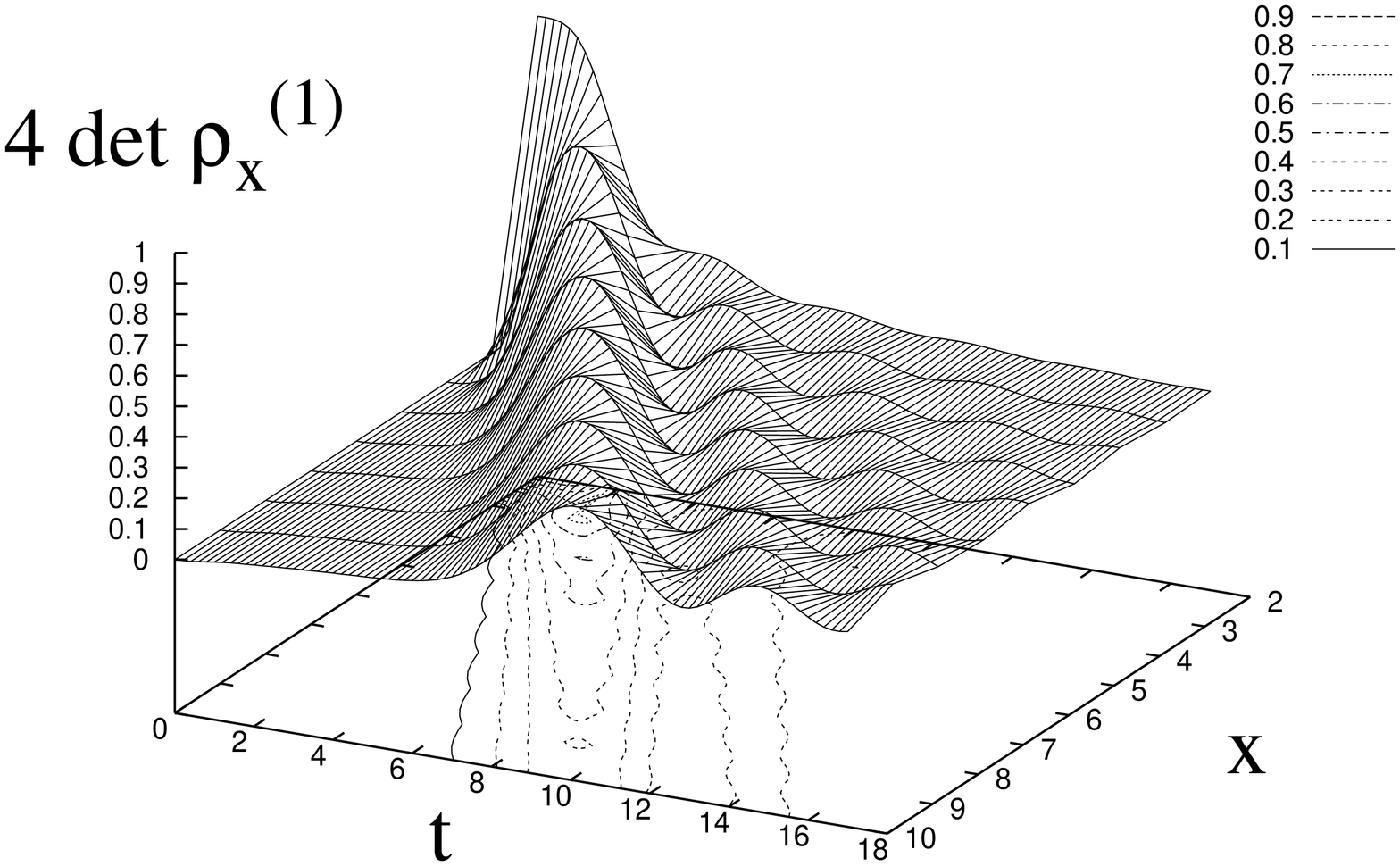}
\includegraphics[width=.30\linewidth,angle=-90]{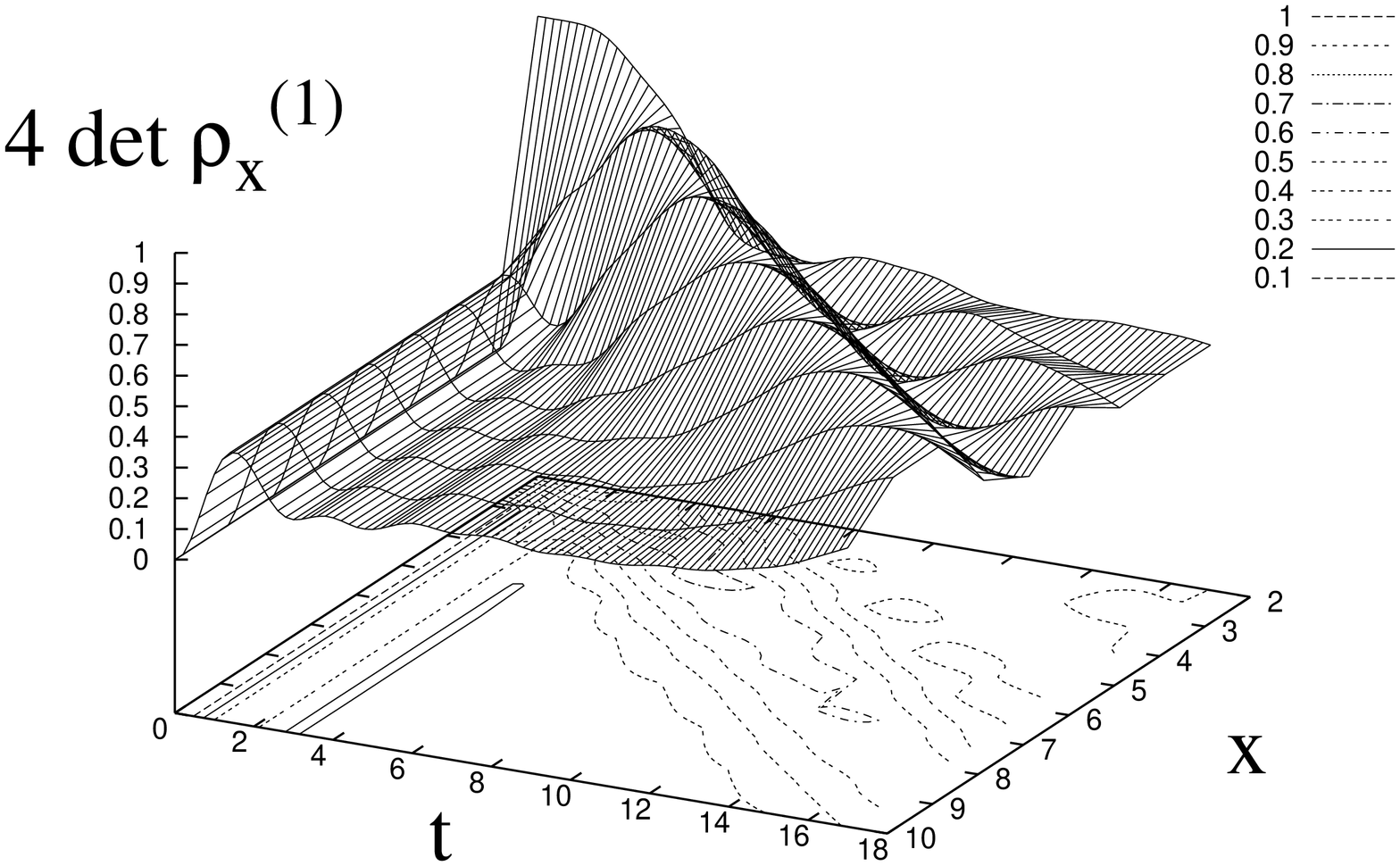}
\includegraphics[width=.30\linewidth,angle=-90]{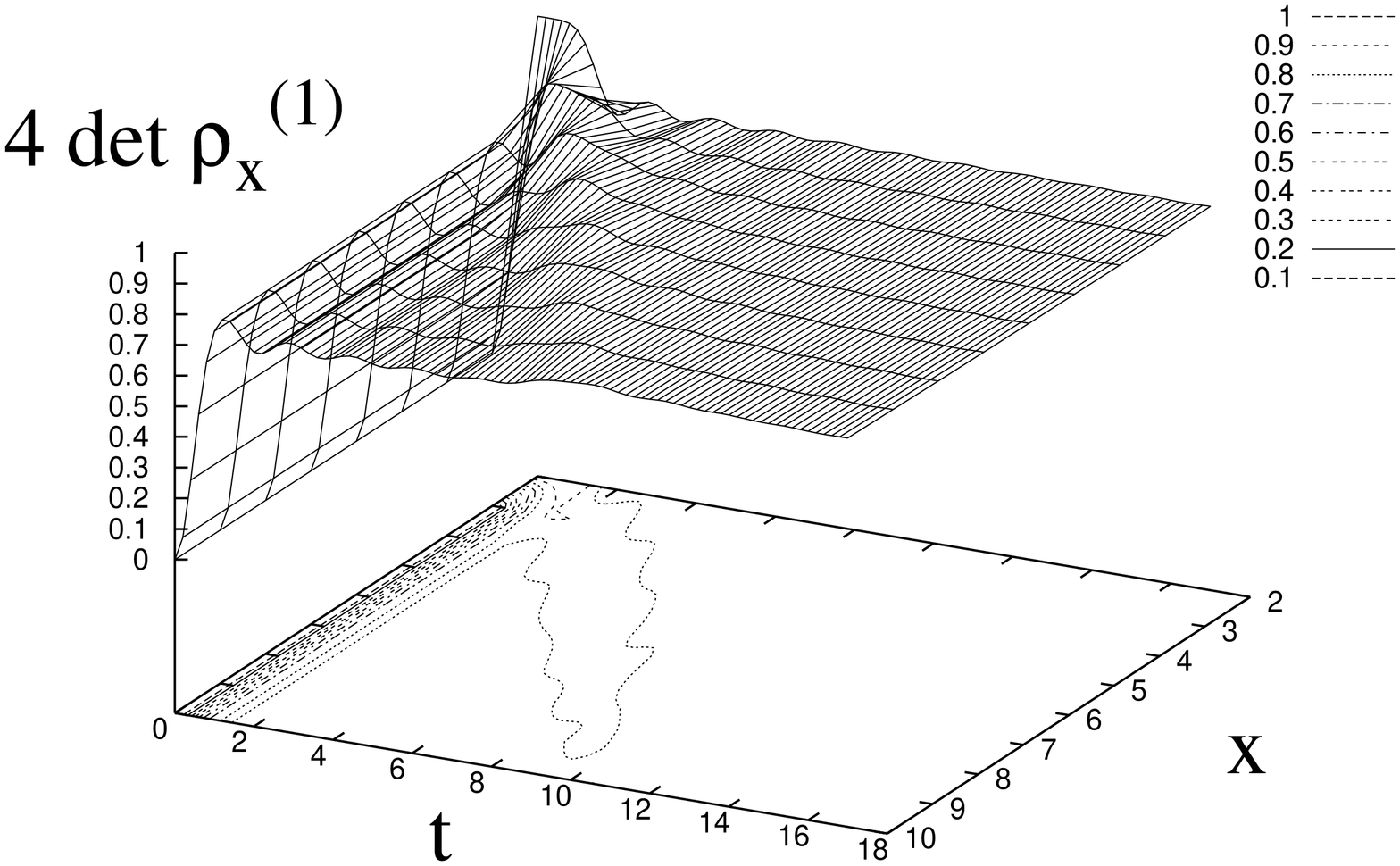}
\caption{The total tangle of site $x$
{\em Upper panel -} for $\gamma=0.1$. the plot qualitatively is very
similar to the nearest neighbor concurrence squared; only the propagation
is enhanced. This holds true for $\lambda=0.5$ (left) and at
critical coupling.
{\em Lower panel -} for $\gamma=1$. 
In contrast to the nearest neighbor concurrence there is a 
clear propagating signal here fro $\lambda=0.5$ (left). It is mounted 
on top of a non-zero background signal coming from the vacuum. 
At critical coupling the vacuum background of the one-tangle is grown
up to about $0.75$, whereas the propagation is hardly visible.}
\label{totaltangle}
\end{figure}
\end{minipage}
\begin{multicols}{2}
For the transverse Ising model, the shoulder on the original singlet position
and the vacuum creation get even more pronounced, but all the signals 
die out much quicker. 
For $\lambda=0.5$ one cannot speak any more of a clearly propagating
entanglement signal (lower panel in Fig. \ref{C1-Vac}).
At the critical coupling, the propagating signal disappeared completely,
small revivals of which can still be seen
near the critical coupling.
Whereas the features in the fidelity concerning the pulse propagation
are essentially the same as described above for $\gamma=\lambda=0.5$,
it is interesting to note that at the critical coupling,
all four fidelities seem to tend to a finite homogeneous value 
(without figure). 
The next-nearest neighbor 
and next-next-nearest neighbor concurrence is shown in Fig. \ref{C2andC3}.
Both show a narrow wall created from the vacuum, which for $C_3:=C_{x-3,x}$ 
is broader.
$C_3$ unveils an additional feature: it shows a large contribution
at $t$ around $4$ at $x=3$. 
It is caused by the original singlet kind of ``splitting up''
into two fragments, which independently propagate in opposite directions.
As the singlet initially was positioned on the sites $1$ and $2$,
these fragments after some time should be observed on sites
$0$ and $3$, corresponding exactly to the peak observed in the 
figure \ref{C2andC3}.
This phenomenon resembles an EPR-type propagation, which
we observed already for the isotropic model (see Fig. \ref{treconc}).
At the critical coupling $C_2$ and $C_3$ identically vanish on the
domain of the demonstrated plots.

\subsubsection{The global tangle}

Is it possible to understand how the entanglement, originally stored
into the singlet, is going to share among all the spins in the chain? 
In order to get an idea about what might happen, we study the one-site reduced 
density matrix $\rho_1$. 
\onecolm
\begin{minipage}[h]{\linewidth}
\begin{figure}\centering
\includegraphics[width=.3\linewidth,angle=-90]{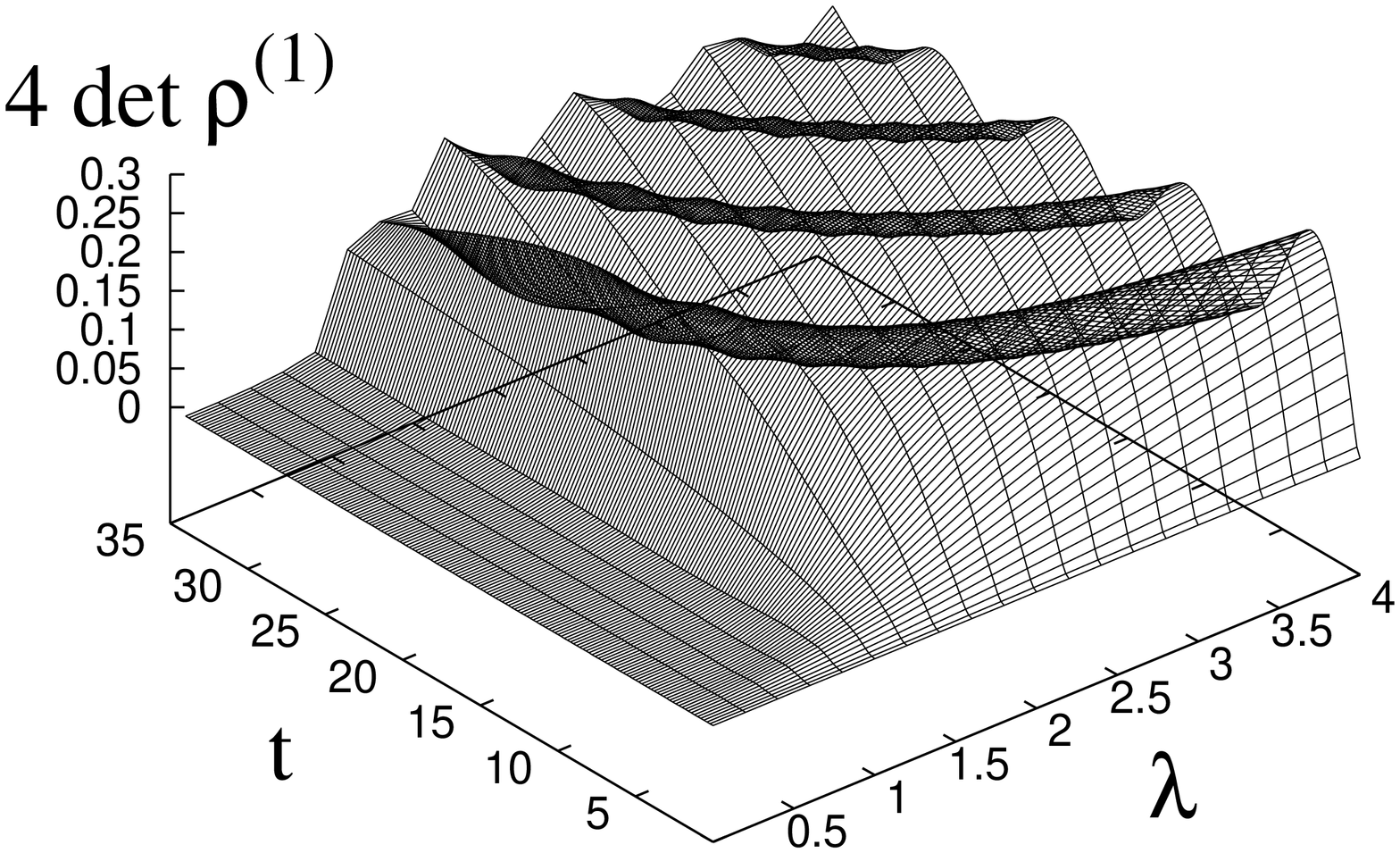}
\includegraphics[width=.3\linewidth,angle=-90]{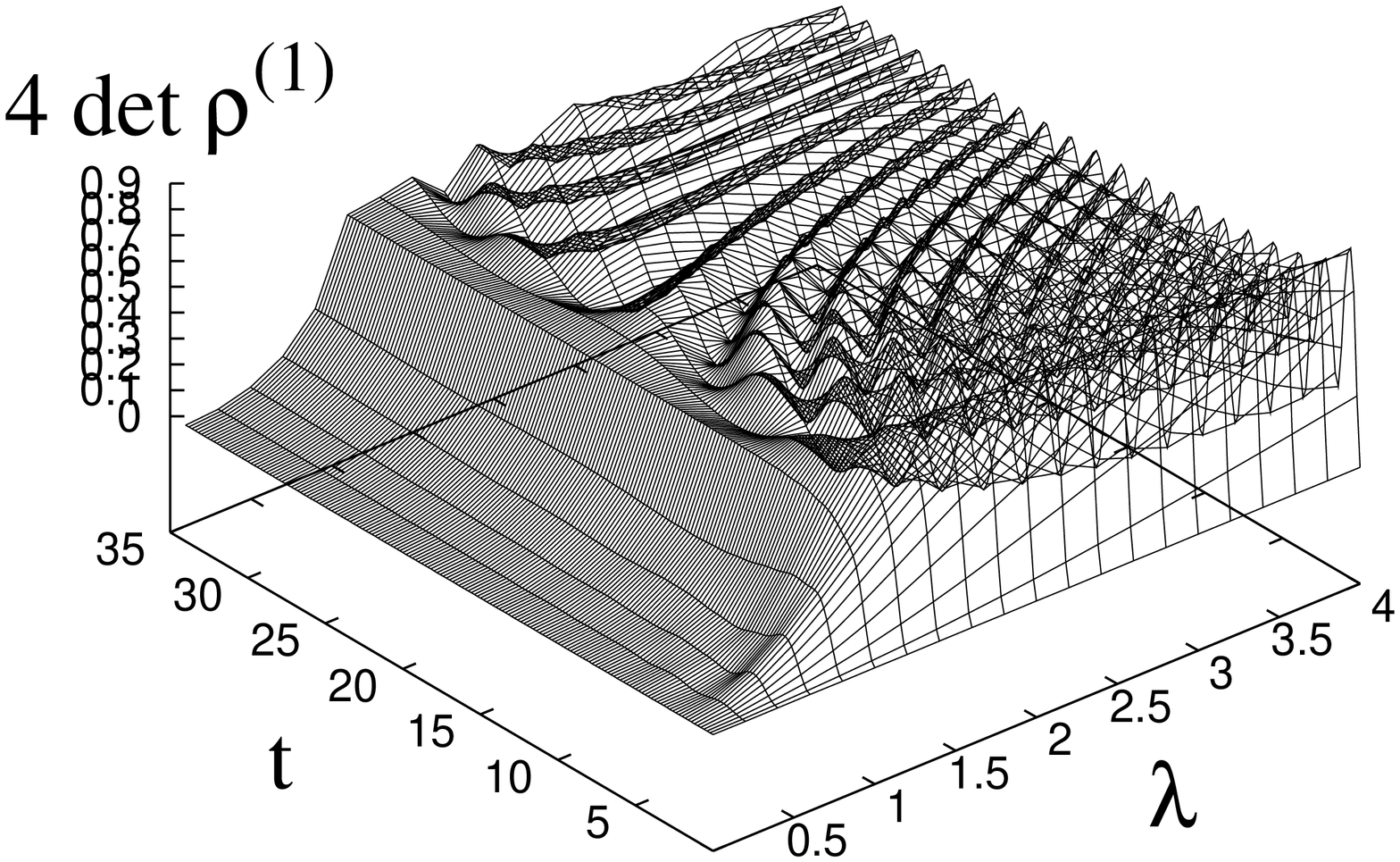}
\includegraphics[width=.3\linewidth,angle=-90]{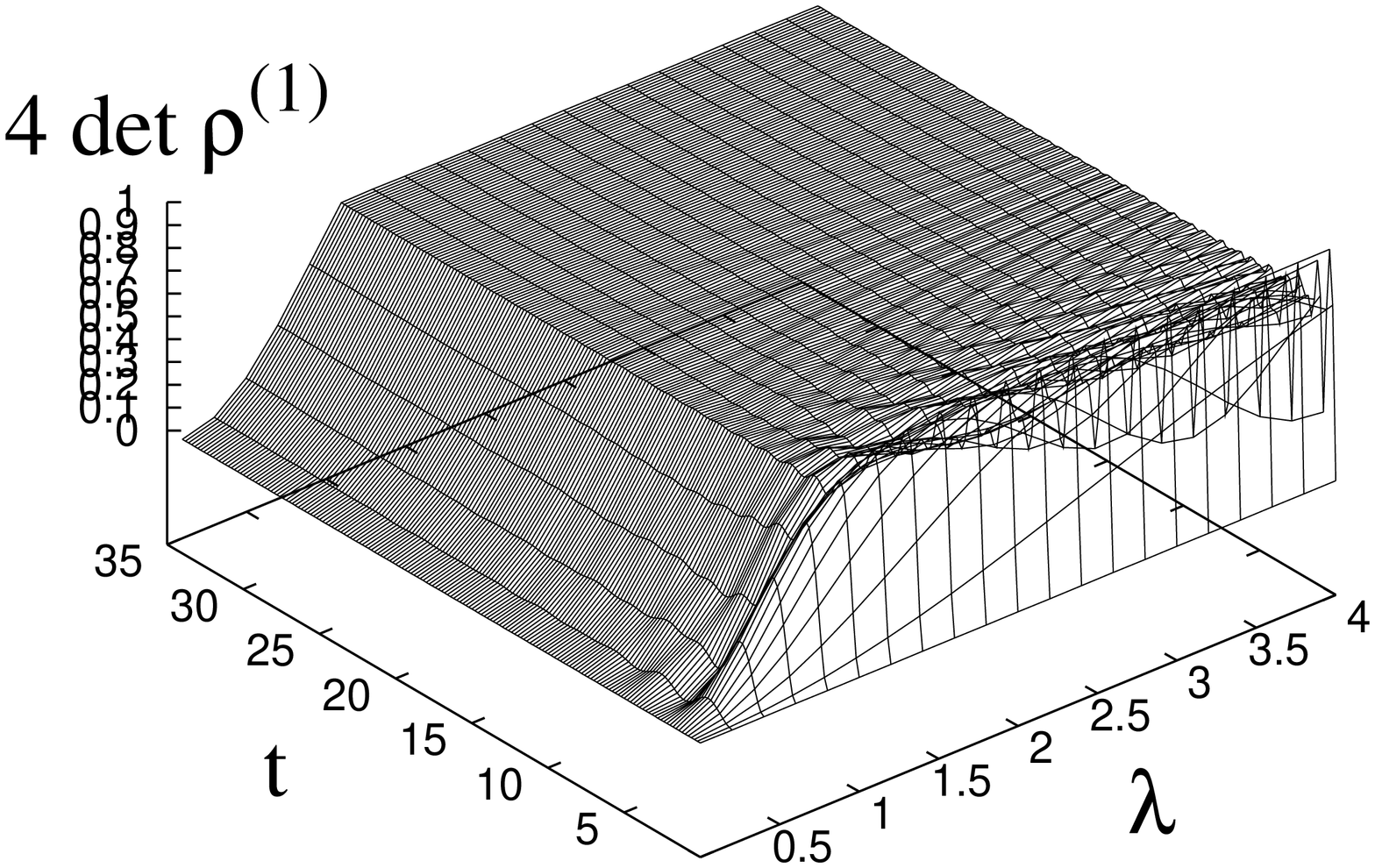}
\caption{The global tangle for the initial state beeing the vacuum and 
anisotropies $\gamma=0.1$, $0.5$, and $1$ from left to the right. 
{\em left :\/ small anisotropy.} For $\lambda<1$ there is hardly any entanglement created 
by the Hamiltonian from the vacuum. This abruptly changes at the critical 
coupling $\lambda=1$. The abruptness manifests itself in the large
exponent in a rough polynomial fit up to $\lambda=1$ at time $t=17$, 
which here gives $0.1\;\lambda^{8.2}$. 
For $\lambda>1$ strong oscillations emerge with an amplitude
of roughly $30\%$ of the average value. About $10\%$ oscillation
amplitude is also present for $\lambda<1$, though unresolved in the plot.
{\em middle:\/medium anisotropy.} Here we observe a much more smooth 
increase of the average entanglement, when $\lambda$ increases.
The corresponding rough fit is $.55\; \lambda^{3.3}$.
As a function of time it very soon saturates. 
Only beyond the critical interaction strength there are oscillation around
this value but with an amplitude of few percent of the average value. 
{\em right:\/quantum Ising model.} 
Towards the Ising model the saturation value of the tangle
increases more and more quickly with $\lambda$. The polynomial fit
is here $0.76\;\lambda^{1.7}$. The oscillations 
for $\lambda>1$ get more and more suppressed with increasing frequency.}
\label{vacuum4detrho}
\end{figure}
\end{minipage}
\begin{multicols}{2}
More precisely, we calculate the quantity 
$4 \det \rho_1(x)$, which is conjectured being a measure of the total entanglement, 
the site number $x$ participates in\cite{Coffman00}.
For three qubits, the difference between this quantity and the sum of
all squared concurrences with site number $x$ is the 3-tangle,
a measure for pure-state three-site entanglement.  
The results are shown in Fig. \ref{totaltangle}.
For $\gamma=0.1$ (upper panel) the result is qualitatively 
very similar to that found for the concurrence: there is an entanglement 
wave, propagating with velocity $\lambda$.
However, for higher $\gamma$'s, such wave is characterized by a much larger 
average value. As we shall see, this is an effect of entanglement dynamically 
generated from the vacuum. 
In order to get an idea of what part of the signal for the total 
entanglement (referring to the conjecture on the amount of 
residual entanglement, Ref.~\cite{Coffman00}) comes from the singlet,
we are next having a look at the total entanglement created solely from 
the vacuum, i.e. the initial state being the vacuum.
The results are shown in figure \ref{vacuum4detrho}.
For the $XY$ model, the vacuum signal identically vanishes.
For {\em small anisotropy} ($\gamma=0.1$) the  entanglement from the vacuum 
is negligible and it increases abruptly for $\lambda$ being considerably
near to the critical coupling; beyond this point,
strong oscillations appear. 
For {\em medium anisotropy} 
the amplitude of the oscillations for $\lambda>1$ is  suppressed and their
frequency increases. 
The created entanglement smoothly increases until  the critical $\lambda$.
At $\gamma=1$ a plateau is rapidly reached at $\lambda=1$ and the 
oscillations are severely suppressed. 
\\
For different $\gamma$ we see that the 
average entanglement increases for $\gamma \rightarrow 1$, whereas
the relative amplitude of its oscillation decreases.
It is worth noticing that the average entanglement is reached
on extremely short timescales, which  decrease with increasing $\gamma$. 

We then analyzed the evolution of the singlet onto the vacuum.
Sufficiently far away from the local perturbation
and for sufficiently short time we expect that the entanglement observed 
should be mainly that of the underlying vacuum. Whereas notable 
deviations should be observed  close to the 
propagating  concurrence. 
\onecolm
\begin{minipage}[h]{\linewidth}
\begin{figure}\centering
\includegraphics[width=.35\linewidth,angle=-90]{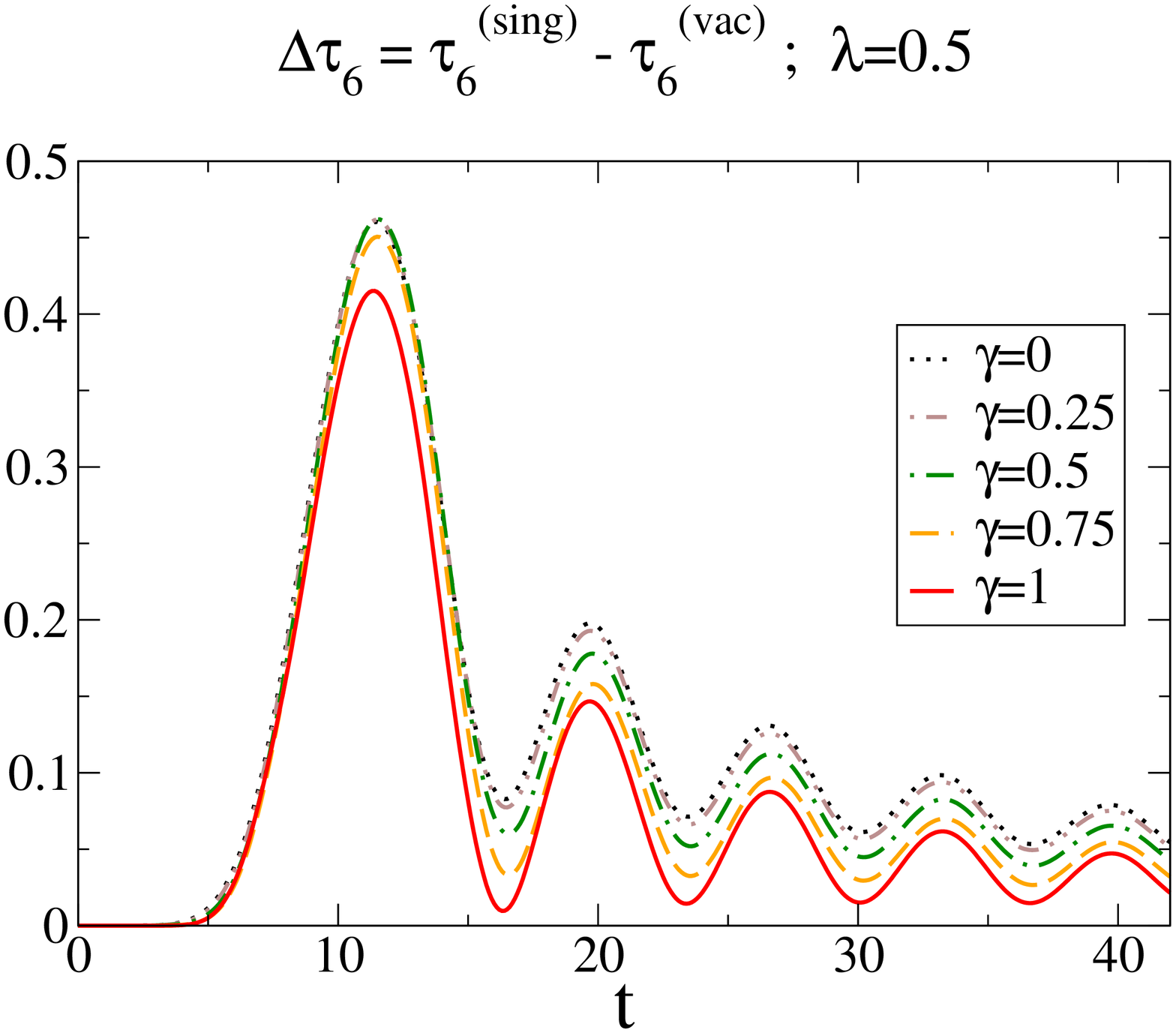}
\includegraphics[width=.35\linewidth,angle=-90]{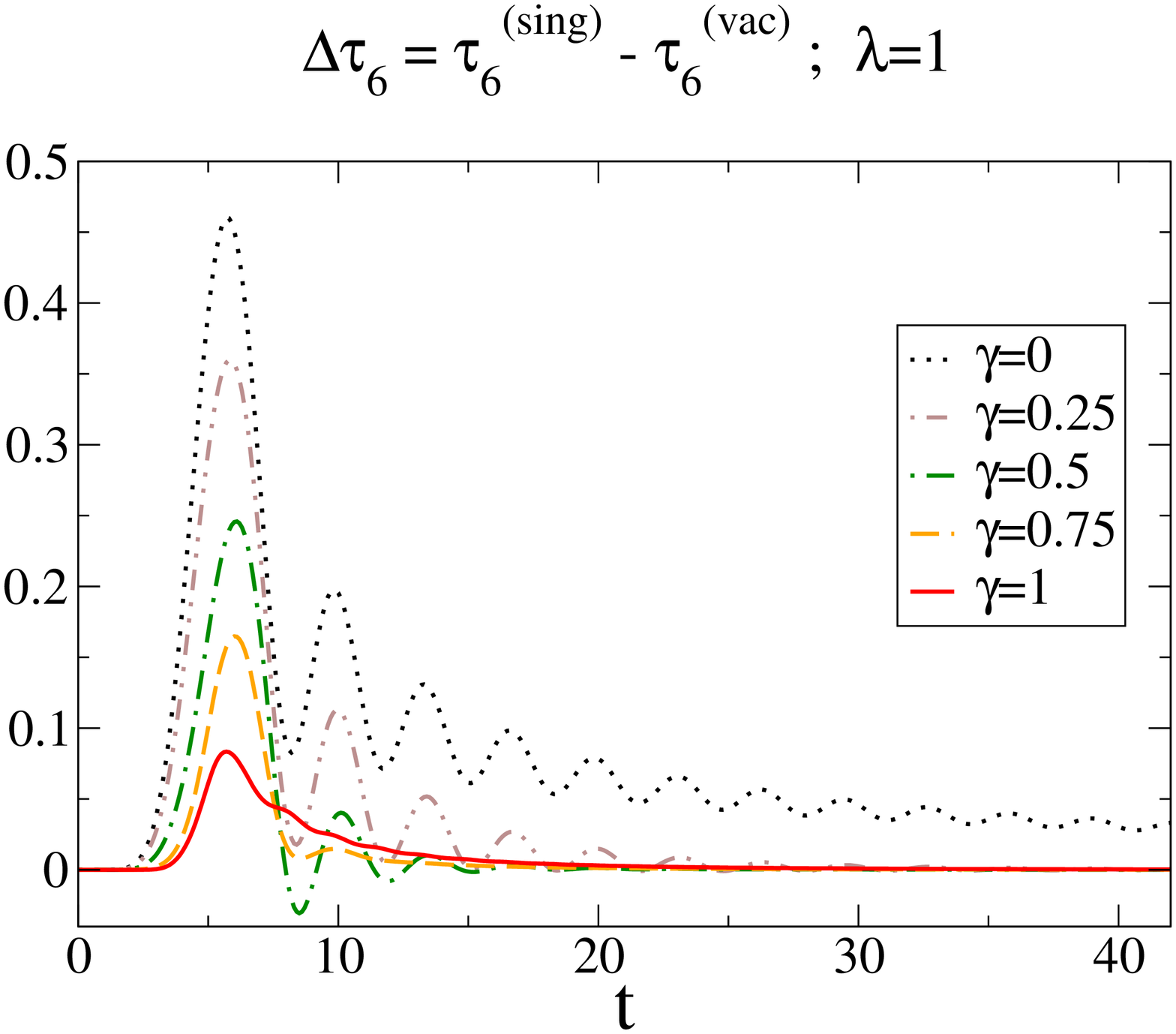}
\includegraphics[width=.35\linewidth,angle=-90]{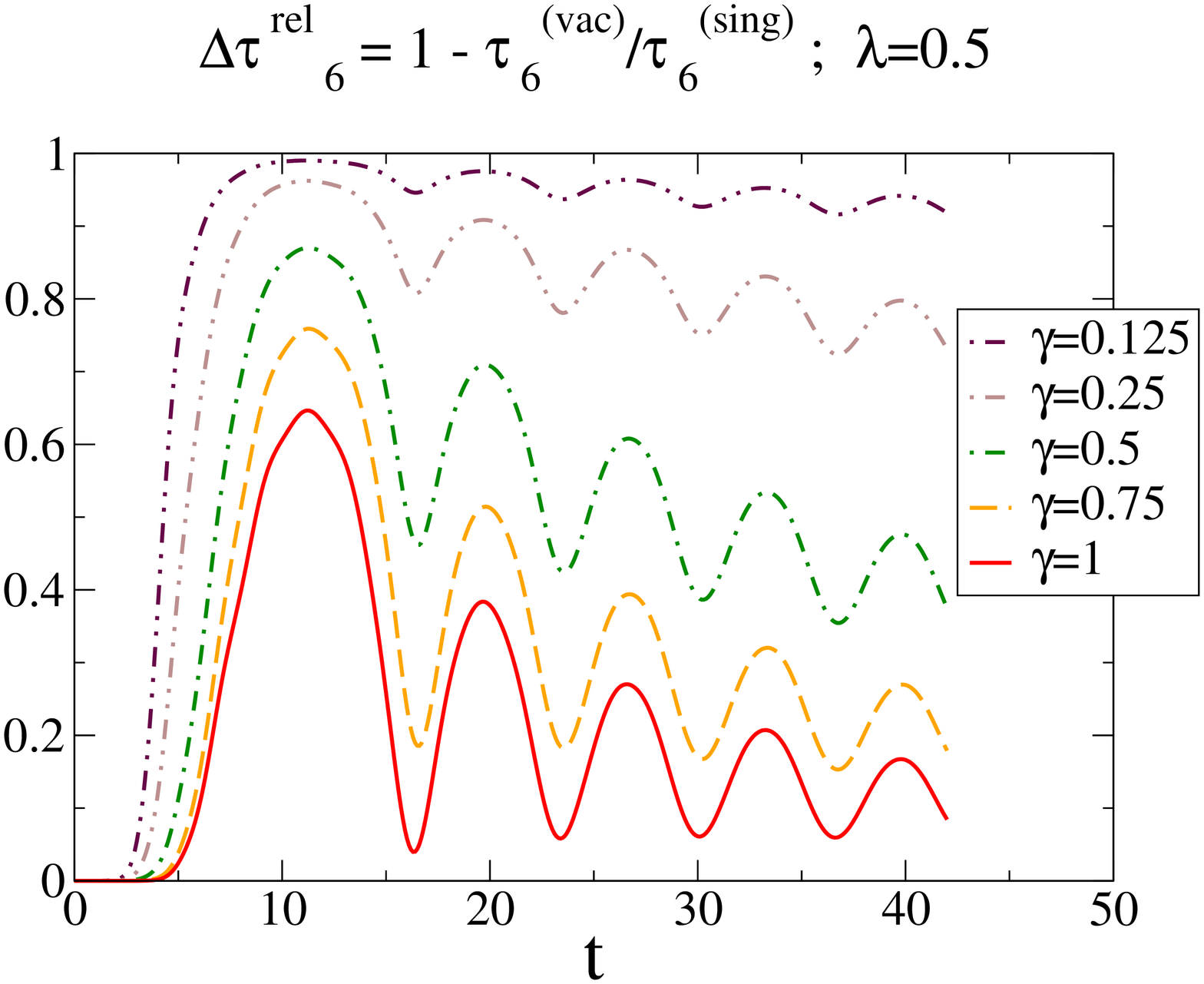}
\includegraphics[width=.35\linewidth,angle=-90]{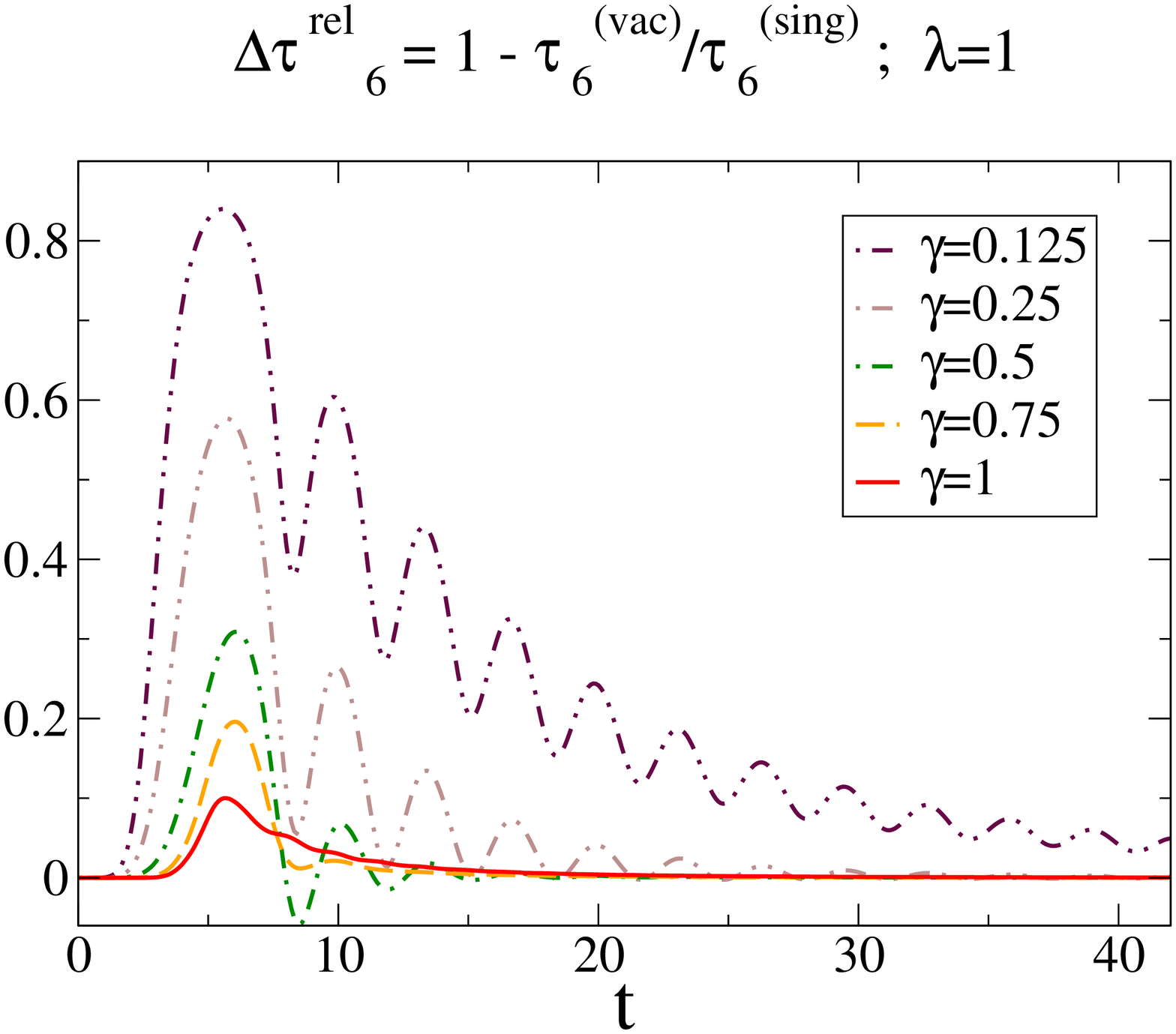}
\caption{The tangle deviation $\Delta\tau^{}_6$ (upper panel) 
and the relative tangle deviation $\Delta\tau^{rel}_6$ from the Vacuum 
tangle at site number $6$ for different values of $\gamma$ for a fixed 
value of $\lambda$: $\lambda=0.5$ (left) and $\lambda=1$ (right). 
It is nicely seen that the oscillation frequency does not depend on $\gamma$,
but their damping does. The damping is enhanced at critical coupling 
$\lambda=1$ due to the increasing relevance of the double spin-flips, which
overwhelm the singlet propagation.}
\label{reldeltangle-Vac-of-g}
\end{figure}
\end{minipage}

\begin{minipage}[h]{\linewidth}
\begin{figure}\centering
\includegraphics[width=.23\linewidth,angle=-90]{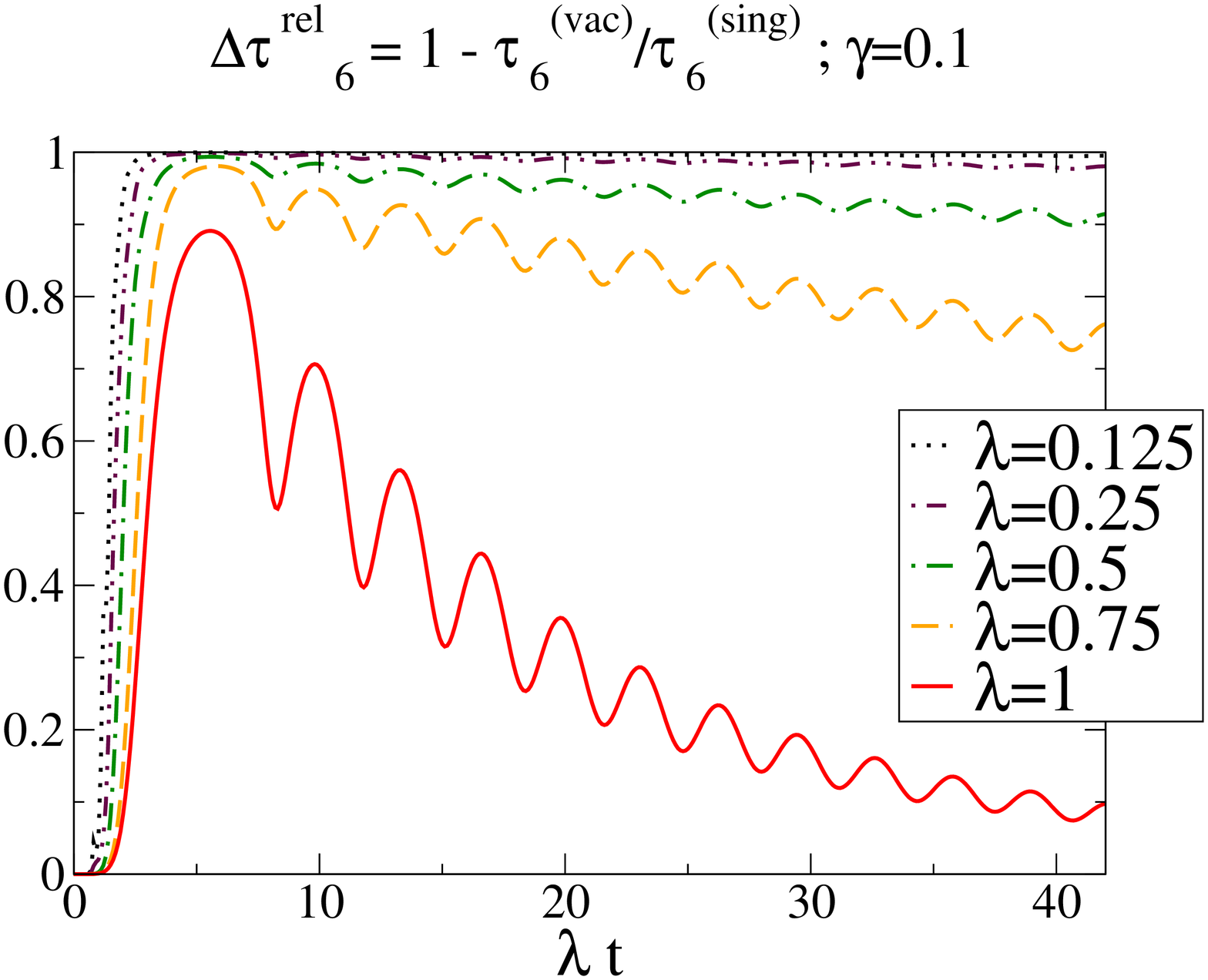}
\includegraphics[width=.23\linewidth,angle=-90]{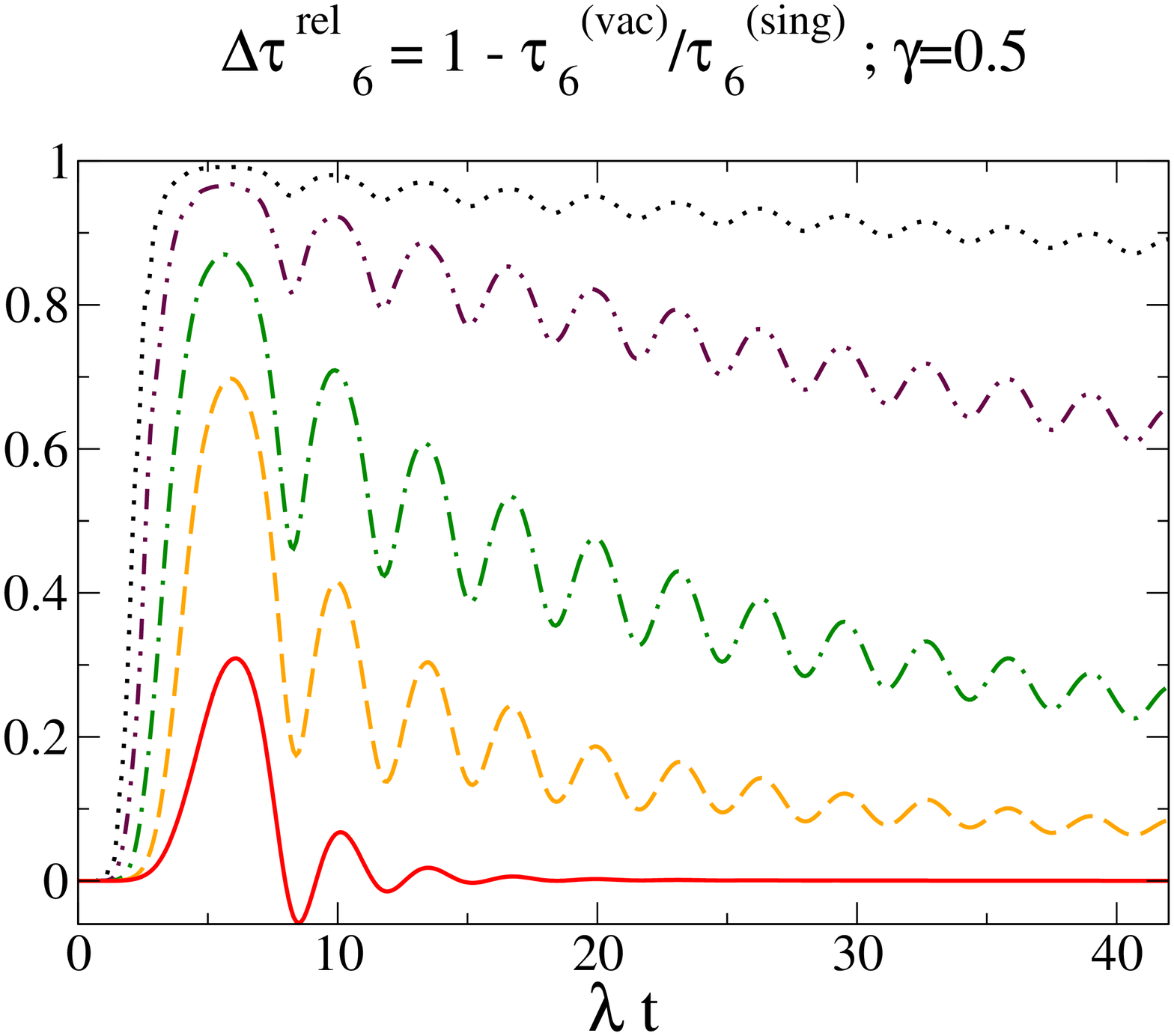}
\includegraphics[width=.23\linewidth,angle=-90]{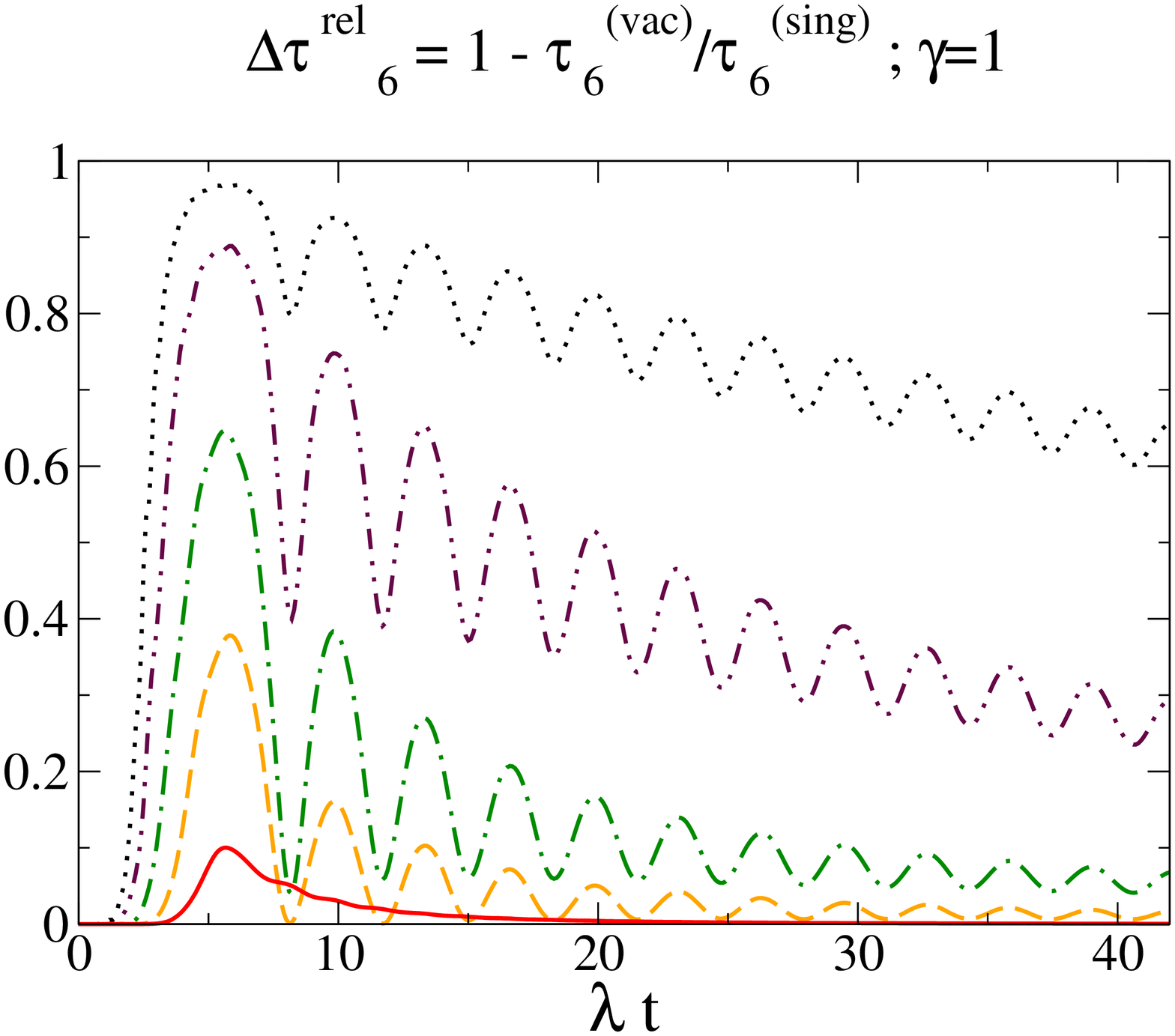}
\caption{The relative tangle deviation $\Delta\tau^{rel}_6$ from the Vacuum 
tangle at site number $6$ for different values of $\lambda$ for a fixed 
value of $\gamma$: $\gamma=0.1$ (left), $\gamma=0.5$, and $\gamma=1$ (right). 
It is plotted as a function of the reduced time $\tau=t/\lambda$;
it is nicely seen that the oscillation frequency grows linearly with 
$\lambda$.}
\label{reldeltangle-Vac-of-l}
\end{figure}
\end{minipage}
\twocolm
In order to separate the deviations due to the propagating perturbation
we subtract the global tangle emerging from the pure vacuum and study:
\beq\label{deltangle}
\Delta \tau_j:=4 ( \det \rho_j^{(1)} - \det \rho_{(vac)}^{(1)})\; .
\eeq  
We want to stress that it is not a priori clear whether the time 
evolution operator preserves  the relative order induced by the entanglement 
measures in the Hilbert space. It is 
visible in negative values of $\Delta \tau_j$ that this does not hold  
in general. Eventually, this is due to the 
fact that superposing (as well as mixing) orthogonal maximally 
entangled states of the same type diminishes the entanglement.
This makes an ``entanglement crossing'' of the vaccum and the singlet possible,
leading to a negative value of their difference.
It is worth noticing that we insert the singlet state into the vacuum;
this is not a superposition of the vacuum and some other state.

Below the critical $\lambda$, $\Delta \tau_j$ increases slightly with 
decreasing $\gamma$ but looks very similar for different values of $\gamma$.
It slowly tends to zero (left-upper part of Fig. \ref{reldeltangle-Vac-of-g}).
At the critical coupling $\lambda=1$ we observe immediate
convergence to zero for large enough $t$, 
(right-upper part of Fig. \ref{reldeltangle-Vac-of-g}).
This is much in contrast to the isotropic case, $\gamma=0$, which is 
also shown in the upper rightmost plot in Fig.~\ref{reldeltangle-Vac-of-g}.
This could indicate that at the critical coupling, though being 
in an excited state, the residual tangle be mainly originated 
by global properties of the initial state. 
A notable contribution from
the singlet is found at very small $t$, which gets even smaller
when $\lambda$ increases.
It is only for small up to medium $\lambda$,
that the singlet contribution survives much longer.
The data is consistent with the propagation velocity being $\lambda$, 
independent of $\gamma$ (Fig.~\ref{reldeltangle-Vac-of-g}). 

The surprising similarity of the difference signal for much different $\gamma$
(upper panel of Fig.\ref{reldeltangle-Vac-of-g})
disappears when looking at the relative deviation, i.e.
$\Delta \tau^{rel}_j:=1 - \det \rho_{(vac)}^{(1)}/\det \rho_j^{(1)}$
(lower panel of Fig.~\ref{reldeltangle-Vac-of-g}; 
Fig.\ref{reldeltangle-Vac-of-l}).
We choose the same parameter range as above.
Along the axis $t=0$ and $\lambda=0$ and for site numbers larger than
two (and smaller than one) we have that $\det \rho_j^{(1)}=0$.
In these cases, also $\det \rho_{(vac)}^{(1)}=0$, and we chose
the plotted value being zero in these cases.
The analysis of $\Delta \tau^{rel}_j$ tells us that for small anisotropy
and sufficiently far from the critical coupling 
the global tangle is dominated by the local perturbation of the
vacuum by the singlet (lower-left plot in Fig.~\ref{reldeltangle-Vac-of-g}). 
Meaning also that the total tangle is concurrence
dominated. For the isotropic $XY$ model, the global tangle 
was given entirely by the sum of the $2$-tangles such that the 
CKW-conjecture would conclude that there is no higher tangle 
contained in the system.
In the presence of a small anisotropy, this is no longer true,
in particular near to the critical coupling $\lambda=1$
(lower-right plot in Fig.~\ref{reldeltangle-Vac-of-g}).
Cranking up the anisotropy enhances the vacuum domination
as well as tuning $\lambda\longrightarrow 1$.

%% file: gammaground.tex
\subsubsection{Singlet-type perturbation of the groundstate}

One could ask the critical question, whether or not the
results in the preceding sections were specific to the vacuum state
or what would happen for different states.
We discuss in this section the propagation of a singlet-like
perturbation onto the groundstate $\ket{GS}$.
With {\em singlet-like perturbation} we mean to study the
time evolution of the initial state
\beq\label{initialstate:GS}
\ket{S}_g:=\bigfrac{1}{\sqrt{2}}(c_1 - c_2)\ket{GS}.
\eeq
We note that this state  
differs from a singlet (i.e. a fully entangled state
on the sites one and two -- see appendix \ref{appendix:GS}
for details) since the operators $c_i$ create global (rather than local) 
excitations. 
For the isotropic model $\ket{GS}\equiv\ket{\Uparrow}$ and hence
the dynamics is the same as on the vacuum.
An indicator for how far away is the 
ground state for general $\gamma>0$ from $\ket{\Uparrow}$ 
is, how much the norm of $\ket{S}_g$ differs from $1$.
We observed that notable deviations from $1$ are present 
either for critical coupling at any $\gamma>0$ or for non-critical coupling
when $\gamma$ is larger than about $1-\lambda$. For $\lambda<1$ at maximum 
$10\%$ of the weight is lost. The situation changes drastically for
$\lambda>1$, where for all $\gamma>0$ we have a weight-loss of about
$50\%$. 
We further want to note that the correspondence between fermion 
creation operators and spin raising operators 
(established by the Jordan-Wigner transformation)
is different for the state $\ket{\Uparrow}$ and the 
vacuum $\ket{\Downarrow}$.
Acting on the state $\ket{\Downarrow}$
it is $S^+_i=c^\dagger_i$; for the state $\ket{\Uparrow}$
we have instead $S^+_i=(-1)^{(i-1)}c^\dagger_i$. 

Apart from the propagation of the pulse, which again goes 
with a velocity about $\lambda$, there are many qualitative differences
for the concurrence. 
One of them is that the propagating pulse and the initial Bell state 
on sites $1$ and $2$ is the triplet with zero magnetization 
(figure \ref{fidel-GS:g0-5l0-5}). 
The concurrence plots (Fig. \ref{C1-different-gamma}) 
show that the inital state, $\ket{S}_g$, 
is almost fully entangled on sites $1$ and $2$. 
This demonstrates how close is
the ground state to $\ket{\Uparrow}$.
Two new features appears in the concurrence signal.
Firstly, the entanglement pulse propagates on top of a nonzero 
background level,
which in very good agreement coincides with the nearest neighbor 
concurrence of the ground state, being around $0.2$ at 
both the critical point (see Fig. \ref{C1-different-gamma}) 
and $\gamma=0.5$\cite{Osterloh02}. This shows that not only is
the ground state very close to the vacuum, but that 
also the state $\ket{S}_g$ is very similar to the ground state
as far as nearest neighbor concurrence is considered.
Secondly, in contrast to the singlet on the vacuum, we here have to deal with
a propagation that eliminates the background concurrence. 
The latter feature gets more pronounced when approaching the Ising model 
and critical coupling (see Fig. \ref{C1GS}).
This elimination is understood from the fact that joining entanglement
of the same type (here: two-site entanglement) in form of states that 
are orthogonal to those forming the entanglement already present, diminishes
the entanglement;
in fact, the fidelity plots demonstrate that the background entanglement
is of the type $\ket{\up\up}+\ket{\down\down}$, whereas the pulse
is of the type $\ket{\up\down}+\ket{\down\up}$.
Whereas here the impact of the propagating concurrence pulse 
coming from the initial $0$-triplet is the stronger the closer we get to
the critical coupling and the quantum Ising model (the opposite of what we
observed for the initial singlet onto the vacuum), the situation
would be the same as far as the total signal along the propagation line
is concerned. This in fact gets more suppressed with growing 
$\lambda$ and $\gamma$. 
For $\gamma$ tending to zero the ground state tends to $\ket{\Uparrow}$,
and hence we have the same situation as for the isotropic model.

\onecolm

\begin{minipage}[h]{\linewidth}
\begin{figure}\centering
\includegraphics[width=.32\linewidth,angle=-90]{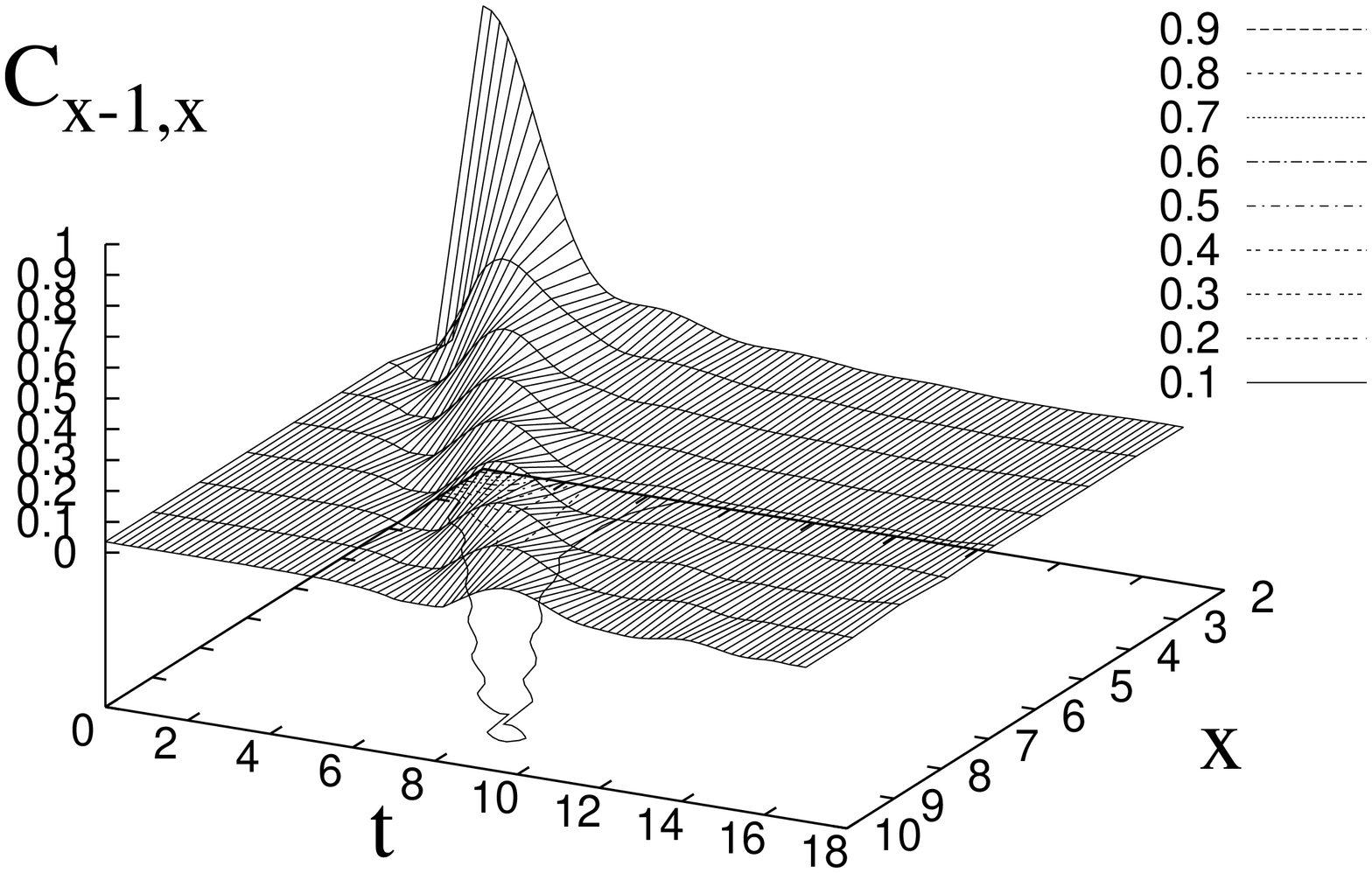}
\includegraphics[width=.32\linewidth,angle=-90]{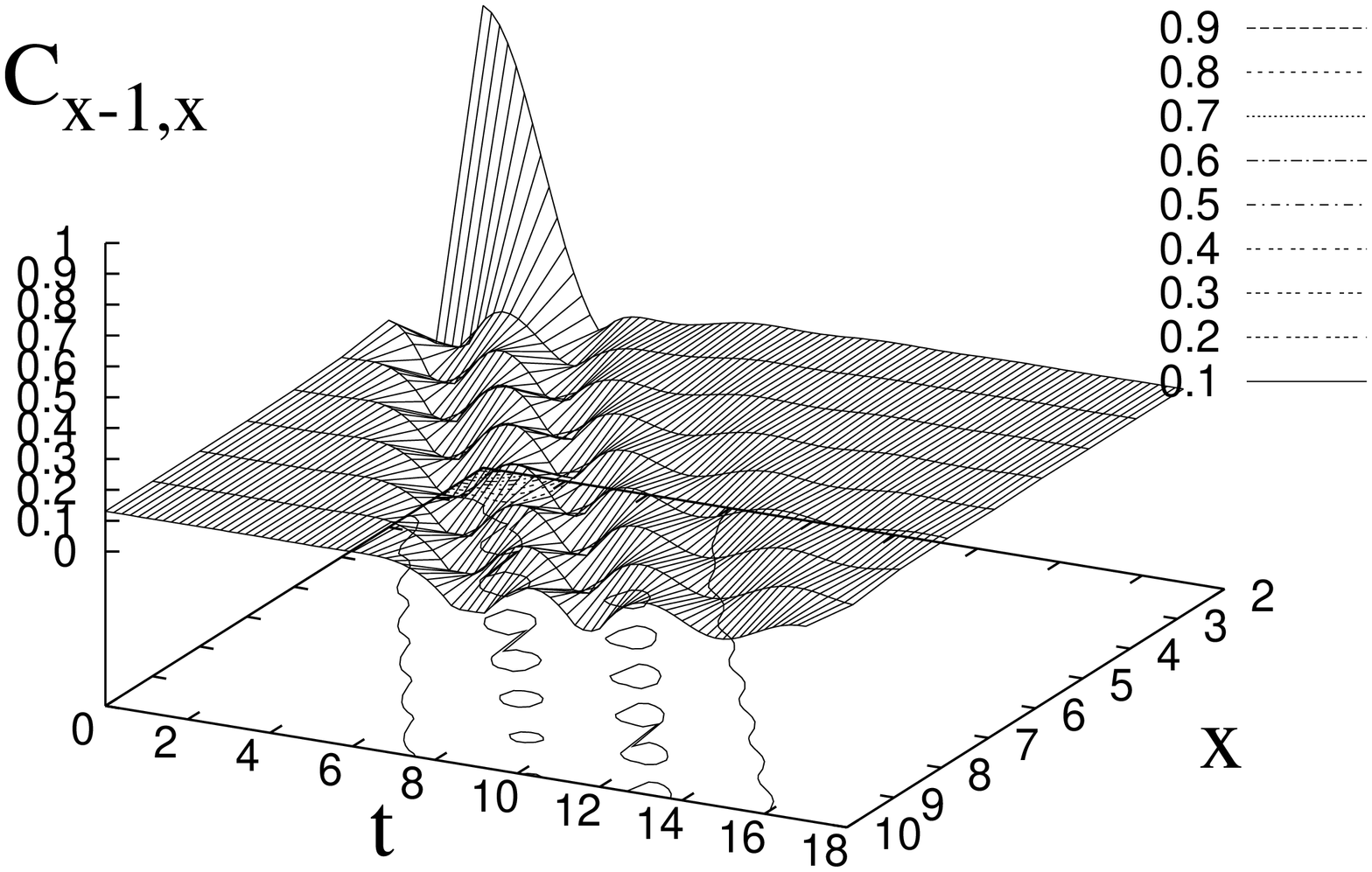}
\includegraphics[width=.32\linewidth,angle=-90]{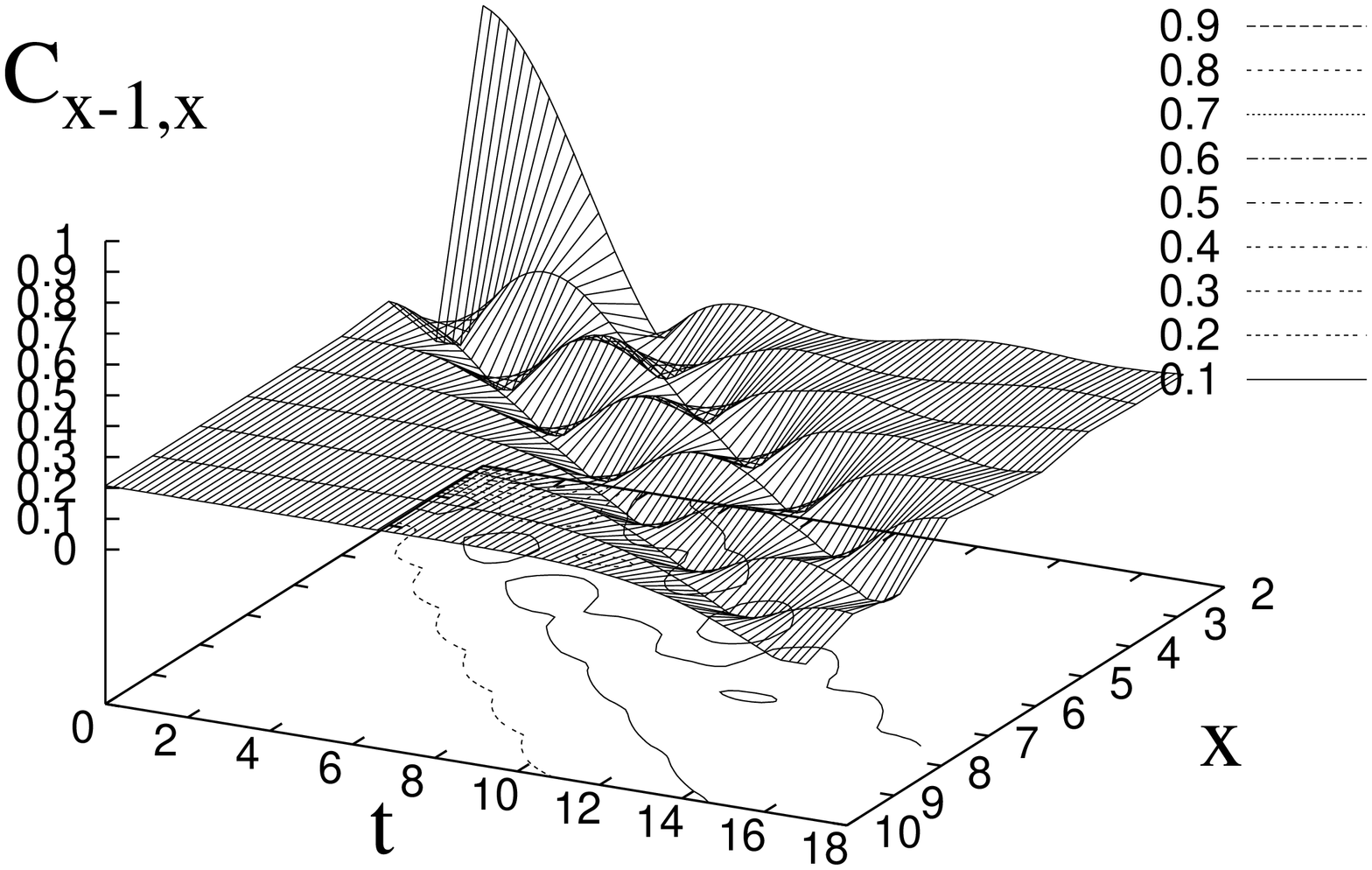}
\includegraphics[width=.32\linewidth,angle=-90]{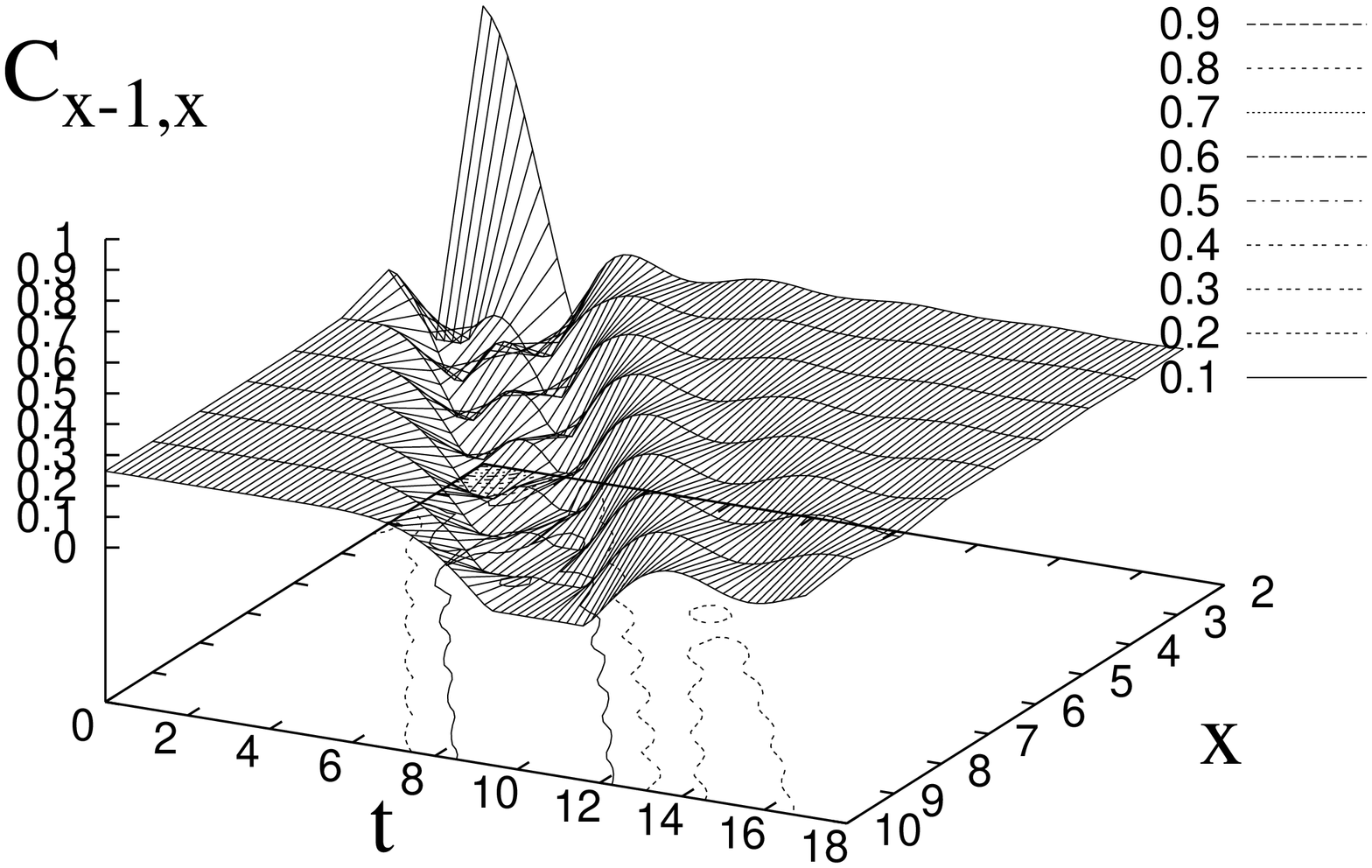}
\caption{The nearest neighbor concurrence for different $\lambda$ 
and $\gamma$.
{\em Upper panel}\/: $\gamma=0.1$ (left) and $\gamma=0.5$ (right)
for critical coupling $\lambda=1$.
{\em Lower panel}\/: Ising model ($\gamma=1$) for $\lambda=0.5$ (left)
and $\lambda=0.9$ (right). It is seen that instead of the enhancement
of concurrence at short times, we here have a background concurrence 
corresponding to the ground state value (see Fig. \ref{C1-different-gamma}
in appendix \ref{GSonly}).
The propagating signal is seen on top of a valley of extinction.}
\label{C1GS}
\end{figure}
\end{minipage}

\begin{minipage}[h]{\linewidth}
\begin{figure}\centering
\includegraphics[width=.32\linewidth,angle=-90]{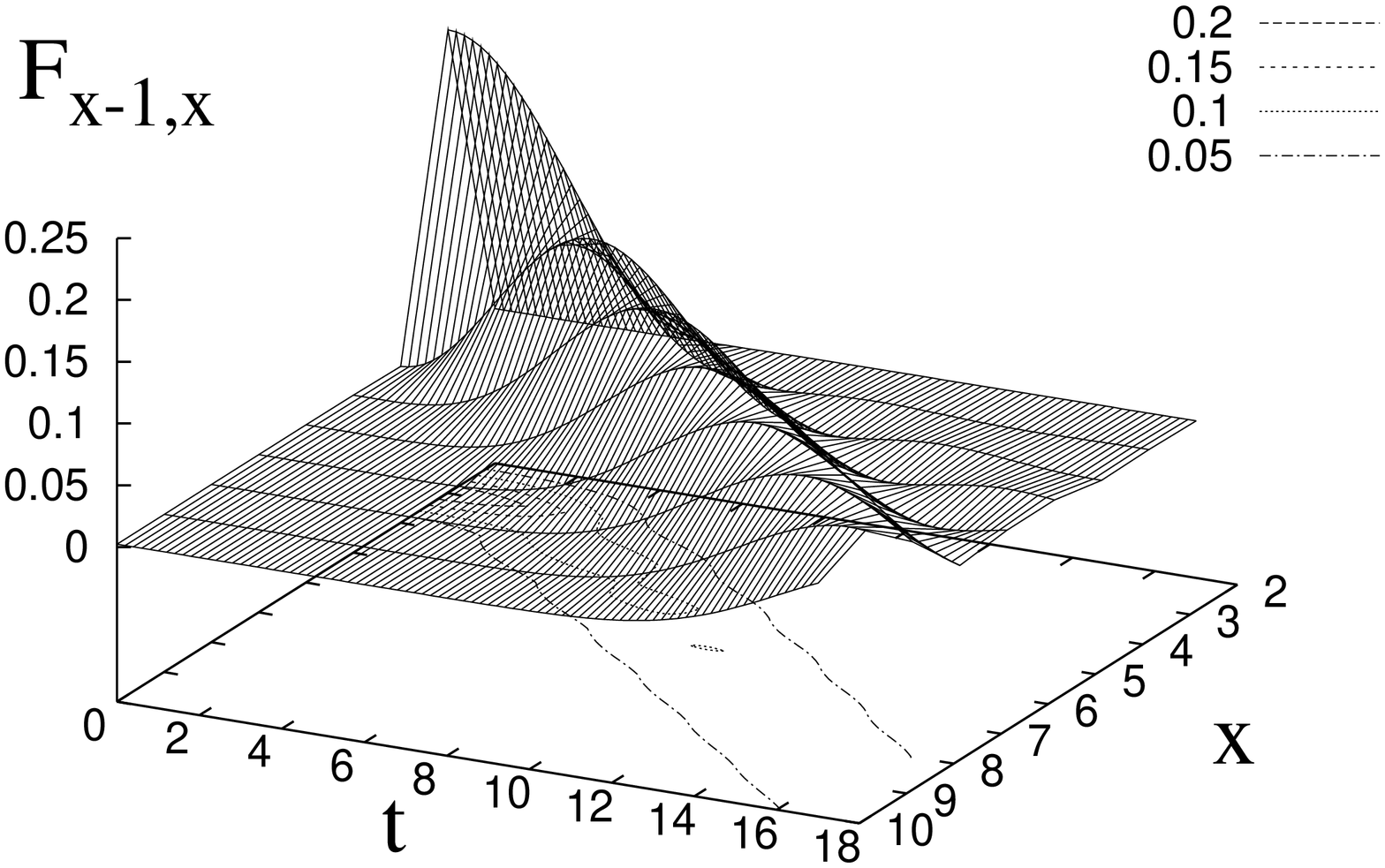}
\includegraphics[width=.32\linewidth,angle=-90]{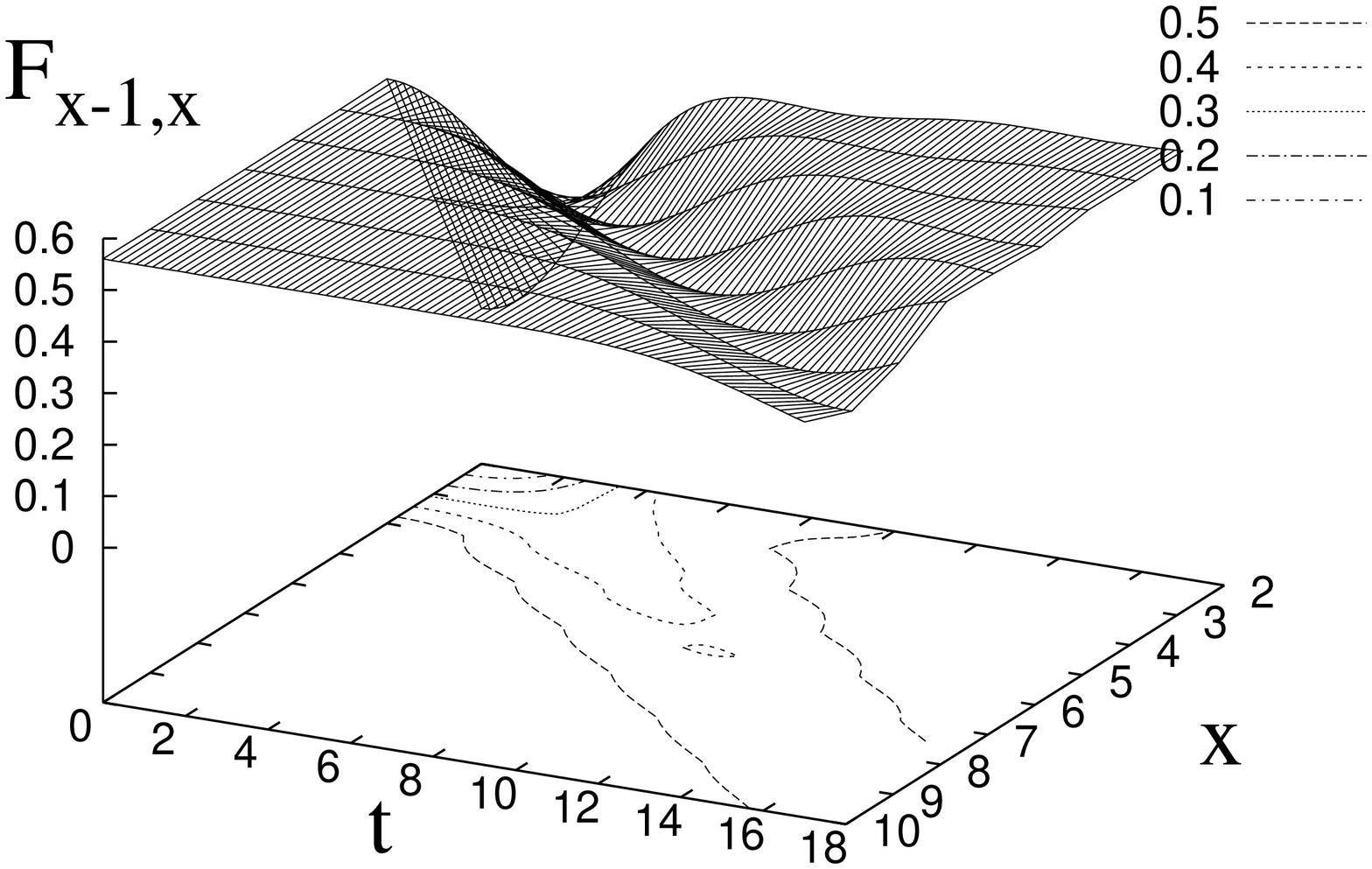}
\includegraphics[width=.32\linewidth,angle=-90]{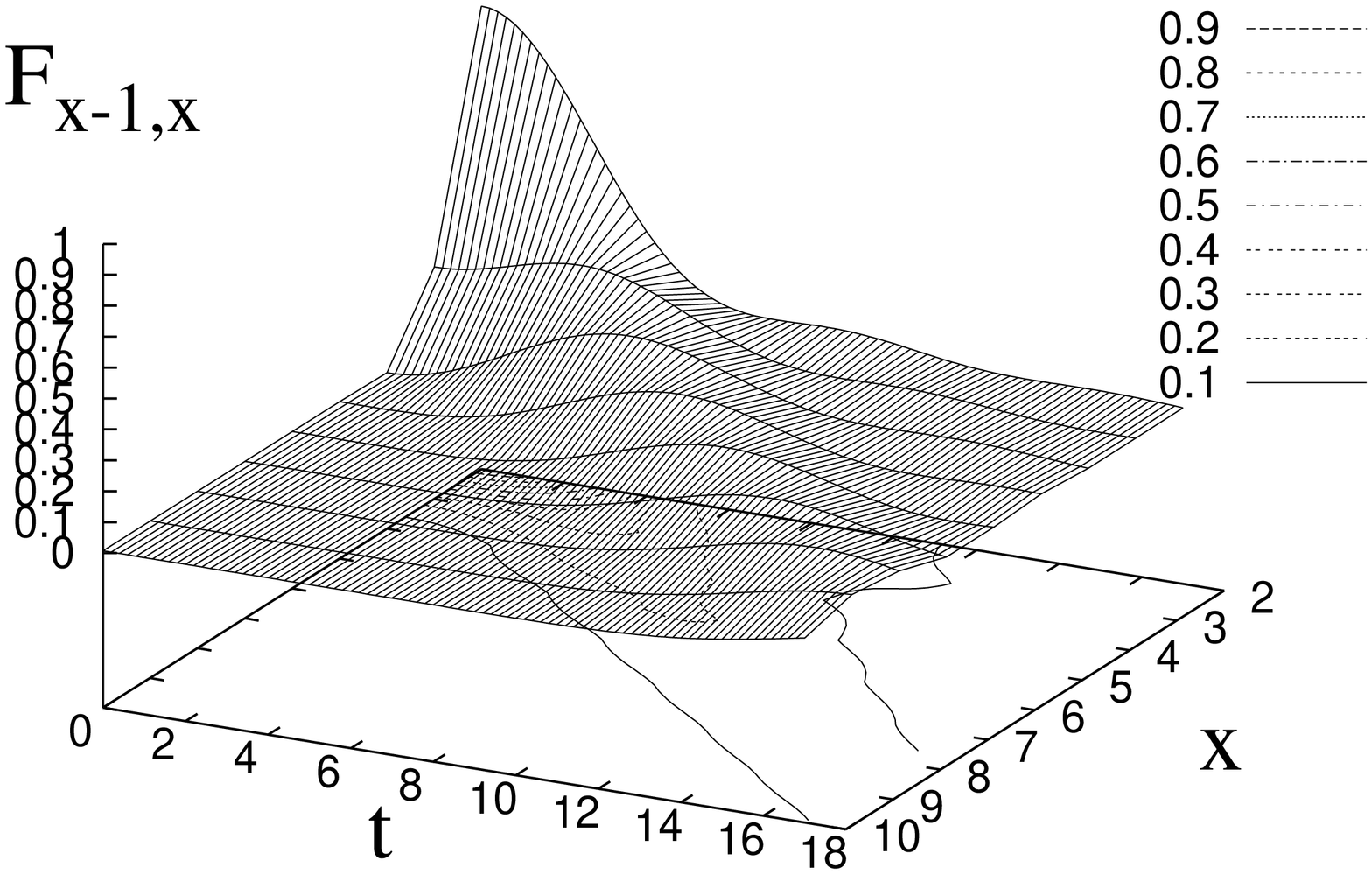}
\includegraphics[width=.32\linewidth,angle=-90]{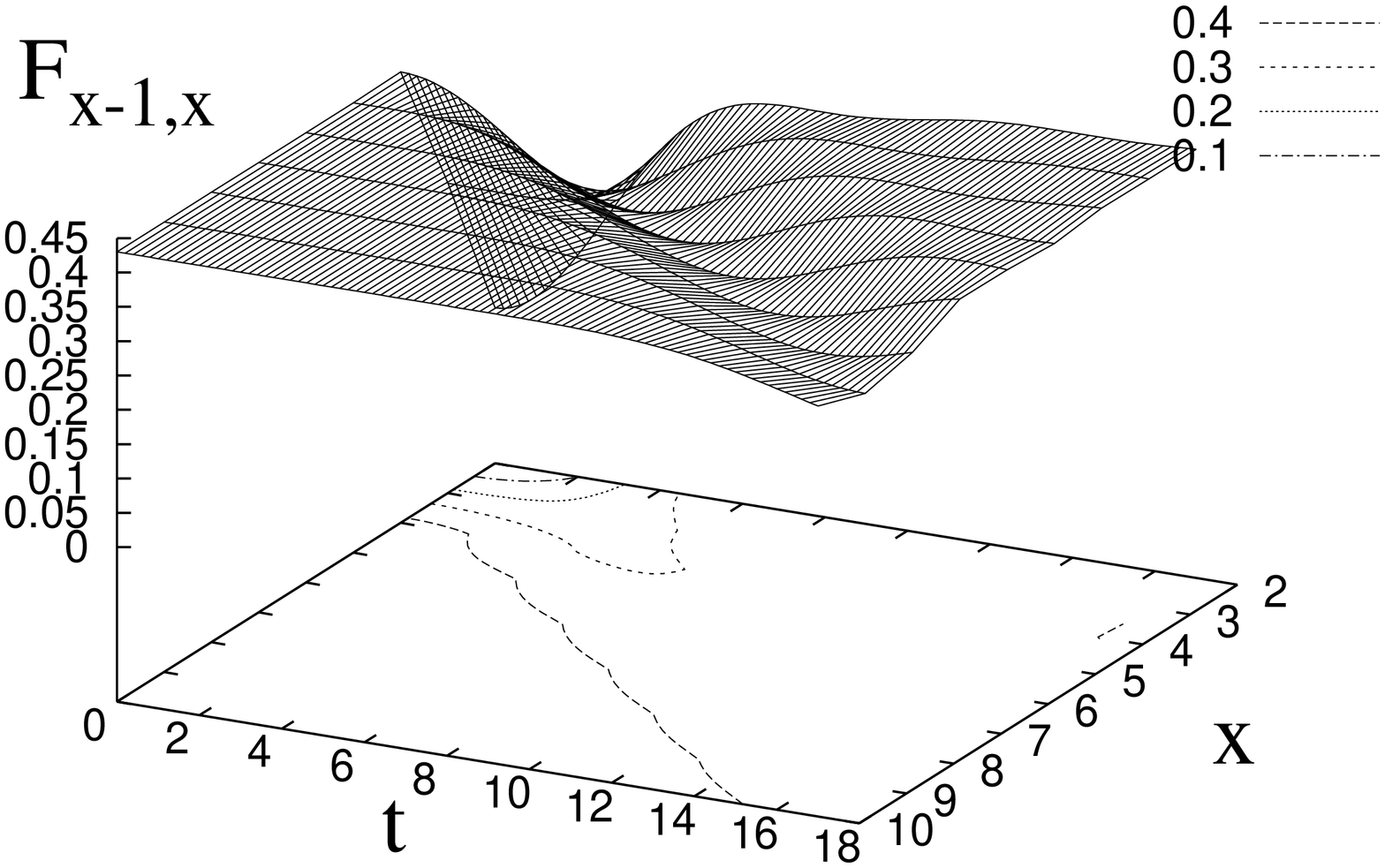}
\caption{
The fidelity for the Bell states in the reduced density matrix is drawn for 
$\gamma=\lambda=0.5$; the initial state is a singlet type perturbation
on top of the ground state. 
Since an equal mixture of two Bell states
is a disentangled state, the pictures show that
the predominant Bell state in the background entanglement is the
triplet $1/\sqrt{2}(\ket{\up\up}+\ket{\down\down}$ (upper rightmost picture). 
For the propagation the situation is opposite to what was found for
the singlet on top of the vacuum: the propagation is dominated
by the zero triplet (lower leftmost picture) and not the singlet 
but the singlet seems to decay more slowly such that it
eventually could become dominant.
}
\label{fidel-GS:g0-5l0-5}
\end{figure}
\end{minipage}

\begin{minipage}[h]{\linewidth}
\begin{figure}\centering
\includegraphics[width=.32\linewidth,angle=-90]{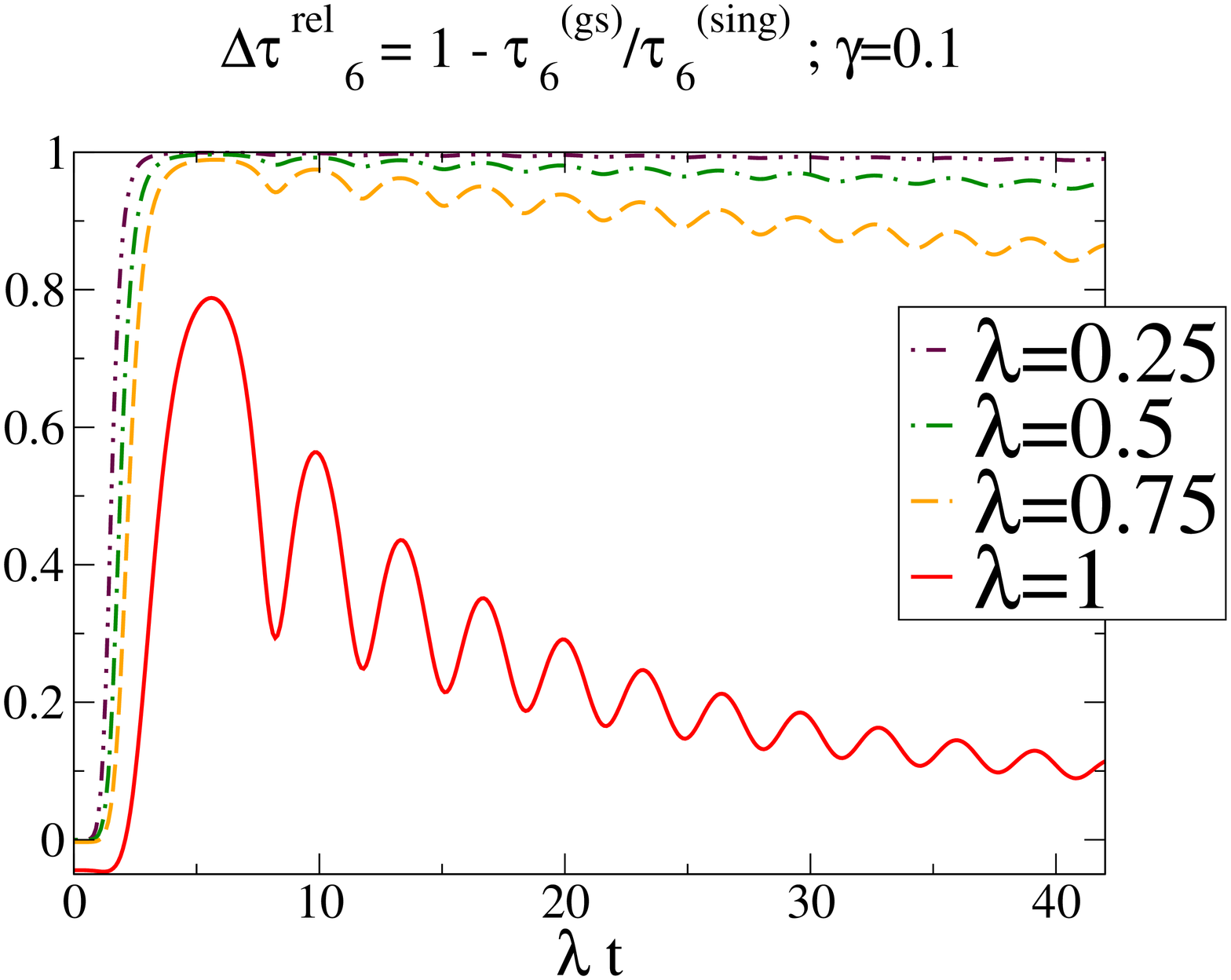}
\includegraphics[width=.32\linewidth,angle=-90]{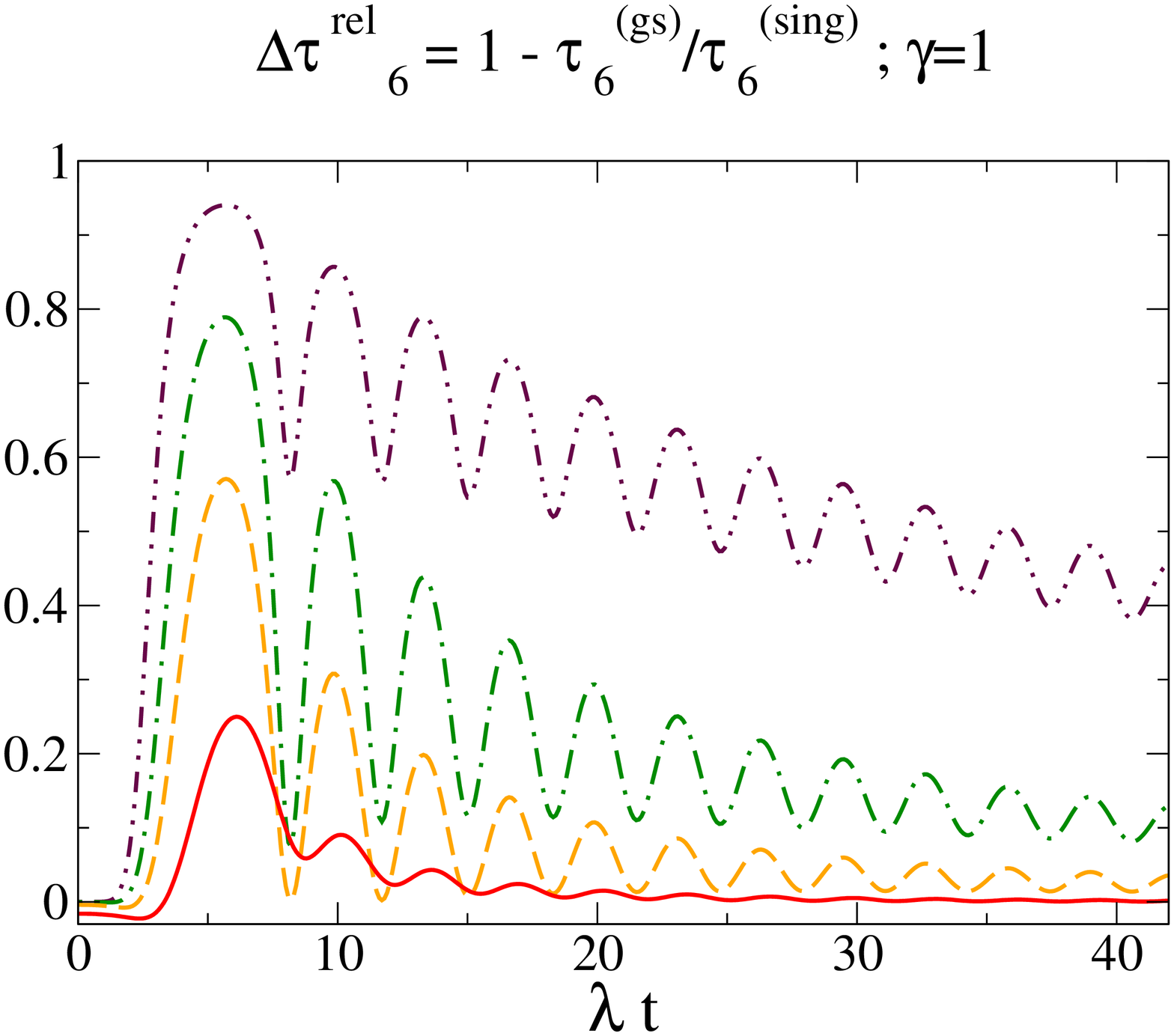}
\caption{The relative tangle deviation $\Delta\tau^{rel}_6$ from the 
ground state tangle at site number $6$ for different values of 
$\lambda$ for a fixed value of $\gamma$: 
$\gamma=0.1$ (left) and $\gamma=1$ (right). 
It is plotted as a function of $\lambda t$;
it can be seen that the oscillation frequency grows linearly with 
$\lambda$.}
\label{reldeltangle-GS-of-l}
\end{figure}
\end{minipage}
\twocolm
For the global tangle we find essentially the same behavior
as for the singlet on the vacuum. 
Only at very short times the 
singlet-type perturbation deminishes the global tangle.
In Fig~\ref{reldeltangle-GS-of-l} we choose two different 
anisotropies $\gamma=$ $0.1$ and $1$ 
and compare $\Delta\tau^{rel}_j$ for different couplings $\lambda$.
In a preliminar analysis, the short-time behavior of $\Delta\tau^{rel}_6$
shows a marked anomaly at the critical coupling that will be discussed 
elsewhere.

%% file: conclusions.tex
\section{Conclusions}

In the present work we studied the effect of a singlet-type
perturbation on the entanglement of an infinite spin system. We
considered quantum $XY$ models for general anisotropy.
The dynamics of entanglement was studied as 
function of the distance to the local perturbation at
$t=0$, and of the reduced interaction strength $\lambda$ (up to the
common quantum critical point of the models at $\lambda=1$).
For this class of models we analized a conjecture formulated
by Coffman, Kundu, and Wootters quantifying the weight of the 
pairwise relatively to the global entanglement, measured by
$4 \det \rho_1 $.

The isotropic model, i.e. anisotropy $\gamma=0$,
can be mapped onto a tight binding model, and consequently
entanglement propagates only, remaining pairwise; 
the velocity of the propagation is $\lambda$. 
A nearest neighbor entanglement at $t=0$ not only propagates
in both directions, remaining of type nearest neighbor;
in addition, there is a notable concurrence signal between
the sites with distance $\pm \lambda t$ to the initial Bell state 
at time $t$ (see Fig.~\ref{treconc}).
This phenomenon can be interpreted as the production of an EPR pair: 
the initial Bell state splits into parts moving with 
the same velocity $\lambda$ in opposite directions of the chain.
We found that singlet-like states are transmitted after a crossing
with higher fidelity than other maximally entangled states in the sense 
that the latter tend to switch into a singlet if the crossing
involves a single site only.
The global tangle and the concurrence (whose square is the $2$-tangle) 
satisfy the Coffman-Kundu-Wootters conjecture with zero residual tangle. 
This means that {\it the system contained only pairwise 
entanglement, measured by the concurrence}. 
The Hamiltonian does not create any entanglement; it distributes the
initially created pairwise entanglement.

For general anisotropy  $\gamma$ we inserted a singlet at $t=0$
in the state $\ket{\Downarrow}$.
Also in this case we found the propagation of entanglement
with velocity $\lambda$ together with the evidence for an EPR-type
propagation (see Fig.~\ref{C2andC3}). 
The propagation is predominantly of the type
($\ket{\up\down}-\ket{\down\up}$)
as is the initial perturbation at $t=0$. A 
triplet state $\ket{\up\down}+\ket{\down\up}$ is also present;
it seems to decay more slowly than the singlet, indicating an eventual swap, 
as the system evolves, of the singlet into a triplet 
(see Fig.~\ref{fidel:g0-5l0-5}).
The propagation is suppressed, compared to the isotropic model.
The suppression is stronger the closer the system is to
the critical coupling and the quantum Ising model
(Fig.\ref{C1-Vac}).
For the latter we found a very rapid damping
of the singlet in the nearest neighbor concurrence.
For all larger distances we considered (up to $7$ lattice spacings) 
the concurrence is zero (in the range of the plots).
A peculiarity of the anisotropic models is the instantaneous
creation of concurrence from the vacuum all over the chain. This
concurrence is due to the presence of pairwise spin-flips $s^+_j
s^+_{j+1}$ and $s^-_j s^-_{j+1}$ and results to be of triplet type. 
It decays very quickly when approaching the quantum
critical coupling and getting close to the Ising model. Neither
effect is of critical origin, though.
Comparing quantitatively  $4 \det \rho_1 $  and concurrence we 
could conclude that {\it  for the anisotropic model  
the propagation of the concurrence is a small effect respect to 
the creation of the global tangle}. 
This  indeed becomes more and more
dominant when approaching the Ising model and the critical
point (Fig.\ref{totaltangle}). Medium interaction strengths
and/or small anisotropy favor the propagation of the singlet.

We also studied the time evolution of a singlet-type perturbation 
of the ground state. We notice that though the perturbation
has the form of a singlet, the initial state is
the triplet $\ket{\up\down}+\ket{\down\up}$ (see Fig~\ref{fidel-GS:g0-5l0-5}). 
Consequently, the propagation involves states of this type;
the propagation velocity is still $\lambda$
but the concurrence signal occurs along a ``valley'' in
the constant background concurrence (Fig.~\ref{C1GS}).
The propagation of the nearest neighbor concurrence (in form
of an extinction) gets more enhanced with growing
anisotropy and approaching the critical coupling; very much the
contrary of what we found for the singlet on the vacuum.
For the critical Ising model we found immediate
extinction of the initial nearest neighbor concurrence
but the valley with a weak propagating signal in it remains.
The background nearest neighbor concurrence coincides with 
that of the ground state, indicating that the nearest neighbor
concurrence distant from the initial perturbation is unaffected. 
The all over creation of entanglement is absent, suggesting that
the Hamiltonian cannot create nearest neighbor concurrence
on top of that already present in the underlying ground state. 
\\
Since the ground state is a highly correlated 
state, a locally created singled on the wavefunction modifies the 
system over a long range.
Nevertheless, the dynamics of the total tangle is basically 
unchanged (with respect to the vacuum case), differing only in minor 
quantitative details.
In the short-time behavior of $\Delta\tau^{rel}_x$
we observed an interesting anomaly at the critical coupling that 
needs to be addressed further.

We would like to mention finally that understanding the dynamics of 
entanglement may be relevant to characterize the behaviour of 
quantum registers. 
The information encoded in a given register can be lost 
either because of decoherence or if (unintentionally) the state  
of the register is not an eigenstate of the Hamiltonian. 
In the latter case, the residual dynamics will let the encoded state evolve 
in time. Due to the spin interactions, fidelity and
the amount of entanglement will consequently vary in time.

In this work we consider exclusively ordered systems, and the role of 
imperfections is studied elsewhere\cite{Montangero03}.

%% file: appendix2.tex
\end{multicols}
\begin{appendix}
\begin{multicols}{2}
\input{pfaffians.short}

\input{reducedrho}
\input{GS}

\input{GSonly}
\end{multicols}
\end{appendix}

%% file: pfaffians.short.tex
\section{Pfaffians}
\label{pfaffians}

In this appendix, in order to make the presentation self-contained,  
we review the results obtained in~\cite{AOXY} 
where the out of equilibrium correlation functions are calculated exactly,
expressed as an expansion of pfaffians. 
The pfaffian for the ``vacuum'' expectation is defined as follows    
\onecolm
\begin{eqnarray}
& \left .
\begin{array}{clllllcll} 
\langle S_l^\alpha S_{l+R}^\beta \rangle= {s(\alpha,\beta)} {\rm pf}\; \left | \right . I^{\alpha\beta}_{1,2} & \dots  & I^{\alpha\beta}_{1,R-1}  &J^{\alpha\beta}_{1} & F^{\alpha\beta}_{1} 
& G^{\alpha\beta}_{1,2} &\phantom{c} .  &  \dots & G^{\alpha\beta}_{1,R}  \\
           & \dots  & \dots &\dots &  \dots & \dots  & \phantom{c} . & \dots & \dots \\
           &        & I^{\alpha\beta}_{R-2,R-1}  & J^{\alpha\beta}_{R-2}         & F^{\alpha\beta}_{R-2}         & G^{\alpha\beta}_{R-2,2} & \phantom{c} .  & \dots & G^{\alpha\beta}_{R-2,R}         \\
           &        &   & J^{\alpha\beta}_{R-1}          & F^{\alpha\beta}_{R-1} &  G^{\alpha\beta}_{R-1,2} & \phantom{c} .& \dots &  G^{\alpha\beta}_{R-1,R}  \\
           &        &   &           & E^{\alpha\beta} &  D^{\alpha\beta}_{2} & \phantom{c} . & \dots &  D^{\alpha\beta}_{R}  \\
           &        &   &           &             & K^{\alpha\beta}_{2} & \phantom{c} .&\dots &  K^{\alpha\beta}_{R} \\  
           &        &   &           &             &                       & H^{\alpha\beta}_{2,3}  &\dots &  H^{\alpha\beta}_{2,R} \\  
           &        &   &           &             &                     & &\dots & \dots         \\  
           &        &   &           &             &                  &   &  & H^{\alpha\beta}_{R-1,R}  

\end{array}
\right |&
\label{pfaffian}
\end{eqnarray}
\twocolm
where $s(x,x)=s(y,y)=1/4 (-)^{R(R+1)/2}$, 
\begin{eqnarray}
 I^{xx}_{\mu,\nu}&&=  \langle A_{l+\mu}(t) A_{l+\nu}(t)\rangle \nonumber \\
J^{xx}_{\mu}  &&=I^{xx}_{\mu,R}\\ 
H^{xx}_{\mu,\nu}&&= \langle B_{l+\mu-1}(t) B_{l+\nu-1}(t)\rangle \nonumber \\
K^{xx}_{\nu}&&= H^{xx}_{1,\nu} \\ \nonumber 
G^{xx}_{\mu,\nu}&&= \langle A_{l+\mu}(t) B_{l+\nu-1}(t)\rangle \label{xx}\\ 
F^{xx}_{\mu}&&=G^{xx}_{\mu,1}  \nonumber \\
E^{xx}&&=G^{xx}_{R,1}  \nonumber \\
D^{xx}_{\nu}&&=G^{xx}_{R,\nu}  \nonumber 
\end{eqnarray}

\begin{eqnarray}
 I^{yy}_{\mu,\nu}&&=  \langle A_{l+\mu-1}(t) A_{l+\nu-1}(t)\rangle \nonumber \\
J^{yy}_{\mu}  &&=I^{yy}_{\mu,R}\\ 
H^{yy}_{\mu,\nu}&&= \langle B_{l+\mu}(t) B_{l+\nu}(t)\rangle \nonumber \\
K^{yy}_{\nu}&&= H^{yy}_{1,\nu} \\ \nonumber 
G^{yy}_{\mu,\nu}&&= \langle A_{l+\mu-1}(t) B_{l+\nu}(t)\rangle \label{yy} \\ 
F^{yy}_{\mu}&&=G^{yy}_{\mu,1}  \nonumber \\
E^{yy}&&=G^{yy}_{R,1}  \nonumber \\
D^{yy}_{\nu}&&=G^{yy}_{R,\nu}  \nonumber 
\end{eqnarray}
and $s(x,y)=s(y,x)= -\i/4 (-)^{R(R-1)/2}$, 
\begin{eqnarray}
I^{xy}_{\mu,\nu}&&=  \langle A_{l+\mu}(t) A_{l+\nu}(t)\rangle \nonumber \\
G^{xy}_{\mu,\nu}&&= \langle A_{l+\mu}(t) B_{l+\nu}(t)\rangle \\
J^{xy}_{\mu}  &&= G^{xy}_{\mu,0} \label{xy} \\ 
F^{xy}_{\mu}&&=G^{xy}_{\mu,1}  \nonumber \\
H^{xy}_{\mu,\nu}&&= \langle B_{l+\mu}(t) B_{l+\nu}(t)\rangle \nonumber \\
E^{xy}&&=H^{xy}_{0,1}  \nonumber \\
D^{xy}_{\nu}&&= H^{xy}_{0,\nu} \\ \nonumber 
K^{xy}_{\nu}&&=H^{xy}_{1,\nu}  \nonumber 
\end{eqnarray}

\begin{eqnarray}
I^{yx}_{\mu,\nu}&&=  \langle A_{l+\mu-1}(t) A_{l+\nu-1}(t)\rangle \nonumber \\
G^{yx}_{\mu,\nu}&&= \langle A_{l+\mu-1}(t) B_{l+\nu-1}(t)\rangle \\
J^{yx}_{\mu}  &&= I^{yx}_{\mu,R} \label{yx} \\ 
F^{yx}_{\mu}&&=I^{yx}_{\mu,R+1}  \nonumber \\
E^{yx}&&=I^{yx}_{R,R+1}  \nonumber \\
D^{yx}_{\nu}&&= G^{yx}_{R,\nu} \\ \nonumber 
K^{yx}_{\nu}&&=G^{yx}_{R+1,\nu}  \\ \nonumber
H^{yx}_{\mu,\nu}&&= \langle B_{l+\mu-1}(t) B_{l+\nu-1}(t)\rangle \nonumber 
\end{eqnarray}

In the case in which the expectation value is taken over the ``vacuum state'' 
$\ket{\Downarrow}$, the two point correlators are 
\begin{eqnarray}
\bra{\Downarrow}A_l(t) &&B_{l+R}(t) \ket{\Downarrow} = \delta_{R,0}  \label{ab-vacuum}\\
&& -\frac{2}{L} \sum_k \left [\cos(kR) (u^o_k)^2+ 
 \sin(kR) 
u^e_k u^o_k \right ]    \nonumber 
\end{eqnarray}
\begin{equation}
\bra{\Downarrow}A_l(t) A_{l+R}(t) \ket{\Downarrow}= \delta_{R,0} -{\rm i} \frac{2}{L} \sum_k \sin(kR) u^o_k v_k \label{aa-vacuum} 
\end{equation}
\begin{equation}
\bra{\Downarrow} B_l(t) B_{l+R}(t) \ket{\Downarrow}= - \bra{\Downarrow} A_l(t) A_{l+R}(t) 
\ket{\Downarrow}^* \label{bb-vacuum}
\end{equation}
where
\begin{eqnarray}\label{uodd}
u^o_k&=&\lambda \gamma \sin k \frac{\sin (\Lambda_k t)}{\Lambda_k}  \\
\label{ueven}
u^e_k&=&(1+\lambda \cos k) \frac{\sin (\Lambda_k t)}{\Lambda_k} \\ 
\label{vu}
v_k&=&\cos (\Lambda_k t)
\end{eqnarray}
We point out that though the translational invariance is not broken,
the Pfaffians do not reduce to determinants because
$\bra{\Downarrow}A_l(t) A_{l+R}(t) \ket{\Downarrow}\neq 0 $.

%% file: reducedrho.tex
\section{Concurrence from Correlation Functions}
\label{app:reducedrho}

The determination of the concurrence is related to the knowledge of the correlation
functions of the model. We derive the structure of the two-qubit
reduced density matrix $\rho^{(2)}$ for the model Hamiltonians
(\ref{model}). These Hamiltonians have all the same quantum
critical point (except for the XY model at $\gamma=0$) and are
solved exactly~\cite{Lieb61,Pfeuty70,Mccoy70}. Certain symmetries
of the Hamiltonian could restrict the structure of the reduced
density matrix as long as they are not broken.

\subsection{The generic case}
The symmetries that can be used to simplify the form of
$\rho^{(2)}$ are the {\em translational invariance}, the {\em
reality} of the Hamiltonian and the {\em parity invariance},
meaning that the Hamiltonian can either leave the value of $S_z$
unchanged or change it in steps of $2$. For periodic boundary
conditions, which we will consider here, we also have a reflection
symmetry with respect to a mirror along a diameter of the ring.
The parity invariance is spontaneously broken in the ordered phase with
$\expect{S_x}\neq 0$. We will consider states with unbroken
symmetry only.

For states out of equilibrium with broken translational invariance,
$\rho_2$ reflects only the parity invariance. Written
in the basis $\{\ket{\uparrow\uparrow},\ket{\uparrow\downarrow},
\ket{\downarrow\uparrow},\ket{\downarrow\downarrow}\}$ it is
\begin{equation}\label{app:rho:non-equilibrium}
\rho^{(2)}\ =\ \Matrix{cccc}{
a&0&0&c\\
0&x&z&0\\
0&z^*&y&0\\
c^*&0&0&b\\
}\, ,
\end{equation}
with real $a, b, x, y$, and complex $c, z$. The concurrence
results to be 
\beqa\label{app:concurrence:non-equilibrium}
C=2\max\{0,C^{(1)},C^{(2)}\}\; ,
\eeqa 
where $C^{(1)}=|c|-\sqrt{xy}$ and $C^{(2)}=|z|-\sqrt{ab}$.
In terms of spin-spin correlation function, using the notations
$g_{\alpha\beta}:=\expect{S^{1}_\alpha S^{2}_\beta}$,
$M_\alpha:=\expect{S^{1}_\alpha + S^{2}_\alpha}/2$, and $\delta
S_\alpha:=\expect{S^{1}_\alpha - S^{2}_\alpha}/2$,
(\ref{app:rho:non-equilibrium}) becomes 
\onecolm 
\beq\label{app:rho2}
\rho^{(2)} = \Matrix{cccc}{ \bigfrac{1}{4}+M_z+g_{zz}&0&
    0&g_{xx}-g_{yy} -\i (g_{xy}+g_{yx})\\
0&\bigfrac{1}{4}-g_{zz}+\delta S_z&
    g_{xx}+g_{yy} +\i (g_{xy}-g_{yx})&0\\
0&g_{xx}+g_{yy} -\i (g_{xy}-g_{yx})&
    \bigfrac{1}{4}-g_{zz}-\delta S_z&0\\
g_{xx}-g_{yy} +\i (g_{xy}+g_{yx})&0&
0&\bigfrac{1}{4}-M_z+g_{zz}
}
\eeq
giving for the concurrence (\ref{app:concurrence:non-equilibrium})
\beqa\label{app:C-of-corrs}
C=2\max\left\{ \phantom{\sqrt{(\bigfrac{1}{4}-g_{zz})^2-\delta S_z^2}}
\hspace*{-3cm}
\right.
0&,&\sqrt{(g_{xx}-g_{yy})^2
    +(g_{xy}+g_{yx})^2}-
    \sqrt{(\bigfrac{1}{4}-g_{zz})^2-\delta S_z^2}\\
&,&\left.   \sqrt{(g_{xx}+g_{yy})^2
    +(g_{xy}-g_{yx})^2}-
    \sqrt{(\bigfrac{1}{4}+g_{zz})^2-M_z^2}\right\}\;.\nonumber
\eeqa
\twocolm
For states with translational invariance $\delta S_z=0$, 
i.e. $x=y$ in Eq. (\ref{app:rho:non-equilibrium}), and in the
presence of reflection symmetry, $g_{xy}=g_{yx}$, leading to real
$z$ in Eq. (\ref{app:rho:non-equilibrium}). As far as eigenstates of
the Hamiltonians are concerned, all above symmetries are present,
making $g_{xy}$ vanish, and we obtain in the equilibrium case
\begin{equation}\label{app:rho:equilibrium}
\Rightarrow \quad \rho^{(2)}\ =\ \Matrix{cccc}{
a&0&0&c\\
0&x&y&0\\
0&y&x&0\\
c&0&0&b\\
}\, , \qquad {\mbox a, b, c, x, y }\in\RR
\end{equation}
Consequently the expression for the concurrence considerably
simplifies in case of equilibrium to \beqa\label{app:C-of-corrs:eq}
C=2\max\left\{
\phantom{\sqrt{(\bigfrac{1}{4}+g_{zz})^2-M_z^2}}\right.
\hspace*{-3cm} 0&,&|g_{xx}-g_{yy}|-
    \bigfrac{1}{4}+g_{zz} \\
&,&\left.   |g_{xx}+g_{yy}|-
    \sqrt{(\bigfrac{1}{4}+g_{zz})^2-M_z^2}\right\} \;.\nonumber
\eeqa In order to calculate the entanglement of one site with the
rest of the chain, we have to consider the one-site reduced
density matrix. It is obtained from (\ref{app:rho2}) \beq \label{app:rho1}
\rho_{j}^{(1)} = \Matrix{cc}{
\bigfrac{1}{2}+\expect{S^z_j}&0\\
0&\bigfrac{1}{2}-\expect{S^z_j} } \eeq and the entanglement, the
site number $j$ participates in, is quantified by the 1-tangle \beq
\tau_1[\rho_{j}^{(1)}]=4\det \rho_{j}^{(1)} =
\bigfrac{1}{4}-\expect{S^z_j}^2 \eeq

\subsection{The isotropic model: $\gamma=0$}

Here the conservation of magnetization, or equivalently of the
total number of particles, simplifies further the structure of the
$2$-site reduced density matrix $\rho^{(2)}$, if the initial
state has a well-defined number of flipped spins. Written in the basis
$\{\ket{\uparrow\uparrow},\ket{\uparrow\downarrow},
\ket{\downarrow\uparrow},\ket{\downarrow\downarrow}\}$ it is
\begin{equation}\label{app:rho:non-equilibrium:gamma0}
\Rightarrow \quad \rho^{(2)}\ =\ \Matrix{cccc}{
a&0&0&0\\
0&x&z&0\\
0&z^*&y&0\\
0&0&0&b\\
}\, ,
\end{equation}
with real $a, b, x, y$, and complex $z$, which is Eq.
(\ref{app:rho:non-equilibrium}) for $c=0$. The concurrence is then
given by \beqa\label{app:concurrence:non-equilibrium:gamma0}
C=2\max\{0,|z|-\sqrt{ab}\}\;, \eeqa because $|c|-\sqrt{xy}$ in Eq.
(\ref{app:concurrence:non-equilibrium}) is always negative here ($c=0$).

For the case of the time evolution of a singlet (one of the Bell
states, i.e. fully entangled) on top of the vacuum, which
corresponds to a one-particle state, the structure of $\rho^{(2)}$
further simplifies to
\begin{equation}\label{app:rho:non-equilibrium:gamma0-1P}
\Rightarrow \quad \rho^{(2)}\ =\ \Matrix{cccc}{
0&0&0&0\\
0&x&z&0\\
0&z^*&y&0\\
0&0&0&b\\
}
\end{equation}
with concurrence
\beqa\label{app:concurrence:non-equilibrium:gamma0-1P}
C=2|z|\; .
\eeqa
For an arbitrary one-particle state $\sum_j w_j \ket{j}$ one
obtains
\beq\label{app:1rho1} \rho_{r,s}^{(2)}\ =\ \Matrix{cccc}{
0&0&0&0\\
0&|w_r|^2&w^{}_rw^*_s&0\\
0&w^*_rw^{}_s&|w_s|^2&0\\
0&0&0&1-|w_r|^2-|w_s|^2\\
}\; ,
\eeq
and consequently $C_{r,s}=2|w_rw_s|$.

%% file: GS.tex
\section{Local singlet insertion into the groudstate}
\label{appendix:GS}
In order to place a singlet into an arbitrary state, 
it is in general inevitable to consider from the
very beginning a density matrix rather than a pure state.
We present the {\em knitting procedure} for the groundstate on
the sites $1$ and $2$; the procedure nevertheless works for
an arbitrary state (and of course for any two sites).

Defining
\beq
\ket{\Phi_{\mu\nu}}:= \ket{s}\braket{\mu\nu}{GS} \; ,
\eeq
with the singlet state 
$$
\ket{s}=1/\sqrt{2} (\ket{\up\down}-\ket{\down\up})=
1/\sqrt{2} (c^\dagger_1-c^\dagger_2) \ket{\down\down}.
$$
They are realized by
\nbeqa
\Phi_{\up\up} &=& 1/\sqrt{2} (c^\dagger_1-c^\dagger_2) c^{}_1 c^{}_2 \ket{GS}\\
\Phi_{\up\down} &=& 1/\sqrt{2} (c^\dagger_1-c^\dagger_2) c^{}_1 c^{\dagger}_2 \ket{GS} \\
\Phi_{\down\up} &=& 1/\sqrt{2} (c^\dagger_1-c^\dagger_2) c^{\dagger}_1 c^{}_2 \ket{GS}\\
\Phi_{\down\down} &=& 1/\sqrt{2} (c^\dagger_1-c^\dagger_2) c^{\dagger}_1 c^{\dagger}_2 \ket{GS}
\neeqa
The density matrix one had to consider, is
\beq\label{singlet-into-GS}
\rho^{singlet}_{GS}:= 
\sum_{\mu,\nu} \ket{\Phi_{\mu\nu}}\bra{\Phi_{\mu\nu}}\; 
\eeq
and gives at $t=0$ a fully entangled singlet state on the sites one and two
and exactly the groundstate concurrence as long
as the sites numbers $1$ and $2$ are not included.
This is, because at $t=0$
$$
\trace_{\{1,2\}}\;\rho^{singlet}_{GS}=\trace_{\{1,2\}}\;\ket{GS}\bra{GS}
$$
i.e. the density matrix for the remaining $N-2$ sites derived 
from $\rho^{singlet}_{GS}$, Eq. (\ref{singlet-into-GS}), 
is identical to that one derived from the ground state itself. 
Consequently, all correlation functions are the same in both.

%% file: GSonly.tex
\section{Addendum for the Ground state}\label{GSonly}

\begin{figure}\centering
\includegraphics[width=.3\linewidth]{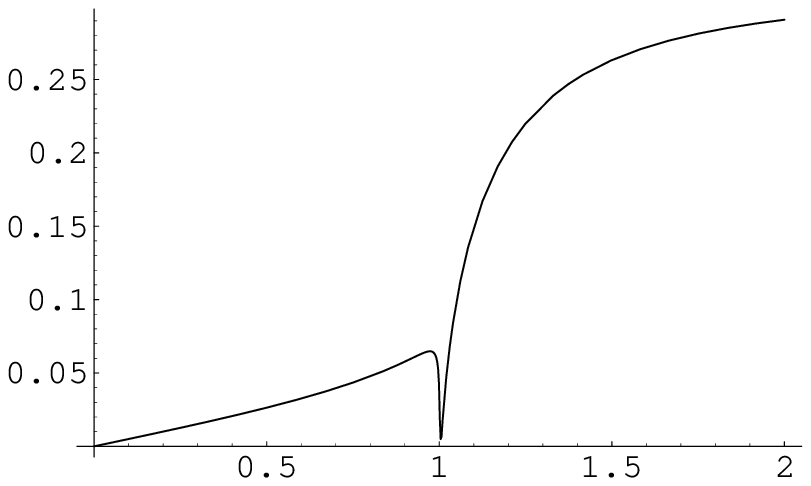}
\includegraphics[width=.3\linewidth]{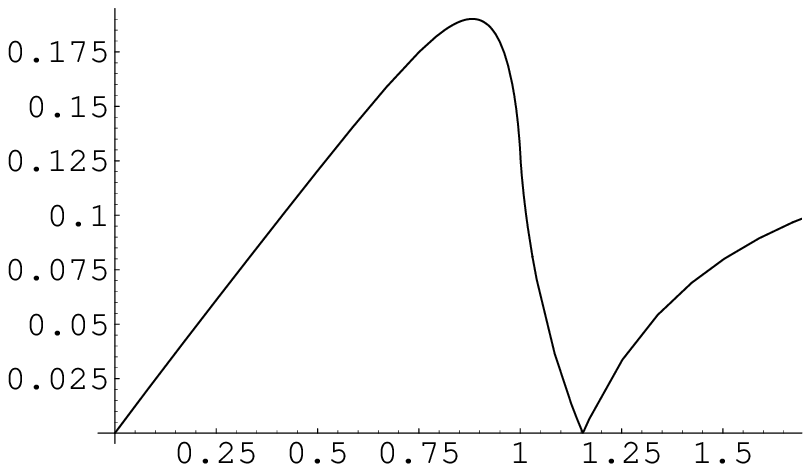}
\includegraphics[width=.3\linewidth]{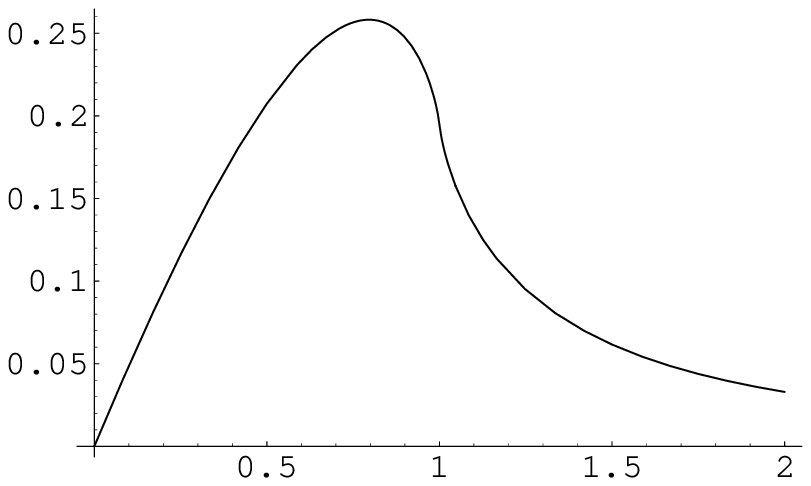}
\caption{The nearest neighbor concurrence $C_1$ for the values of $\gamma=0.1$,
$0.5$, and $1$ from the left to the right. At $\lambda=0.5$ and
$\lambda=1$, $C_1$ has roughly the same value: $0.0264$ and $0.0337$ for
$\gamma=0.1$, $0.1204$ and $0.1285$ for $\gamma=0.5$; $0.2074$ and $0.1946$
for $\gamma=1$, respectively. For the Ising model at $\lambda=0.9$
(the coupling shown in Fig.~\ref{C1GS}) we have $C_1=0.2475$.
These values appear as uniform background in the concurrence
$C_1(x,t)$ for the singlet-type perturbation of the ground state
(Fig.~\ref{C1GS}).}
\label{C1-different-gamma}
\end{figure}

In this section we recall some of the results from Ref.~\cite{Osterloh02}
and add some that have not been explicitely presented there.
In Fig.~\ref{C1-different-gamma} the nearest neighbor concurrence 
for the ground state (infinite chain) is shown for different 
values of $\gamma$ as a function of the reduced coupling $\lambda$.
All these functions exhibit a logarithmic divergence of the first 
derivative respect to $\lambda$ at the quantum critical point 
$\lambda_c=1$.  For finite chains $\partial_\lambda C_1$
scales according to finite size scaling theory with the system size.
This observation also applies to the first (non-zero) derivative
of the concurrence for larger distances.
The cusp, where $C_1$ vanishes is the point where the large eigenvalues
of $R=\rho\sigma_y\otimes\sigma_y\rho^*\sigma_y\otimes\sigma_y$
for the invariant sectors $|M_z|=0$ and $|M_z|=1/2$ are equal.
This means that the the type of Bell state responsible for
the entanglement changes from one regime to the other.
The invariance of these blocks is due to the parity symmetry.
The value of $\lambda$ where this degeneracy happens converges
from above to the critical coupling $\lambda_c=1$ for 
$\gamma\longrightarrow 0$. 
Details will be discussed elsewhere.
For the purpose of this work it is worth noticing that the groundstate
values agree very well with the uniform background concurrence.

\onecolm

\twocolm

\begin{figure}\centering
\includegraphics[width=.32\linewidth,angle=-90]{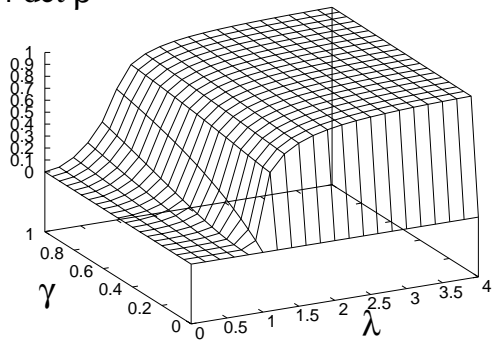}
\includegraphics[width=.32\linewidth,angle=-90]{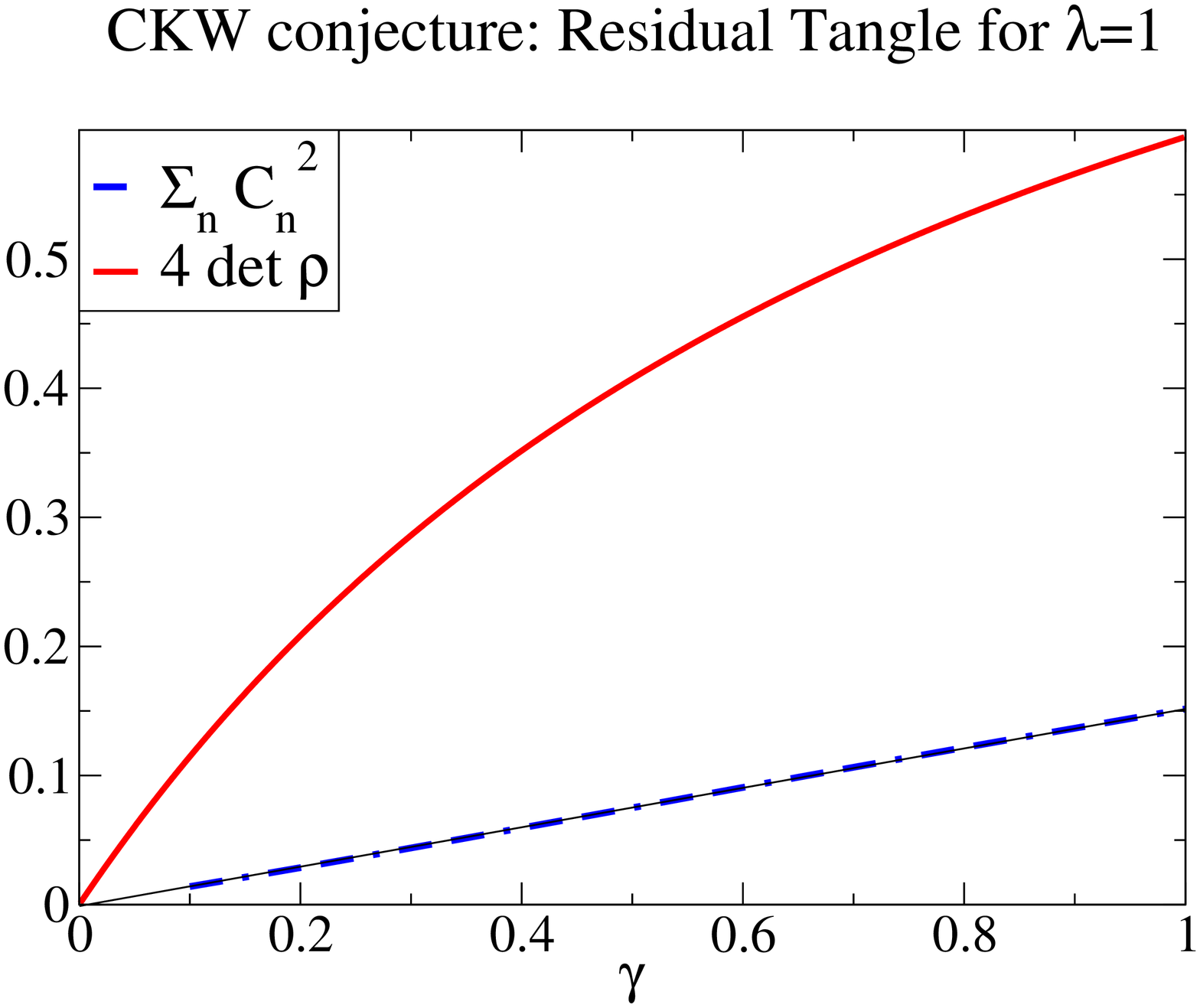}
\caption{The global tangle for the groundstate (left). 
For $0.1\leq\gamma\leq 1$
we could verify the CKW conjecture for critical coupling (right).
The one-tangle (thick line) is much larger than 
$\sum_n C_n^2$ (thick dash-dotted line) suggesting that the major part of 
entanglement should be stored in higher than two-qubit entanglement. 
The thin line is a guide to the eye showing that the sum of the
two-tangles linearly tends to zero as $\gamma\longrightarrow 0$.}
\label{4det-GS-g0upto1}
\end{figure}

The one-tangle for the ground state is shown on the left of figure
\ref{4det-GS-g0upto1}. This quantity was suggested to being an
additive measure of entanglement\cite{Coffman00} in the sense that
the sum of all $n$-tangles ($n$ from $2$ to $\infty$)
should give exactly the one-tangle, being then the total entanglement 
present in the system (in the case of a translational invariant state,
which is the ground state).

\onecolm

\twocolm

At the critical point for $\gamma\in [0.1\, ,\, 1]$ we could demonstrate
that the sum of the two-tangles is much smaller than the one-tangle.
The result is shown in the right plot of figure \ref{4det-GS-g0upto1}.
If the CKW conjecture holds, this indicated that
far the most entanglement is stored in higher tangles (yet to be quantified).
This indication recently found support in Ref.~\cite{Vidal02},
where the scaling of the von Neumann entropy of a compact block of spins
with the size of the block was studied.

%% file: dynamics.bbl
\begin{thebibliography}{100}
\bibitem{footnote1}We would like to point out
    that the time dependence of the correlation functions calculated in
    Ref.~\protect{\cite{Mccoy70}} was due to a step-like time dependence in
    $\lambda$ and the system was taken to be initially in equilibrium.
\bibitem{footnote2}
    A further  peculiarity of Eq.(\protect{\ref{correlators}}) as 
    compared to the
    calculations done in Refs.~\protect{\cite{Lieb61,Pfeuty70,Mccoy70}}
    is that in the present case the Wick theorem leads to time 
    dependent Pfaffians.
\bibitem{note:A-entropy} The entropy of 
	a block of $L$ spins and the scaling of it as a function of $L$
	was discussed in Ref.~\cite{Vidal02}.
	Following the line of thought in Ref.~\cite{Coffman00},
	this quantity contained the more different 
	types of entanglement the larger is $L$.
	We consider the case where $A$ contains only a single site
	in order to study the conjecture in Ref.~\cite{Coffman00}.
\bibitem{FOOTNOTETANGLE}
    This relation can be found by notng the relation between the
    tangle and one of the eigenvalues $\lambda$ of $\rho_1$
        $
    \lambda=:=\bigfrac{1}{2}\left(1+\sqrt{1-\tau_1[\rho_1]}\right)
    $.
\bibitem{note:noindex} We omit the indices $n$ and $m$ here, indicating that
    the considered Hilbert space consists of that two sites only.
\bibitem{remark:QPT}
    This is in contrast to what
    should be observed, if the damping was related to the
    quantum critical point: it should depend on the size of the
    system in a way prescribed by the critical exponents of the system.
\end{thebibliography}
